\documentclass[aps,prd,10pt,onecolumn,nofootinbib,showpacs,showkeys,notitlepage]{revtex4-1}

\usepackage{graphicx,bm}
\usepackage{amsmath}
\usepackage{hyperref}
\usepackage{amssymb}
\usepackage{caption}
\usepackage{subcaption}
\usepackage{slashed}
\usepackage{color}
\usepackage{soul}
\usepackage{amsfonts}

\newcommand{\beq}{\begin{equation}}
\newcommand{\eeq}{\end{equation}}
\newcommand{\bea}{\begin{eqnarray}}
\newcommand{\eea}{\end{eqnarray}}

\newcommand{\nn}{\nonumber\\}

\renewcommand{\d}{\partial}

\newcommand{\sgn}{\mathop{\textrm{sgn}}}

\newcommand \be {\begin{equation}}
\newcommand \ee {\end{equation}}
\newcommand \bed {\begin{displaymath}}
\newcommand \eed {\end{displaymath}}

\newcommand{\bit}{\begin{itemize}}
\newcommand{\eit}{\end{itemize}}

\def\eq#1{(\ref{#1})}
\def\s0#1#2{\mbox{\small{$ \frac{#1}{#2} $}}}
\def\0#1#2{\frac{#1}{#2}}

\def\mr#1{{\mathrm{#1}}}

\def\eq#1{(\ref{#1})}

\begin{document}

\title{Vanishing beta function curves from the functional renormalisation group} 
\author{P. Mati}

\affiliation{MTA-DE Particle Physics Research Group, P.O.Box 51, H-4001 Debrecen, Hungary}
\affiliation{Budapest University of Technology and Economics, H-1111 Budapest, Hungary}
\affiliation{E\"otv\"os University, H-1117 Budapest, Hungary}
\email{matip@edu.bme.hu, matipeti@gmail.com} 

\begin{abstract}
In this paper we will discuss the derivation of the so-called vanishing beta function curves which can be used to explore the fixed point structure
of the theory under consideration. This can be applied to the O($N$) symmetric theories, essentially, for arbitrary dimensions ($D$) and field
component ($N$). We will show the restoration of the Mermin-Wagner theorem for theories defined in $D\leq2$ and the presence of the Wilson-Fisher fixed point in $2<D<4$. Triviality is found  in $D>4$. Interestingly, one needs to make an excursion to the complex plane to see the triviality of
the four-dimensional O($N$) theories. The large\,-$N$ analysis shows a new fixed point candidate in $4<D<6$ dimensions which turns out to define
an unbounded fixed point potential supporting the recent results by R. Percacci and G. P. Vacca in: "Are there scaling solutions in the O($N$)-models for large\,-$N$ in $D>4$?" [Phys. Rev. D 90, 107702 (2014)].
\end{abstract}


\maketitle

%\flushbottom
%\tableofcontents

%------------------------------------------------------------------ 
\section{Introduction} 
\label{intro}
%------------------------------------------------------------------ 
There have been many investigations on the phase structure of the scalar O($N$) models
\cite{o(n),scheme_o(n),pms_o(n),bmw_o(n),low_d_o(n)} 
using functional renormalisation group (FRG) techniques \cite{We1993,tetra1,Mo1994,rg} in various dimensions. Critical exponents 
were computed with high precision, fixed points were found, and many physical aspects seem to be well understood. Hence, we can 
safely state: the FRG method is a reliable tool for non-perturbative calculations which can be used to consider the phase structure 
of a model and consequently to study e.g. the appearance of spontaneous symmetry breaking (SSB). For instance, according to 
the Mermin-Wagner (MW) theorem \cite{Mermin66,Hohenberg67,Coleman73} a continuous symmetry cannot be broken spontaneously 
below a lower critical dimension, which is $D=2$ for the O($N$) symmetric  theory for $N\geq2$.
This has been recently 
shown in the framework of FRG \cite{ssb}. Numerical evidence for the MW theorem is presented in \cite{cog}. 
The only exception is in $D=2$  for $N=2$ (the XY model), when an infinite order phase transition is 
present \cite{ktb,ktb_1}. That is called the Berezinskii-Kosterlitz-Thouless (BTK) phase 
transition. To detect this kind of symmetry breaking the local potential approximation (LPA) is not sensitive enough \cite{ssb}. Throughout this paper we will not consider the special case when $N=2$ in $D=2$ for these aforementioned reasons. Beyond LPA the BTK phase transition can be observed in the framework of FRG \cite{frg_ktb_1,frg_ktb_2,frg_ktb_3}.
In summary, we can say that the O($N$) scalar
model represents an excellent playground for testing new techniques.
The goal of this paper is to introduce the so-called vanishing beta function (VBF) method in the framework of the O($N$) 
symmetric scalar field theory, where the Wetterich FRG equation \cite{We1993,tetra1} is used in the LPA assuming that the potential is analytic around the vanishing field. There is an extensive study about the 
radius of convergence of the Taylor-expanded effective potential in the large\,-$N$ limit of the O($N$) model in \cite{Edo}. Expanding 
around vanishing field enables us to provide a one-parameter representation of the fixed point solution which actually defines 
the VBF curves through the couplings. This single parameter is the quadratic coupling of the theory ($m^2$) and the roots of the VBF curve will define the position
of the fixed points at a given truncation level. In this approach we will lay down general requirements that a root must satisfy in order to define a stable
fixed point potential. Using this polynomial expression for the couplings will reveal an interesting root structure, which depends highly on
dimensionality but there is a dependence on $N$, too, which is rather quantitative.
As a consequence of the approximations used in the VBF technique could lead us to potentially artificial fixed points even in the
higher order of truncation. However, it can be shown that these "spurious" fixed point solutions are excluded by taking the infinite limit in the 
order of the expansion, although strictly mathematically they must be considered as a valid fixed point at finite truncation.
We will show, from a different aspect that has been used in \cite{ssb} and \cite{cog}, that the
Mermin-Wagner theorem is not violated, although at finite truncation order one would draw the opposite conclusion. An equally interesting finding is
the direct observation of triviality above four Euclidean dimensions; however for $D=4$ triviality can be found only if we track down the root
structure on the complex plane, too. In theories $4<D<6$ a new fixed point candidate is found in the large\,-$N$ limit, a detailed analysis is shown
for the $D=5$ case. However, it turns out that this fixed point defines an unbounded
potential from below in agreement with the findings in \cite{Percacci}.\\
In the last two sections we will analyse the $N$ dependence and some fractal-dimensional results are presented, too.

%%%%%%%%%%%%%%%%Subsection%%%%%%%%%%%%%%%%%%
 
\subsection{The functional renormalisation group}
The effective average action of an O($N$) symmetric scalar field theory in $D$ Euclidean dimensions using the LPA 
approximation will have the following form:
\begin{equation}\label{Gamma}
\Gamma_k=\int d^D x \left[ \frac{1}{2}(\partial\bar\phi)^2 + U_{k} (\bar\phi^2) \right].\end{equation}
$U_k$ is the dimensionful potential that depends only on the O(N) invariant terms of the theory, namely on $\bar\phi^2$, where $\bar\phi$ is the
vacuum expectation value (VEV) of the field. The subscript $k$ stands for the RG scale (i.e. the Wilsonian cutoff), on which the effective theory is
being defined. The dependence of Eq.~(\ref{Gamma}) on the RG scale parameter $k$ is governed by an exact functional differential equation
\cite{RiWe1990,We1993,tetra1,Mo1994}
\begin{equation}\label{flow}
\partial_t\Gamma_k=\frac{1}{2}\text{Tr}\left(\Gamma _k ^{(2)} + R_k \right)^{-1} \partial_t R_k. \end{equation}
Here, we have introduced the logarithmic flow parameter $t=\ln ({k}/{\Lambda})$, and a momentum dependent regulating function $R_k(q^2)$,
which ensures that only the fluctuations above the Wilsonian cutoff scale are being integrated out. The $\Lambda$ is the momentum scale on
which our theory was initially defined. $\Gamma _k^{(2)} [\bar\phi]$ is a shorthand notation for the second functional derivative 
$\delta^2\Gamma_k/\delta \bar\phi \delta \bar\phi$, and the trace denotes an integration over the momentum. Alternatively one can introduce
a $\bar\rho\equiv\frac{1}{2}\bar\phi^2$, and from now on we are going to use this as our variable.
Let us note that in the LPA, the two-dimensional Wegner-Houghton RG equation (i.e. the sharp cutoff regulator) is mathematically 
equivalent (see e.g. \cite{sg_scheme}) to the effective average action RG equation \cite{RiWe1990,We1993,tetra1} with the power-law regulator
$R_k(q^2)\equiv q^2 (q^2/k^2)^{-b}$ \cite{Mo1994} with $b=1$ and the functional Callan-Symanzik RG equation \cite{internal}. By inserting  Eq.~(\ref{Gamma}) into Eq.~(\ref{flow}) one will obtain the flow equation for the effective potential
\begin{eqnarray}\label{loop}
\partial_t U_k&=& \frac{1}{2}\int \frac{d^D q}{2\pi^D} \partial_t R_k\left(\frac{N-1}{q^2+R_k+U'_k}+\frac{1}{q^2+R_k+U'_k+2\bar\rho U_k''}\right).
\end{eqnarray}
This equation on the right-hand side defines a loop-integral structure with the propagators of the $N-1$ Goldstone modes and the single 
massive radial mode. The integration by the momentum $q$ can be performed by choosing the $R_k(q^2)$ regulator function in such a way that
it satisfies all the following requirements: $\Gamma_k$ approaches the bare action in the limit $k\rightarrow\Lambda$ and the full 
quantum effective action when $k\rightarrow 0$ \cite{rg}. Various types of regulator functions can be chosen, but a more 
general choice is the so-called Compactly Supported Smooth regulator (CSS) \cite{css,css_pms,css_sg} which recovers all major types of regulators in its appropriate limits.
By using a particular normalisation \cite{css_sg,css_pms} and the notation $y=q^2/k^2$, the dimensionless CSS regulator ($r_k(q^2)\equiv
R_k(q^2)/q^2$) has the following form
\begin{align}
\label{css_norm}
r_{\mr{css}}^{\mr{norm}}(y) =& \;
\frac{\exp[\ln(2) c]-1}{\exp\left[\frac{\ln(2) c y^{b}}{1 -h y^{b}}\right] -1}  
\theta(1-h y^b), 
\end{align}
with the Heaviside step function $\theta(y)$ where the limits are
\begin{subequations}
\label{css_norm_limits}
\begin{align}
\label{opt_lim}
\lim_{c\to 0,h\to 1} r_{\mr{css}}^{\mr{norm}} = & \;
\left(\frac{1}{y^b} -1\right) \theta(1-y^b), \\
\label{pow_lim}
\lim_{c\to 0, h \to 0} r_{\mr{css}}^{\mr{norm}} = & \;
\frac{1}{y^b}, \\ 
\label{exp_lim}
\lim_{c \to 1, h \to 0} r_{\mr{css}}^{\mr{norm}} = & \;
\frac{1}{\exp[\ln(2) y^b]-1}.
\end{align}
\end{subequations}
Thus, the CSS regulator has indeed the property to recover all major types of regulators: 
the Litim \cite{opt_rg}, the power-law \cite{Mo1994} and the exponential \cite{We1993,tetra1} ones. 
Let us note that in the LPA the momentum integral of the RG equation can be performed 
analytically in some cases, e.g. by using the Litim, the sharp, and the $b=1,2$ power-law cutoffs. By using the optimised (or Litim) regulator \cite{opt_rg}, $R_{k}(q^2)=(k^2-q^2)\,\theta(k^2-q^2)$, one can evaluate the integral in Eq.~(\ref{loop})
analytically. In the present paper we are always going to use this regulator. In the meantime we introduce the following dimensionless variables in order to be able to extract the fixed point structure of the
Wetterich equation Eq.~(\ref{flow}):
\begin{equation}
\begin{array}{rl}
u(\rho)&=U/k^D,\\
\rho&=\frac12 \bar\phi^2\,k^{2-D}.
\end{array}
\end{equation} 
At the end of the procedure one will obtain the flow equation for the dimensionless effective potential in $D$ Euclidean dimensions:
\begin{eqnarray}\label{FeE}
\partial_t u&=& - D u +(D-2)\rho u'+(N-1) \frac{A_D}{1+u'} + \frac{A_D}{1+u'+2\rho u ''},\\
A_D& =& \frac{1}{2^{D-1}\pi^{D/2}\Gamma(D/2) D}.\nonumber
\end{eqnarray}
The first two terms arise due to the canonical dimension of the potential and of the fields, whereas the third and fourth terms are a consequence of
the fluctuations of the $N-1$ Goldstone modes and the radial mode, respectively. The numerical factor $A_D$ comes from the angular integration
of the $D$-dimensional integral in Eq.~(\ref{loop}), and it could be absorbed by using a rescaling $\rho\to \rho/A_D$ and $u\to u/A_D$, but we will
choose not to do that, to show the explicit dimension dependence. ($A_D$ can be set to 1 at any point.) In the following section we are going to
study the scaling solutions of Eq.~(\ref{FeE}).

%%%%%%%%%%% NEW SECTION %%%%%%%%%%%%%%%%%%%%%%%%%%%%%%%%%%%%%%%%%%%%%%%%%%%%%

\section{The vanishing beta function curves}\label{vbfc}
In the following, we are going to use the most common ansatz for solving Eq.~(\ref{FeE}). Namely, it
is the Taylor expansion of the effective potential in the field variable. 
First we are assuming that we can expand the potential in a Taylor series around vanishing field. That is:
\begin{eqnarray}
u(\rho)=\lim\limits_{n\to\infty}\sum\limits_{i=1}^{n} \frac{u^{(i)}}{i!}\rho^i.
\end{eqnarray}
For the sake of simplicity we are going to use the following notations for the coefficients $\lambda_i\equiv u^{(i)}(0)$. The scale dependence is encoded into these dimensionless couplings. Most of the time we are
also going to use for the quadratic coupling $\lambda_1\equiv m^2$, sometimes for the quartic coupling $\lambda_2\equiv \lambda$ and 
$\lambda_3\equiv \tau$ for the sextic coupling. Keeping this in mind, we can look at the flow equation of the effective potential and differentiate it
once, then evaluate it at $\rho=0$. So, we get the flow equation for the mass
\begin{eqnarray}\label{mflow}
\partial_t m^2=(\text{D}-2) m^2-\text{D} m^2-\frac{3 \lambda  A_D}{\left(1+m^2\right)^2}-\frac{(N-1) \lambda  A_D}{\left(1+m^2\right)^2}.
\end{eqnarray}
If we are looking for the scale independent solutions (i.e. the fixed point solutions) of this partial differential equation, one can take $\partial_t
m^2=0$. By doing this, we can express $\lambda$ by using only the mass term
\begin{eqnarray}\label{lamgen}
\lambda=-\frac{2 m^2 \left(1+m^2\right)^2}{(2+N) A_D}.
\end{eqnarray}
This curve defines the value of $\lambda$, provided $\partial_t m^2=0$; i.e. this relation is only true when the mass stopped running \cite{ssb, Codello}.
If one does not have quartic coupling, i.e. $\lambda=0$, then the solutions for this equation are just the roots of $\lambda(m^2)$; that is, $m^2=-1$
or $m^2=0$. Now we derive the flow equation for the quartic coupling, too. To do so, we need to perform the same idea as before, but now we
need to differentiate the flow equation of the effective potential twice with respect to $\rho$ and only then evaluate it at $\rho=0$. One will have then
\begin{eqnarray}\label{lflow}
\partial_t\lambda&=&2 (\text{D}-2) \lambda -\text{D} \lambda +A _D\left[\left(\frac{18 \lambda ^2}{\left(1+m^2\right)^3}-\frac{5 \tau }{\left(1+m^2\right)^2}\right)+(N-1) \left(\frac{2 \lambda ^2}{\left(1+m^2\right)^3}-\frac{\tau }{\left(1+m^2\right)^2}\right)\right].
\end{eqnarray}
Again, if we are interested in the fixed point of the equation we need to take $\partial_t \lambda=0$, which enables us to express the sextic
coupling as $\tau=\tau(\lambda,m^2)$. If we are looking for the fixed points of both equations Eq.~(\ref{lflow}) and Eq.~(\ref{mflow}) we can express
the sextic coupling by only using the mass as an explicit parameter: $\tau=\tau(\lambda(m^2),m^2)=\tau(m^2)$. This looks like
\begin{eqnarray}\label{taugen}
\tau=-\frac{2 m^2 \left(1+m^2\right)^3 \left(\text{D} \left(1+m^2\right) (2+N)-4 \left(2+N+2 m^2 (5+N)\right)\right)}{(2+N)^2 (4+N) A_D^2}.
\end{eqnarray}
This function defines the value of $\tau$, provided $\partial_t m^2=0$ \textit{and} $\partial_t \lambda=0$. When we are looking for a fixed point for
the whole system of equations we need to set the lhs of Eq.~(\ref{taugen}) to zero (which sets $\partial_t\tau=0$ automatically), thus
providing the values for $m^2$ where the fixed points are found.   
The general statement is the following: one can express the $n$th coupling simply by using the mass term $m^2$ as an explicit parameter. This
nested formula reads
\begin{eqnarray}\label{nestedf}
\lambda_n=\lambda_n(\lambda_{n-1}(\lambda_{n-2}(...\lambda_2(m^2))),m^2)=\lambda_n(m^2).
\end{eqnarray}
It is straightforward to prove this using induction (see the Appendix). As a consequence, one can find a formula which tells the general form for the $n$th coupling for $n\geq
2$:
\begin{equation}\label{genvbf}
\lambda_n=(-1)^{n+1}\frac{2^{[n/2]}}{A_D^{n-1}\prod\limits_{i=1}^{n-1}\left( 2 i + N \right)^{[(n-1)/i]}}m^2(1+m^2)^n\sum\limits_{i=0}^{n-2}\sum
\limits_{j=0}^{n-2}f_{i,j}(N^\alpha)(m^2)^i D^j.
\end{equation} 
Here the notation $[.]$ means the integer part of its argument and $f_{i,j}(N)$ is an integer valued function of $N^{\alpha}$, where $\alpha$ is an 
integer, too. From Eq.~(\ref{genvbf}) we can conclude that (apart from the prefactor which depends only on $N$ and $D$) $\lambda_n$ is a 
polynomial of $m^2$ over the integers, which has roots at $m^2=0$ and $m^2=-1$ for every $n\geq 2$. Another interesting consideration is that 
$\lambda_2$ is the greatest common divisor of all the $\lambda_n$ VBF polynomials, $n\geq 2$.
Of course, there are many more roots in the complex numbers domain in general  (actually the number of roots is growing with $n$), but we are 
only going to consider the real ones, for which the physics is meaningful.
However, in special cases like in $D = 4$ and $4<D<6$ in the large\,-$N$, one can observe a convergent series of the unphysical roots on the complex plane from which the physically sensible results can be extracted. It is worthwhile to
emphasise that Eq.~(\ref{genvbf}) is the most general form of the 
couplings $\lambda_n(m^2)$; their qualitative behaviour highly depends on the dimensionality, as we are going to see later on. It also depends on 
the number of the fields $N$, of course, but this dependence is rather quantitative.
\par The coupling $\lambda_n=\lambda_n(m^2)$ in Eq.~(\ref{genvbf}) defines a curve in terms of $m^2$ on which $\partial_t \lambda_{n-1}=0$, 
i.e. the beta function of the $(n-1)$th coupling, vanishes. But we know that $\lambda_{n-1}=\lambda_{n-1}(m^2)$ [which we have already in 
hand, otherwise we could not build up $\lambda_n(m^2)$] defines a curve on which $\partial_t\lambda_{n-2}=0$, and we can continue this till
$n=1$. For this reason from now on we will refer to the curves defined by Eq.~(\ref{genvbf}) as the VBF curves.\\
As a next step we would like to extract the fixed points of the theory. The VBF curves of course define the possible values of the couplings for
the fixed points. In order to find a fixed point we need to perform the following procedure. The VBF $\lambda_n$ defines a curve where $\partial_t
\lambda_{n-1}$ vanishes, but it does not say anything about $\partial_t\lambda_n$ itself. The curve which on $\partial_t \lambda_n$ vanishes is
defined by $\lambda_{n+1}=\lambda_{n+1}(m^2)$, and so on. But we need to cut our Taylor expansion at some order, to be able to carry out real
computations. Let us say we truncate it at the $n$th order, but then again we would need the beta function of $\lambda_n$ to be zero, which is encoded in the 
VBF defined by $\lambda_{n+1}$. Since we expanded the effective potential in Taylor series until the $n$th order, clearly we cannot construct that curve. The only 
way to get $\partial_t\lambda_n=0$ is to set $\lambda_{n}(m^2) \equiv 0$ which gives zero for its beta function automatically. In other words, assume that we would like 
to have an $(n-1)$ order expansion, but before we do so we do not set $\lambda_n=0$ for the moment. If $\lambda_n$ were not zero we would need its beta 
function to vanish, too, in order to find a fixed point. But we do not want to compute its beta function, because at the end we are satisfied with an $n-1$ order expansion. 
Since we did not set $\lambda_n$ to zero yet, we can express it through the vanishing beta function of the $(n-1)$th coupling because its fixed point equation has 
the form $\partial_t\lambda_{n-1}=0=F(m^2)+ \lambda_n$, where $F(m^2)$ is a polynomial in $m^2$. Now, $\lambda_n=-F(m^2)$, which just defines its VBF 
curve. Since we did not want to have this term (i.e. the $n$th), we can set $F(m^2)=0$, which is satisfied at its roots $m_0^2$s; hence $\lambda_n=0$, too. Thus, 
although practically we expanded the effective potential till the order $n$ and we were able to construct the VBF curves all the way till order $n$, we must find the roots 
of the $n$th VBF curve to set the $n$th coupling to zero, which would imply by construction that $\partial_t \lambda_n\equiv0$. To make a long story short, we have to 
solve the following equation,
\begin{eqnarray}
\lambda_n(m_0^2)=0,
\end{eqnarray}        
where $m_0^2$ represents the roots of the VBF curve $\lambda_n(m^2)$. We need to evaluate the other $n-1$ VBF curves at these
$m^2=m_0^2$ points to obtain the values of all the dimensionless couplings at the fixed point, and thus define the truncated effective fixed point
potential. So far so good, but let us suppose we find a fixed point potential which is unbounded from below, i.e. one that defines an unstable theory. That obviously 
must be wrong; hence we need to bring into play another restriction: since the potential is a polynomial, the asymptotics (i.e. the
boundedness from below for large field values) depends on the highest degree term in the polynomial. As a consequence, we need to exclude the
$m_0^2$ roots that give $\lambda_{n-1}(m_0^2)<0$ coupling, which is the coefficient of the highest degree non-zero term in the polynomial expansion.
\par In general we can sum up all these requirements in the following way. Let us define the set 
$M=\left\{m_0^2\in \mathbb{R}| \lambda_n (m_0^2)=0\right\}$. 
A stable fixed point effective potential can be found by substituting all ${m_0^2}\in M$ into the $n-1$ VBF curves $\left\{\lambda_{n-1}(m^2),
\lambda_{n-2}(m^2),...,\lambda_{2}(m^2),\lambda_1\equiv m^2\right\}$ provided $\lambda_{n-1}(m_0^2)\geq0$. [If it happens to be zero, too, one
needs to apply this rule for the VBF $\lambda_{n-2}(m^2)$, and so on.] In other words, the set of $m_0^2$'s which defines a true fixed point is
\begin{eqnarray}
M^*=\left\{ m_0^2 \in \mathbb{R} | \lambda_n (m_0^2)=0 \wedge \lambda_{n-1}(m_0^2)\geq 0\right\}.
\end{eqnarray}
In the following, some example are presented individually for $D\leq2$, $2<D<4$ and $D\geq4$, starting with $D\leq 2$.

%%%%%%%%%%%%%%%%%NEW subsection%%%%%%%%%%%%%%%%%%

\section {VBF curves for $D\leq 2$}\label{MW}
\subsection{Continuous symmetries ($N\geq2$)}
The Mermin-Wagner theorem essentially states that no spontaneous breaking of continuous symmetry is present in systems of $D\leq 2$. In \cite{cog} numerical evidence was given in the framework of the FRG that the MW theorem indeed holds; here a beyond-LPA scheme was used, where the wave function renormalisation was taken into account, too.
In \cite{ssb} it was shown, using analytical considerations, that the MW theorem is not violated even at the LPA level. We will verify the MW theorem for the case of the expanded potential (where this statement cannot be seen directly) by applying the rules that have been settled above. Although the title of this section suggests that we include the case of a system with $D=2$ and $N=2$, we will perform our analysis using $D=2$ with $N\geq3$, since in 
the former case an infinite order BTK phase transition is present (\cite{ktb,ktb_1}), as it was mentioned above. 
To detect this kind of symmetry breaking the LPA is not sensitive enough (cf. \cite{ssb}). However, beyond LPA the BTK phase transition can be shown in the framework of FRG \cite{frg_ktb_1,frg_ktb_2,frg_ktb_3}.
\par Using Eq.~(\ref{genvbf}) for $D=1$ and $D=2$ case one can find a simplified expression for the $n$th coupling  
\begin{equation}\label{12dvbf}
\lambda_n \propto(-1)^{n+1} m^2(1+m^2)^n\sum\limits_{i=0}^{n-2}g_{i}(N,D)(m^2)^i.
\end{equation} 
Here we only indicated the polynomial structure in $m^2$, although the prefactors are slightly modified, too. The coefficient functions $g_{i}(N,D)$
are defined as follows:
\begin{eqnarray}\label{l2}
g_{i}(N,D)=\sum\limits_{j=0}^{n-2}f_{i,j}(N^\alpha)D^j.
\end{eqnarray}
Interestingly, when setting $D\leq2$ the finding is that all the $g_i$'s are positive for every term, at least till the highest order ($n=45$) that has been considered in the expansion; this is the observation. This fact suggests that the roots (which are not complex) must be either negative or zero.
\par
\begin{figure}[h!]
  \centering
      \includegraphics[width=0.6\textwidth]{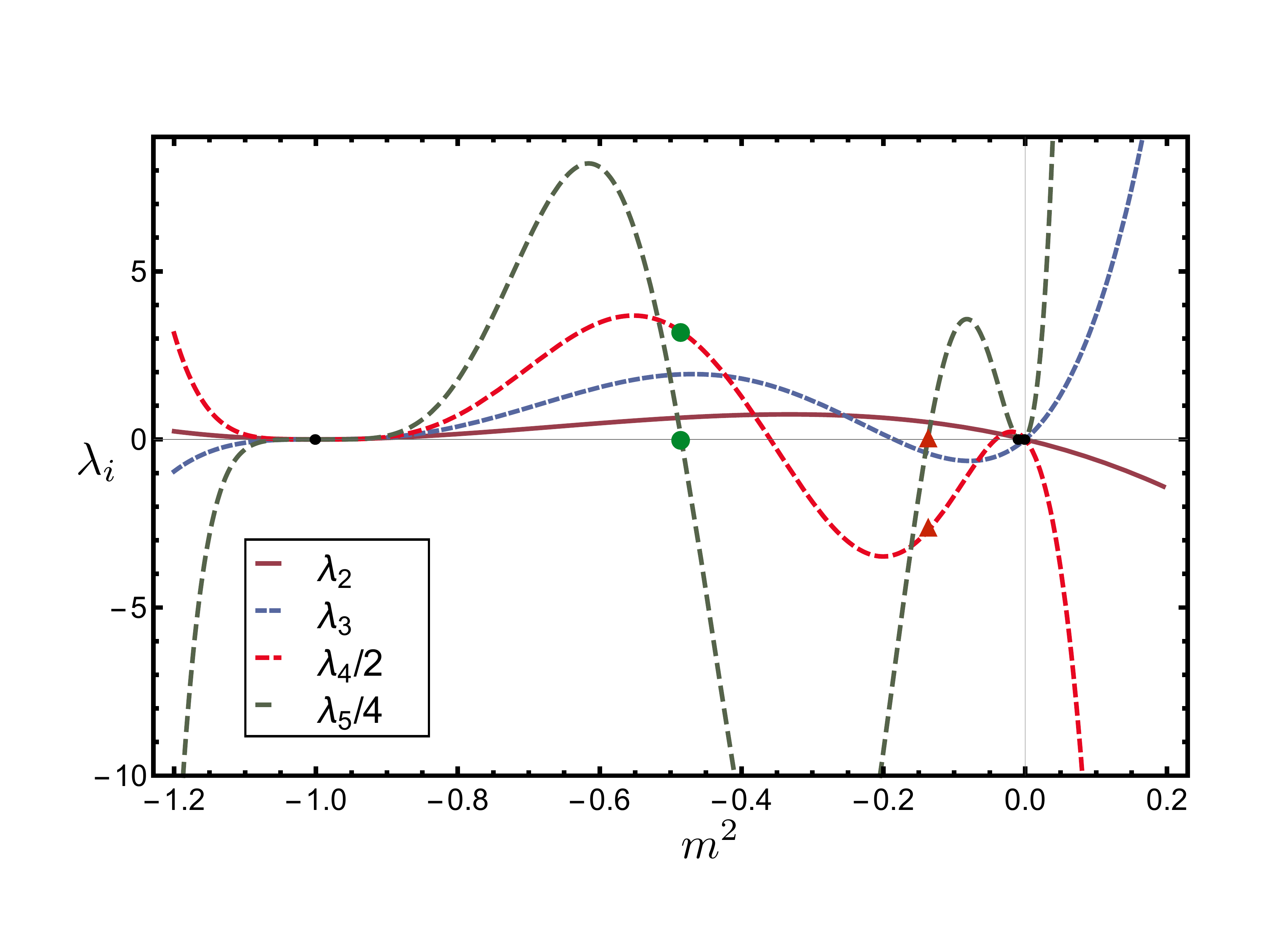}\vspace{-.8cm}
  \caption{The VBF curves for the couplings $\lambda_2,\lambda_3,\lambda_4$ and $\lambda_5$ in the O($3$), $D=2$ model . Each curve is defined by the vanishing beta 
  functions: $\partial_t m^2=\partial_t \lambda_2=\partial_t \lambda_3=\partial_t \lambda_4=0$ respectively. The roots for $\lambda_5(m^2)$ are
  indicated by black dots. These roots are going to define the fixed point potentials at the truncation level $n=4$. Two of its roots are indicated 
  by a green dot and a red triangle to demonstrate a valid and a false fixed point, respectively. At the green dot ($m^2\approx-0.484$) we find that $\lambda_{4}>0$, and evaluating 
  $\lambda_4$ at the red triangle ($m^2=-0.137$) we get $\lambda_4<0$, hence defining an unbounded potential (see Fig.~\ref{pts}). On $\lambda_4$ and
  on $\lambda_5$ a scaling is performed for the sake of the presentation.}\label{VBFs}
\end{figure}

 Let us find the fixed points according to the rule that has been established in the previous section, using $N=3$ and $D=2$. The following
VBF curves are found:
\begin{eqnarray}
\lambda_{2}&=&-\frac{8\pi}{5} m^2 (1+m^2)^2 ,\nn
\lambda_{3}&=&\frac{64\pi^2}{175} m^2 (1+m^2)^3 (5+27 m^2),\nn
\lambda_{4}&=&-\frac{512 \pi ^3} {7875}m^2 (1+m^2)^4 \left(25+670 m^2+1671 (m^2)^2\right),\nn
\lambda_{5}&=&\frac{4096\pi ^4}{606375} m^2 (1+m^2)^5 \left(175+18595 m^2+161115 (m^2)^2+254799 (m^2)^3\right),\nn
...
\end{eqnarray}  
Considering only the real roots, we can find that they are all situated in the interval $[-1,0]$. In Fig.~\ref{VBFs}, one can see the plot of the VBF
curves versus $m^2$. The mere fact that we can find roots is against the Mermin-Wagner theorem, because for that to hold true, we should
have no roots at all, except at the ending points of the interval $[-1,0]$.
%%%%%%%%%%%%%%%%%%%%%%%%%%%%%%%%

%%%%%%%%%%%%%%%%%%%%%%%%%%%%%%%%

We found true fixed point potentials at a finite truncation level according to Fig.~\ref{VBFs}, which should not be there as a consequence of MW theorem. The situation is getting worse when we go to higher order in the truncation: indeed, in this case we are going to have more zeros; hence
more and more fixed point potentials emerge. How can we resolve this contradiction? One way to overcome this situation would be that if all the roots
between $-1$ and $0$ turned into complex valued roots as we went to higher order.\\

%%%%%%%%%%%%%%%%%%%%%%%%%%%%%%%%%
\begin{figure}[h!]
  \centering
      \includegraphics[width=0.4\textwidth]{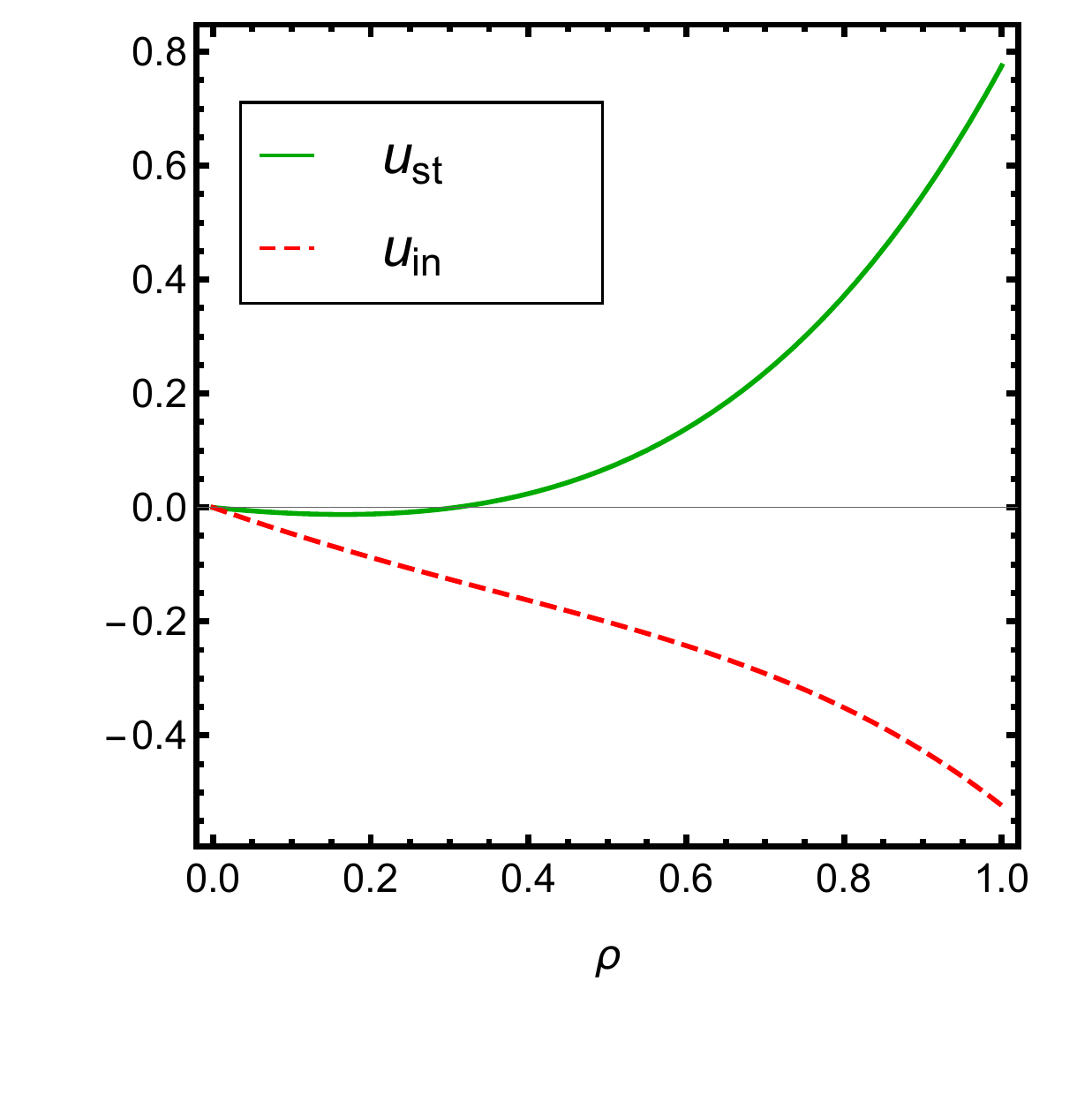}\vspace{-.8cm}
  \caption{The fixed point potentials evaluated at the fixed point discussed in Fig.~\ref{VBFs}. A stable potential can be obtained for $m^2\approx-0.484$
  and an instable one for $m^2=-0.137$, which are indicated on the figure by $u_{st}$, $u_{in}$, respectively.}\label{pts}
\end{figure}
%%%%%%%%%%%%%%%%%%%%%%%%%%%%%%%%%

\begin{figure}[h!]
  \centering
      \includegraphics[width=0.6\textwidth]{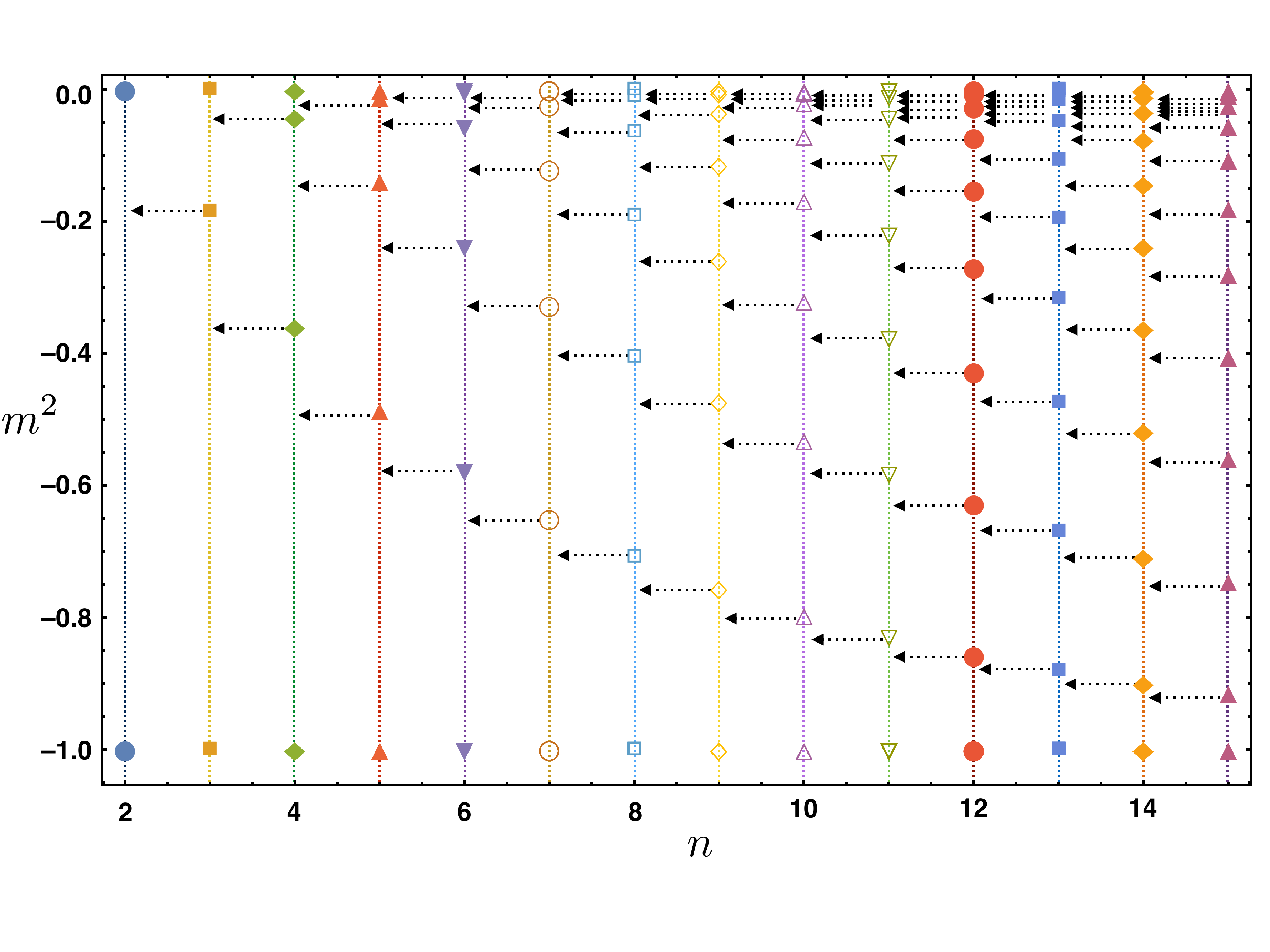}\vspace{-.8cm}
  \caption{The roots of the VBF curves $\lambda_2,...,\lambda_{15}$. Observe the pattern which is given by the following rule: any root of 
  $\lambda_n$ is between the roots of $\lambda_{n-1}$ (except for $-1$ and $0$ of course).}\label{root2d}
\end{figure}

However, from the order of the expansion we used, we cannot put our hope in
this. There is another scenario, too. Let us analyse the VBF curves more carefully.
Apparently, one can find a pattern of the roots for the VBF curves: evaluating the $\lambda_{n-1}$ VBF at each of the roots of $\lambda_{n}$,
one will find real numbers alternating in sign starting from the closest root to $-1$ with $\sgn \lambda_{n-1}(m_1^{2})=+1$ (here the 1 in the
subscript indicates the closest root of $\lambda_n$ to $-1$). For instance in our example in Fig.~\ref{VBFs} that was $m_1^2\approx-0.484$ corresponding to the
stable potential (Fig.~\ref{pts}). 
In other words, from the set $M$ defined above, we will consider only $M\setminus\{-1,0\}$ and the claim is as follows: if we make an ordering from the 
smallest value to the largest in this set, then every odd element of $M\setminus\{-1,0\}$ is in the set of $M^*\setminus\{-1,0\}$, hence for every
such element $\lambda_{n-1}>0$ thus defining a stable potential. So, we are still going to have fixed points even though the number of them is
halved by taking into account only the fixed points which define stable potential. It seems that this procedure does not help too much, but, actually, by doing this,
we can extract some useful information: each root of the $\lambda_{n}$ VBF is surrounded by the roots of the $\lambda_{n-1}$; otherwise the
alternating signs that were explained above could not be possible. So, it means that by deriving higher and higher order VBF curves the position 
and the number of roots change in the way that all of the previous
roots are around the new ones; see Fig.~\ref{root2d}. Let us call this interesting pattern of the set of roots the $M^*$ {\emph{pattern}} for future use. We can do one thing with this without knowing anything about the structure but the root pattern statistics: 
we can simulate a sequence of sets of points which behave in this way.
 
%%%%%%%%%%%%%%%%%%%%%%%%%%%%%%%%
%%%%%%%%%%%%%%%%%%%%%%%%%%%%%%%%
%%%%%%%%%%%%%%%%%%%%%%%%%%%%%%%%%
\begin{figure}
\begin{subfigure}{.54\textwidth}
\centering
  \includegraphics[width=.6\linewidth]{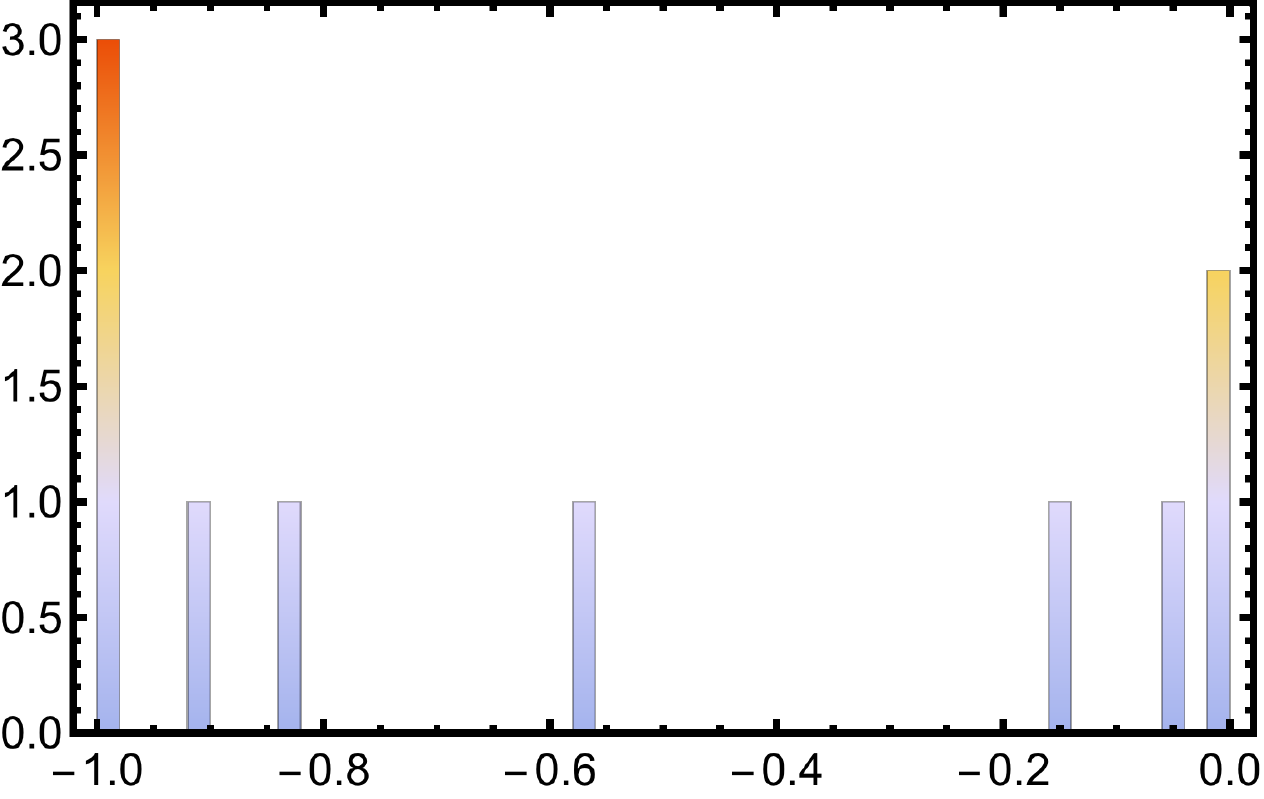}
  \caption{$n=10$}
  \label{fig:sfig1}
\end{subfigure}%
\begin{subfigure}{.54\textwidth}
 \centering
  \includegraphics[width=.6\linewidth]{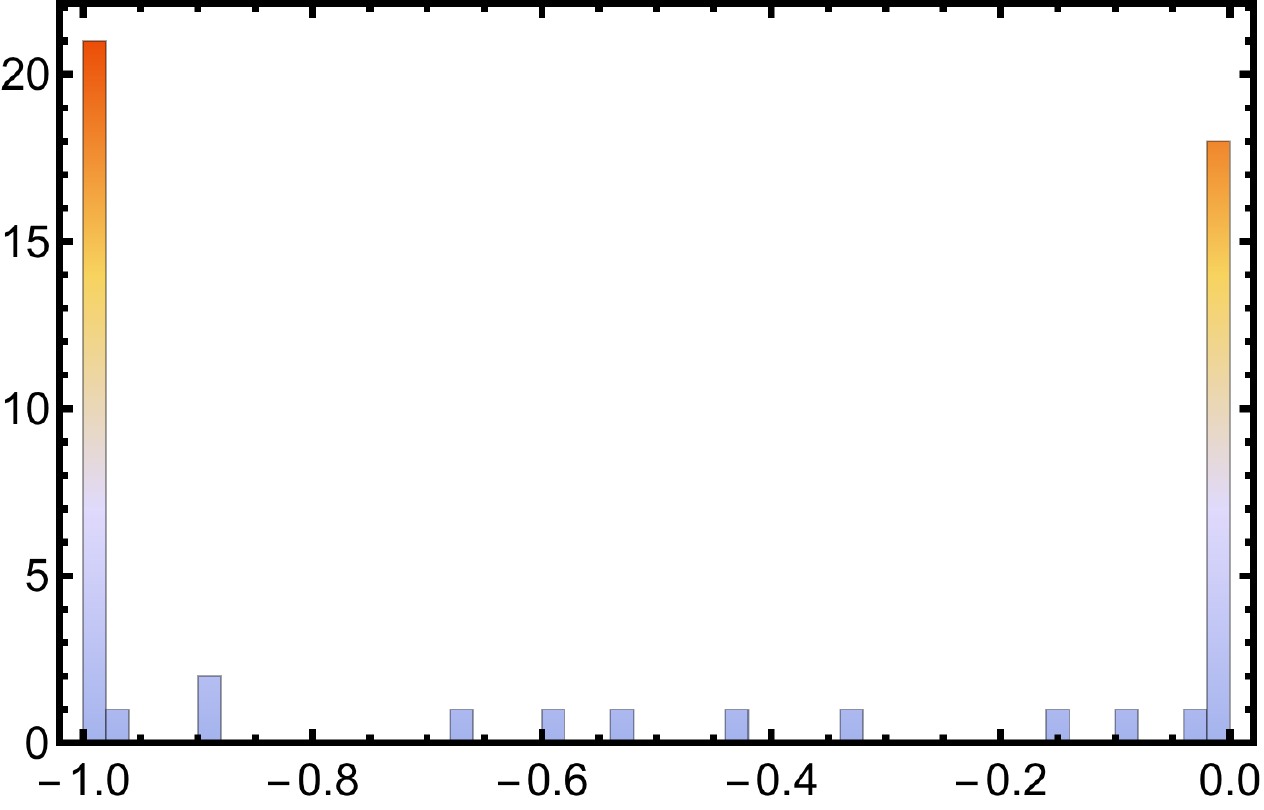}
  \caption{$n=50$}
  \label{fig:sfig2}
\end{subfigure}\vspace{.2cm}
\centering
\begin{subfigure}{.54\textwidth}
\centering
\includegraphics[width=.61\linewidth]{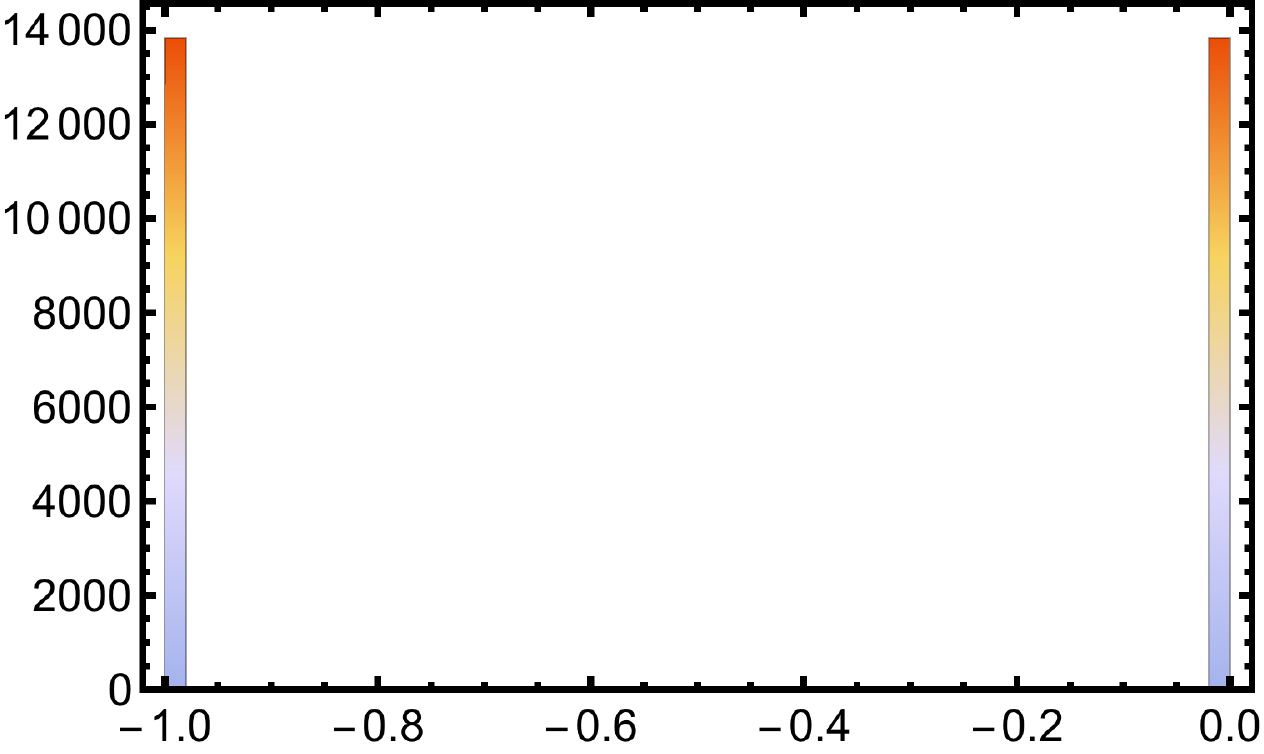} 
  \caption{$n=28000$}
  \label{fig:sfig3}
\end{subfigure}
\caption{Histograms of the distribution of the points generated in $X_n$: (a) the distribution of $X_{10}$, (b) the
distribution of $X_{50}$, and (c) the distribution of $X_{28000}$. Note that the positions of the points are tending to
$-1$ and $0$ and at very high $n$ the number of points between the two end points is negligible; in other words, the probability of finding a point
between the two end points is tending to zero as $n\rightarrow \infty$.}
\label{fig:dis}
\end{figure}
%%%%%%%%%%%%%%%%%%%%%%%%%%%%%%%%%%

\begin{figure}[h!]
  \centering
      \includegraphics[width=0.65\textwidth]{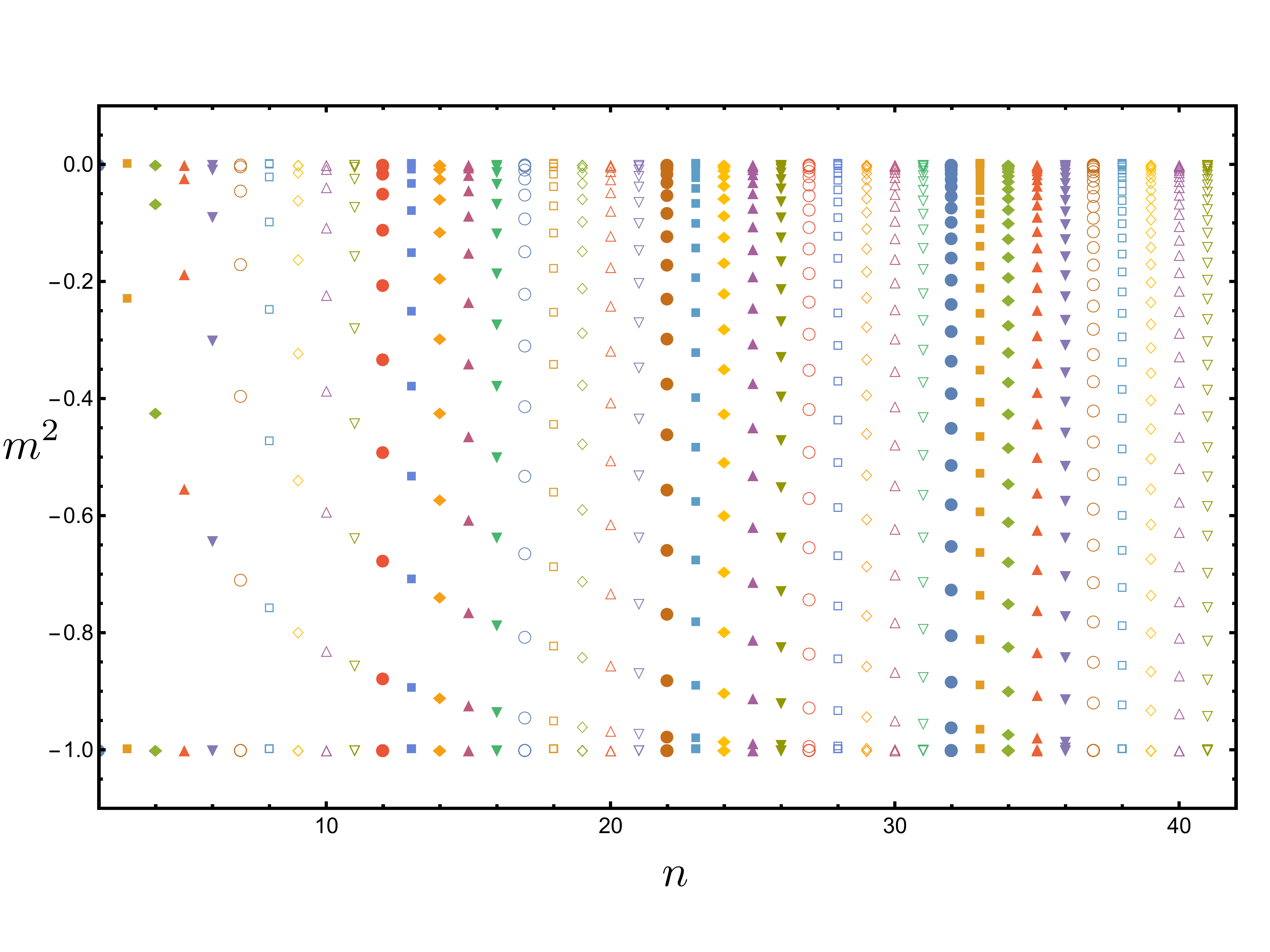}\vspace{-.5cm}
  \caption{Roots of a $D=1$-dimensional theory with $N=3$ field components. A similar structure can be observed as in Fig.~\ref{root2d}.}\label{D1N3}
\end{figure}

Let us consider a randomly generated number $X_1$ which can take a value in the interval $(-1,0)$. We generate this number and then we
consider two new random numbers: $X_2^1$ and $X_2^2$. The first one can take any value in the interval $(-1,X_1)$ and the second in 
$(X_1,0)$. After we generate values for $X_2^1$ and $X_2^2$, they are going to be the new ending points of the intervals where we define again
random numbers but this time three: $X_3^1$, $X_3^2$ and $X_3^3$. We continue this procedure with the random numbers $X_n^i$ with
$i=1,2,...,n$ and $n\rightarrow \infty$, where $n$ indicates that we defined them in the $n$th step in the interval $(X_{n-1}^{i-1},X_{n-1}^{i+1})$,
including $-1$ and $0$, too. By increasing $n$ we can obtain the distribution of the points created in the way described above; this is shown in Fig.~\ref{fig:dis}.
From this construction one can see that the distribution of the randomly generated points is changing in a way that   
they are accumulating at the two ending points of the interval. This suggests that the limit of the probability density for $X_n$ (at least in
distributional sense) is
\begin{equation}\label{dist}
\lim_{n\rightarrow \infty}f_{X_n}(x)=\frac{1}{2}(\delta(x+1)+\delta(x)).
\end{equation}
The aim was to simulate the zeros of the VBF polynomials $\lambda_n$, and we found that the roots are accumulating at $-1$ and $0$. Thus, if there
exists a limit for the VBFs ($\lim_{n\rightarrow \infty} \lambda_n=\lambda$), then this will give zero only at $-1$ and $0$, even if this limit is not a
continuous function. But this is physically well justified since this would indicate that we do not find any roots other than the two ending points of
the interval, which leads us to the conclusion that no SSB is present in dimensions $D=2$ with $N=3$. Indeed, this is a seemingly paradoxical
result, namely that we would expect infinitely many fixed points as the degree of the VBF polynomials grows, but at the end we will find 
two actual fixed points defining only one phase, in agreement with the Mermin-Wagner theorem that we wanted to show. 
We can make another remark on the role of the convexity (or IR) fixed point $m^2=-1$. From the VBFs one can see that every higher coupling
vanishes at this point, i.e. $\lambda_2(-1)=\lambda_3(-1)=...=0$; hence the potential, which is defined by this fixed point, is an unstable one, corresponding to the
dimensionless potential $u=-\rho^2$. One can now speculate whether this fixed point is a real one or if it is just an artifact of the approximations that we use in the FRG. 
Similar results can be obtained for the $D=1$ (see Fig.~\ref{D1N3}). For $N$ dependence and fractional dimensions, see Secs. \ref{Ndepp} and
\ref{fractal}, respectively.
%%%%%%%%%%%%%%%%%%%%%%%%%%%%%%%%
%%%%%%%%%%%%%%%%%%%%%%%%%%%%%%%%

%%%%%%%%%%New subsubsection%%%%%%%%%%%%%%

\subsection{$\mathbb{Z}_2$ symmetry ($N=1$)}
In the case of the discrete symmetry the O($N$) theory is equivalent to the Ising model, and as the symmetry is being non-continuous, the MW theorem
does not necessarily hold for $N=1$. Indeed, it is well known that in the Ising model we can find a spontaneous symmetry breaking even in the
$D=2$ case, which was carried out in an exact calculation by Onsager \cite{Onsager}. Here, we are going to see that we can reproduce this result
using the technique introduced above.

%%%%%%%%%%%%%%%%%%%%%%%%%%%
\begin{figure}[h!]
\centering
\begin{subfigure}{.5\textwidth}
 \centering
  \includegraphics[scale=.25]{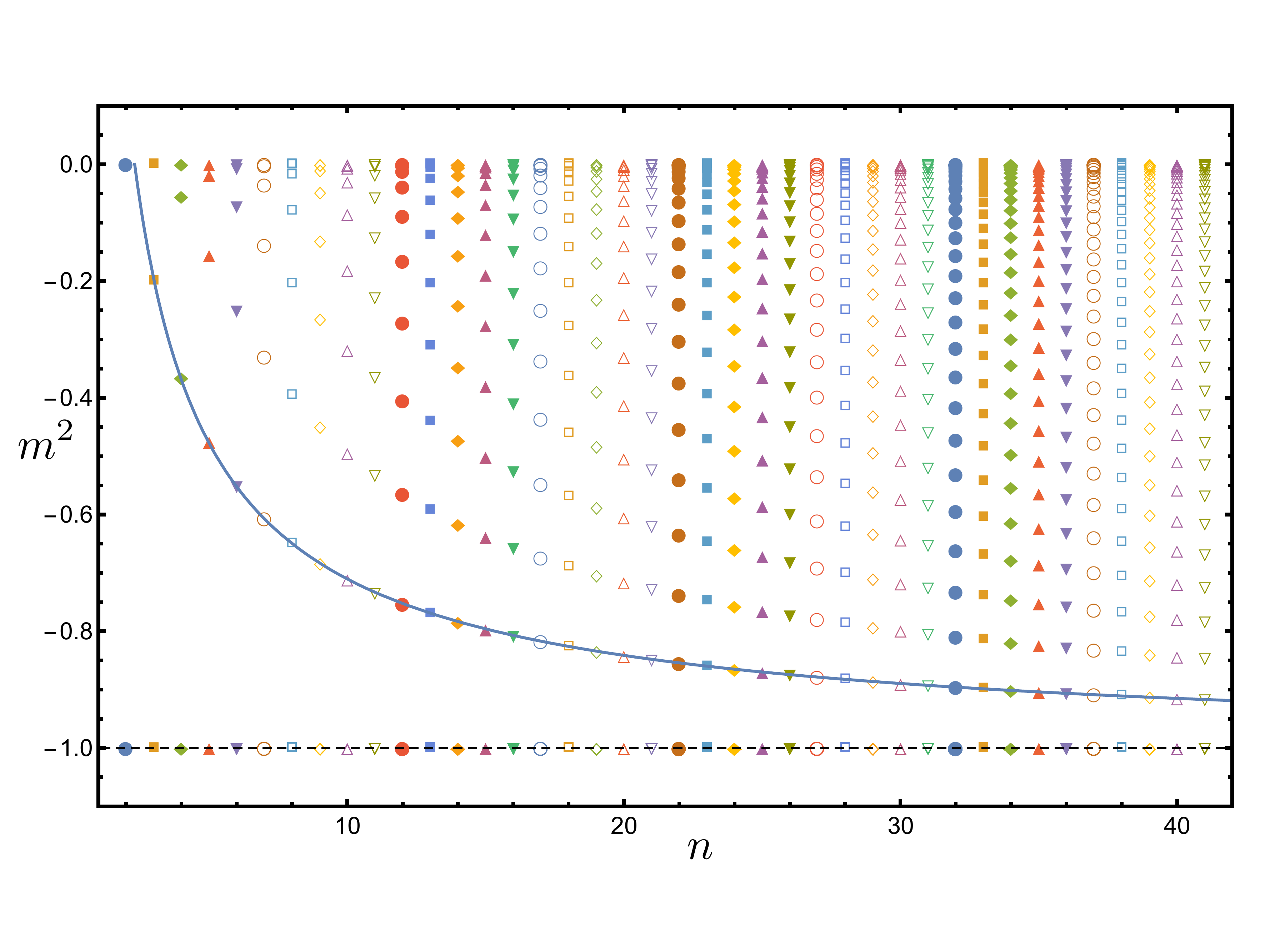}
   \label{fig:sfig21}
   \vspace*{-1cm}
\end{subfigure}
\begin{subfigure}{.5\textwidth}
  \centering
  \includegraphics[scale=.25]{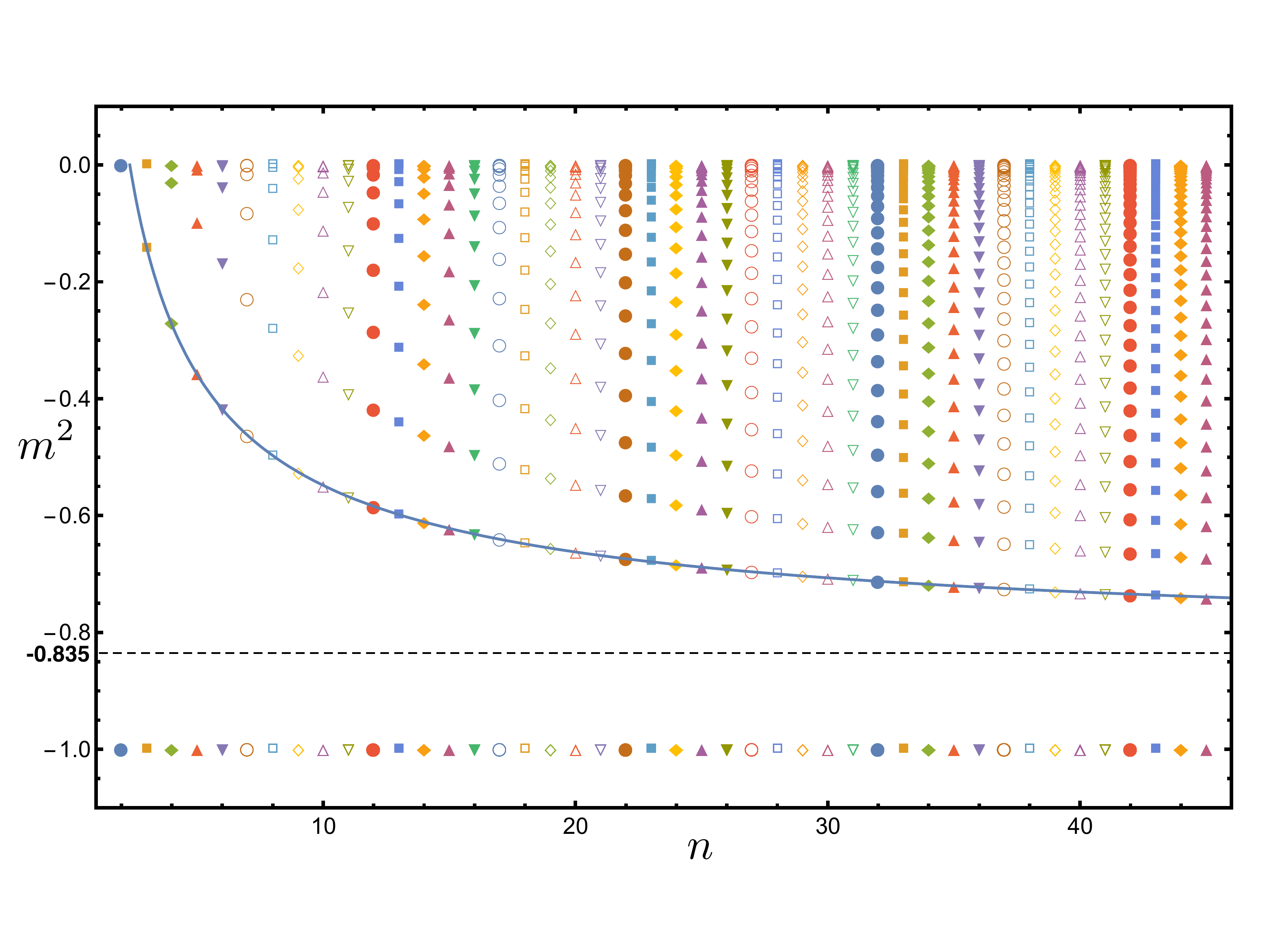}\vspace*{.4cm}
  \end{subfigure}\vspace*{-.5cm}
\caption{In the upper (lower) panel the position of the roots of the VBF curves are shown for $N=1$ in $D=1$ ($D=2$). In the one-dimensional case one finds
the same pattern that was shown above for $N=3$ with a convergence to $m^2=-1$. In two dimensions the roots are converging towards $m^2=-0.835$
signaling a SSB phase.}\label{isingroots} 
\end{figure}
%%%%%%%%%%%%%%%%%%%%%%%%%%%%

In Fig.~\ref{isingroots} we can see the position of the roots for $D=1$ and $D=2$ in a separate plot. The VBF polynomials are described by Eq.
(\ref{12dvbf}) with different $g_i$ coefficients, of course. We can see qualitative difference between the two figures; namely, in the one-dimensional case, 
the first root is converging to $m^2=-1$ and for the two-dimensional case the asymptotic goes to $m^2=-0.835\pm0.001$, defining a threshold for all the roots in higher 
orders. One can fit a function of the form of $f(x)=a+b x^c$ on the positions of the first roots, with fit parameters $a=-1.00\pm0.0002$, $b=2.08\pm0.002$ 
and $c=0.85\pm0.0006$ in $D=1$; and $a=-0.835\pm0.001$, $b=1.56\pm0.004$ and $c=0.738\pm0.003$ in $D=2$. Now, since the root position pattern is 
the same as we observed above, we can apply the random number procedure for $D=1$ in the interval $[-1,0]$ and for $D=2$ in the interval $
[-0.835,0]$. The limiting result is the following: in one dimension the roots are accumulating at $m^2=-1$ and $m^2=0$; thus we can conclude only the 
stable Gaussian fixed point potential is well defined at the latter. In the case of the two dimensions we will find two fixed points, the Gaussian one ($m^2=0$) 
and another one which is at $m^2=-0.835$. Now, it is a question whether this fixed point is stable or not. The finding is the following: for every truncation in the 
present computation (where the highest order was $n=45$) $\lambda_n(m^2=-0.835)>0$, thus we can safely state that it defines a stable fixed point 
potential, hence indicating a spontaneous symmetry breaking phase.
It is interesting that even at LPA level one can observe the SSB for the two-dimensional Ising model contrary to the finding in \cite{ssb}, where even with 
the numerical spike plot technique \cite{Morris94,Codello12} it is undetectable, due to its dependence on the numerical precision.
 
%%%%%%%%%%%%%%%%%NEW SECTION%%%%%%%%%%%%%%%%%%%%%%%%%%%%%%%%%%%%%%%%%%

\section{VBF curves for $2<D<4$}\label{wf}
\begin{figure}[h!]
  \centering
      \includegraphics[width=0.6\textwidth]{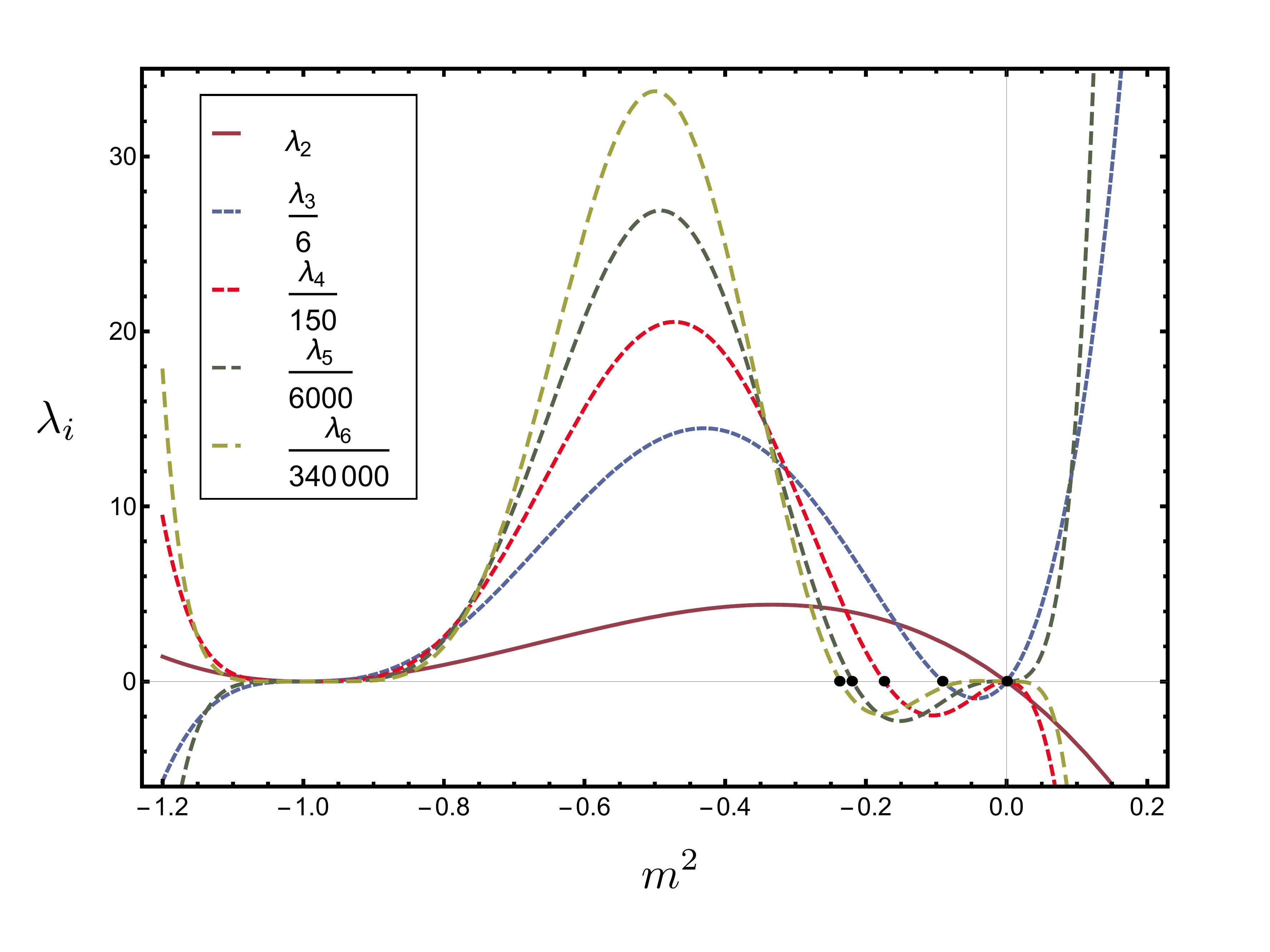}\vspace{-.5cm}
  \caption{In this plot the VBFs up to $\lambda_6$ are shown for $D=3$, $N=2$. For each $\lambda_i$ the root that is closest to $-1$ (from
  above) is indicated with a black dot. Note that if we consider the position of these points as a sequence then they will converge to a finite value
  $m^2=-0.23$; see also Fig.~\ref{3DROOTSEP}.}\label{VBFd3}
  \end{figure}

\begin{figure}[h!]
  \centering
      \includegraphics[width=0.55\textwidth]{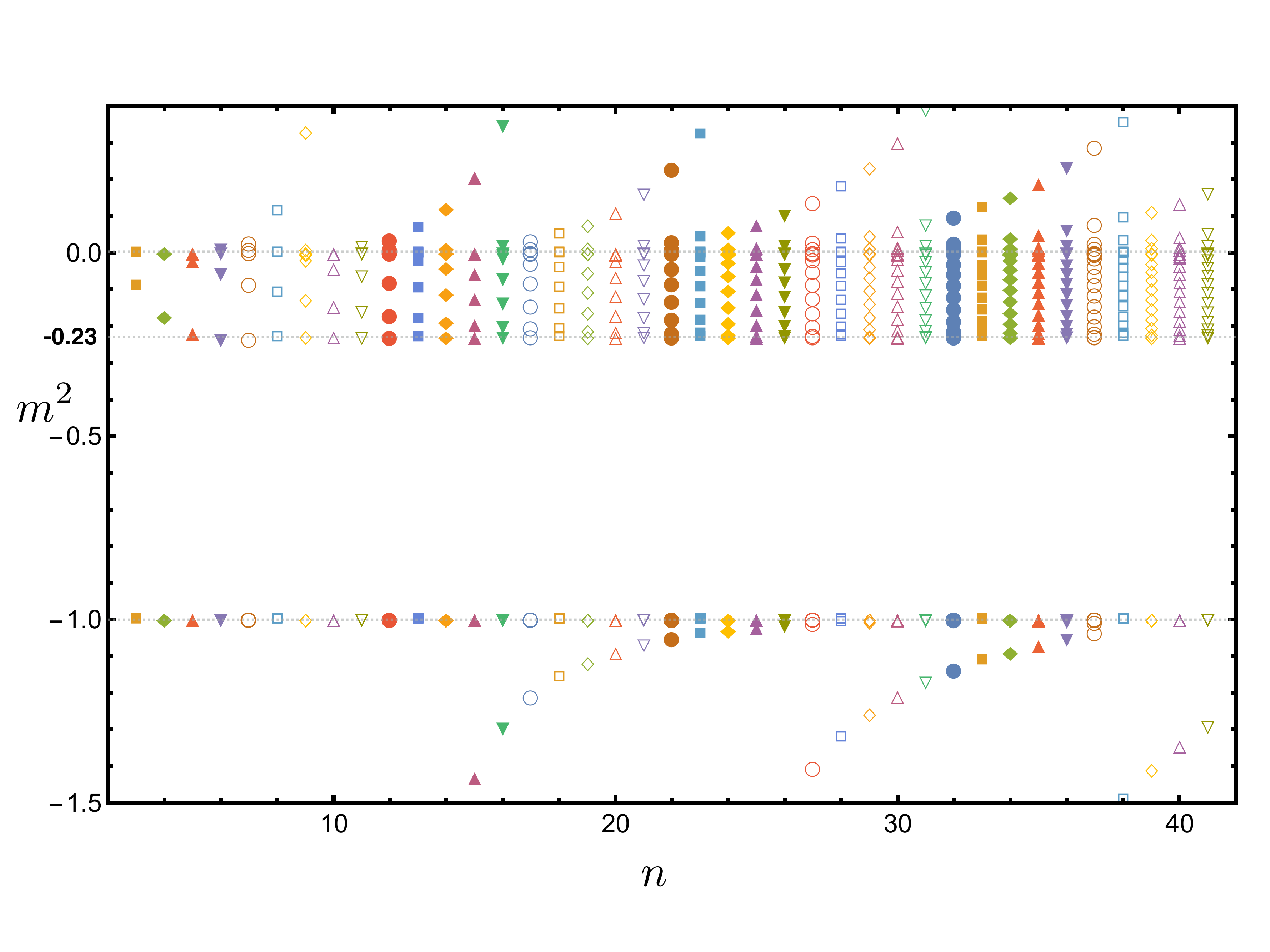}\vspace{-.5cm}
  \caption{Here we present the root structure of the VBF curves up to order $n=41$, in the O(3), $D=3$ model. Observe that we can distinguish three regions where the roots are
positioned: $m^2>0$, $m^2<-1$ and $m^2\in[-1,0]$. For a detailed picture see Fig.~\ref{3DROOTSEP}.}\label{rootsd3}
\end{figure}

The O($N$) models which belong to the class $2<D<4$ have the richest fixed point structure; hence studying them is the most challenging. We are
going to perform a detailed analysis for the only integer dimension found in this interval, $D=3$ restricting ourselves to $N=2$. We expect here to obtain
the well-known Wilson-Fisher fixed point in the broken regime, however, we are going to see that there is no clear root structure to be observed for
the roots $m^2>0$. In this case we have the following generic form for the VBFs in $2<D<4$, similarly to Eq.~(\ref{12dvbf}):
\begin{equation}
\lambda_n \propto(-1)^{n+1} (m^2)^{1+\Theta(n-4)}(1+m^2)^n\sum\limits_{i=0}^{n-2}g_{i}(N,D)(m^2)^{i-\Theta(n-4)}.
\end{equation}
Here, $g_{i}(N,D)$ is defined in a similar way as above in Eq.~(\ref{l2}), but now, as we can see, the exponent of $m^2$ has changed, hence in the sum Eq.
(\ref{l2}) there must be $D^{\alpha(j)}$ rather than $D^j$, where $\alpha(j)$ represents the remaining exponents, after the prefactor of the sum is factored out. The $\theta$ function in the exponent is just the Heaviside step function: $\Theta(n-4)=1$ if $n\geq4$ and $0$ otherwise.
The first few polynomials from the VBFs for $D=3$ and $N=2$ are
\begin{eqnarray}
\lambda_{2}&=&-3 \pi ^2 m^2 (m^2+1)^2,\nn
\lambda_{3}&=&3 \pi ^4 m^2 (m^2+1)^3 (11 m^2+1),\nn
\lambda_{4}&=&-27 \pi ^6 (m^2)^2 (m^2+1)^4 (23 m^2+4),\nn
\lambda_{5}&=&\frac{27 \pi ^8}{5} (m^2)^2 (m^2+1)^5 \left(2993 (m^2)^2+719 m^2+14\right),\nn
\lambda_{6}&=&-\frac{27 \pi ^{10}}{5} (m^2)^2 (m^2+1)^6 \left(97167 (m^2)^3+27418 (m^2)^2+997m^2-14\right),\nn
...
\end{eqnarray}
Contrary to the cases of $D\leq 2$, the coefficients $g_i$ can take now negative values, too. From these considerations one can already expect a
different root structure from that which we had for $D\leq2$. In Fig.~\ref{VBFd3} we see the VBF curves up to $\lambda_6(m^2)$ and in Fig.~\ref{rootsd3} the
root structure up to order $n=41$. For our analysis we are going to separate the real line of $m^2$ into three regions.
First, we will consider the roots in the interval $[-1,0]$; then we will turn to the complement set, separately for the intervals $(-\infty,-1)$ and
$(0,\infty)$.

%%%%%%%%%%%%%%%%%%%%%%%%%%%%%%%
%%%%%%%%%%%%%%%%%%%%%%%%%%%%%%%%

\begin{figure}
\begin{subfigure}{.45\textwidth}
  \centering
  \includegraphics[width=1\linewidth]{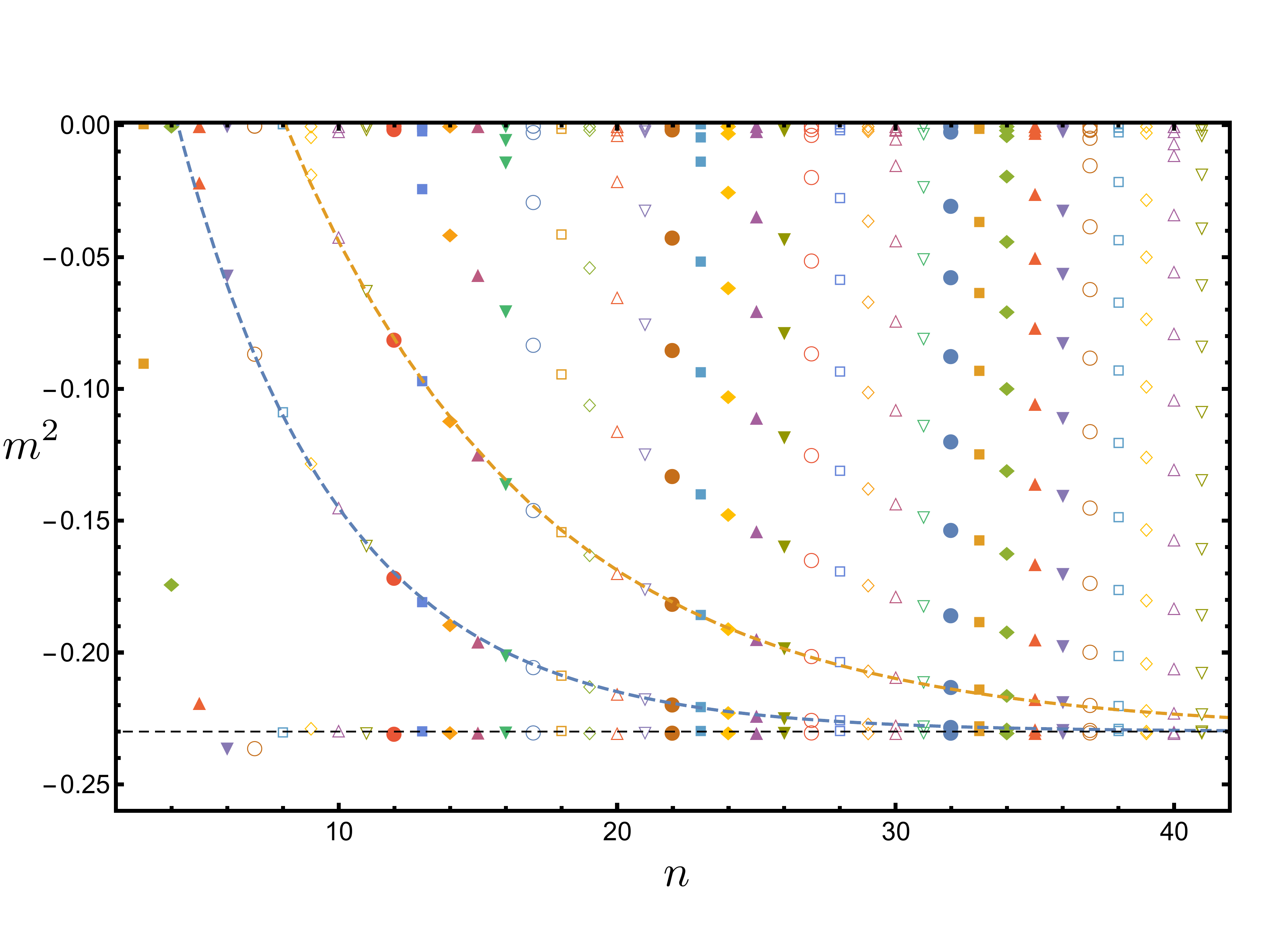}\vspace{-.5cm}
  \caption{The region $-1<m^2<0$.}
  \label{D3N2WF}
\end{subfigure}%
\begin{subfigure}{.45\textwidth}
  \centering
  \includegraphics[width=1\linewidth]{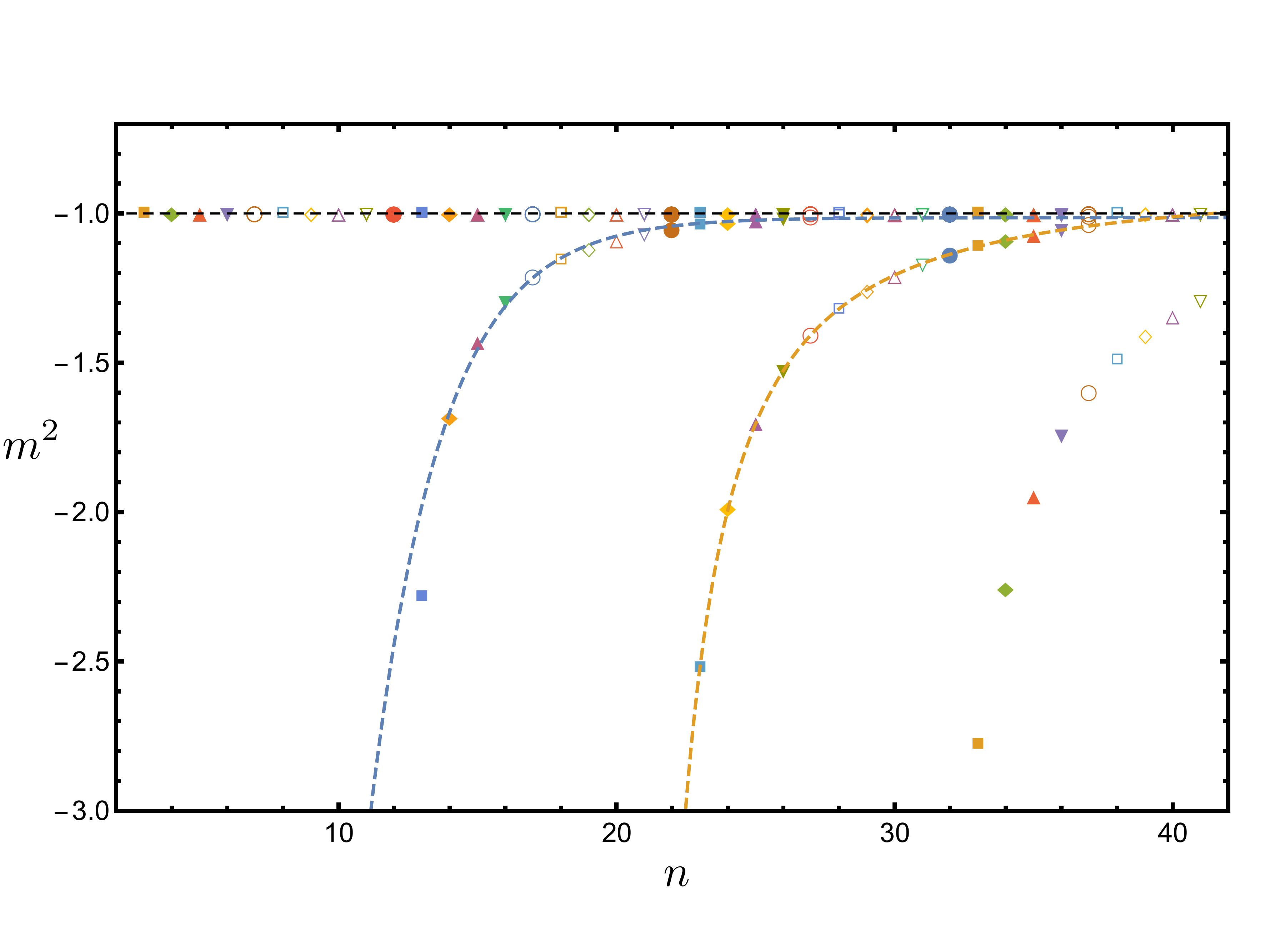}\vspace{-.5cm}
  \caption{The region $m^2<-1$.}
  \label{D3N2CO}
\end{subfigure}
\centering
\begin{subfigure}{.45\textwidth}
  \centering
  \includegraphics[width=1\linewidth]{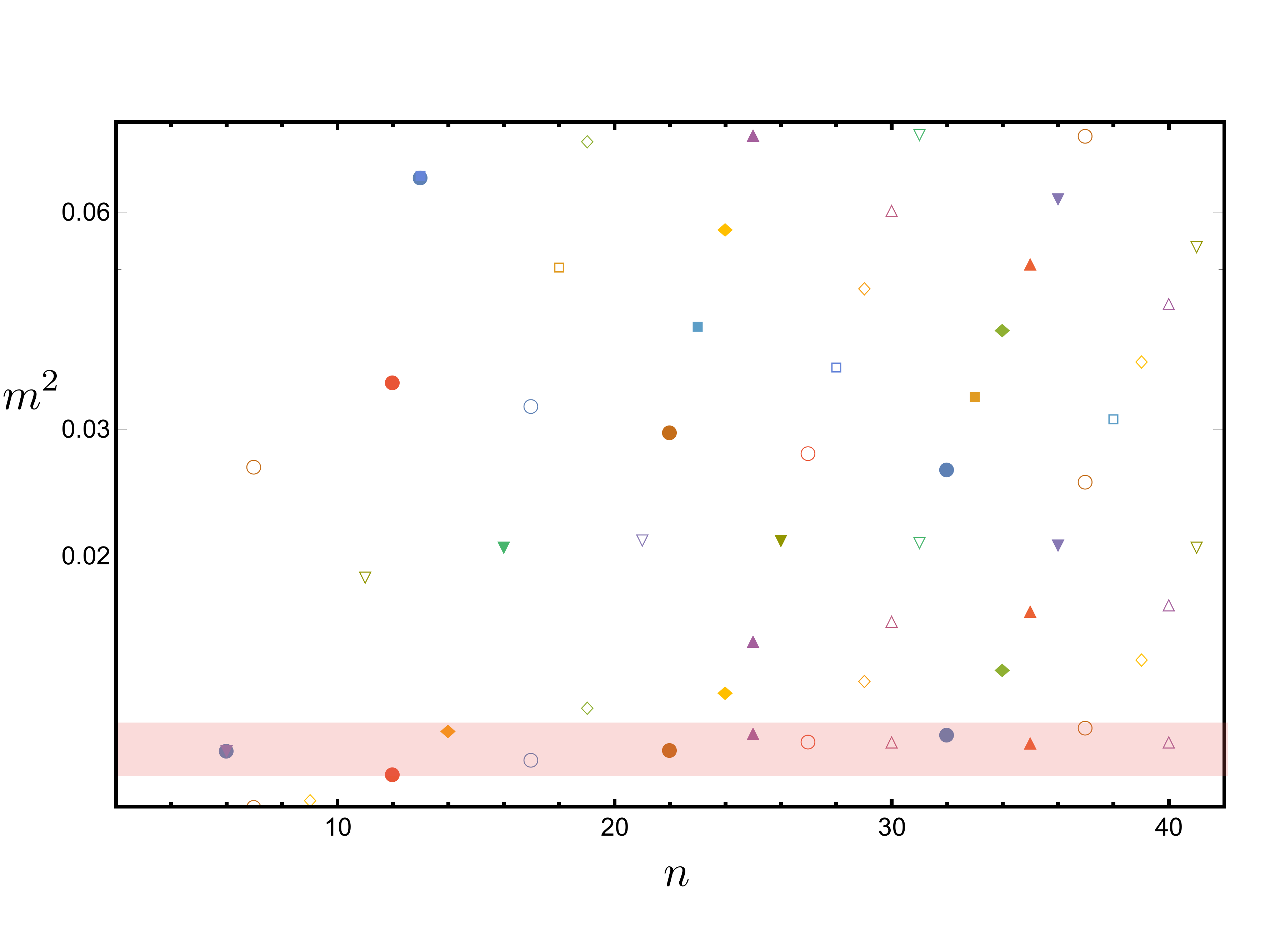}\vspace{-.5cm}
  \caption{The region $m^2>0$}
  \label{D3POS}
\end{subfigure}
\caption{The root structure of the $D=3$, $N=2$ theory separated into different regions. In (a) and (b) we can conclude for the $n\to\infty$ limit; that is,
we will have the following fixed points: Gaussian, Wilson-Fisher and the IR. These suggestions are also confirmed by functions. Regarding (c) for $m^2>0$,
no clear answer is found: the red lane indicates the last real root for $m^2>0$. For details see the text.}
\label{3DROOTSEP}
\end{figure}

From Fig.~\ref{3DROOTSEP} it is clear that in the interval $[-1,0]$ we have a very similar pattern for the roots (above some $n$) to the one that we had in the two-dimensional case (see Fig.~\ref{root2d}). However, there is a striking difference between the $D\leq2$ case and the present $D=3$ theory: the roots seem to be accumulating around the value $m^2\approx-0.23$, rather than $-1$. Although this is of course an approximated value, for the sake of simplicity, 
we will use the equal sign in the following whenever we consider this value: $m^2=-0.23$. If one restricts the pattern to the $[-0.23,0]$ interval, then the random number generating model for the root pattern statistics becomes available again. Since the distribution of the points signals the accumulation at $m^2=0$ and $m^2=-0.23$ the probability
density of finding a root in this interval will have the same form as in Eq.~(\ref{dist}); the only difference is that now the Dirac deltas are centred at $m^2=-0.23$
and $m^2=0$ rather than $m^2=-1$ and $m^2=0$. We can also check the convergence of the roots by establishing an ordering among them: $m^2_1$,
$m^2_2$ and $m^2_3$, where the first indicates the root that is closest, the second is the second closest and the third is the third closest root to $m^2=-1$ at
each order. The root labeled by $m^2_1$ starts to go towards $-1$ but at order $n=6$ it stops and starts to converge to $m^2=-0.23$ in an oscillatory way. 

%%%%%%%%%%%%%%%%%%%%%%%%%%%%%%%%%
%%%%%%%%%%%%%%%%%%%%%%%%%%%%%%%%%

%%%%%%%%%%%%%%%%%%%%%%%%%%%%%%%%%   
\begin{figure}[h!]
  \centering
      \includegraphics[width=0.4\textwidth]{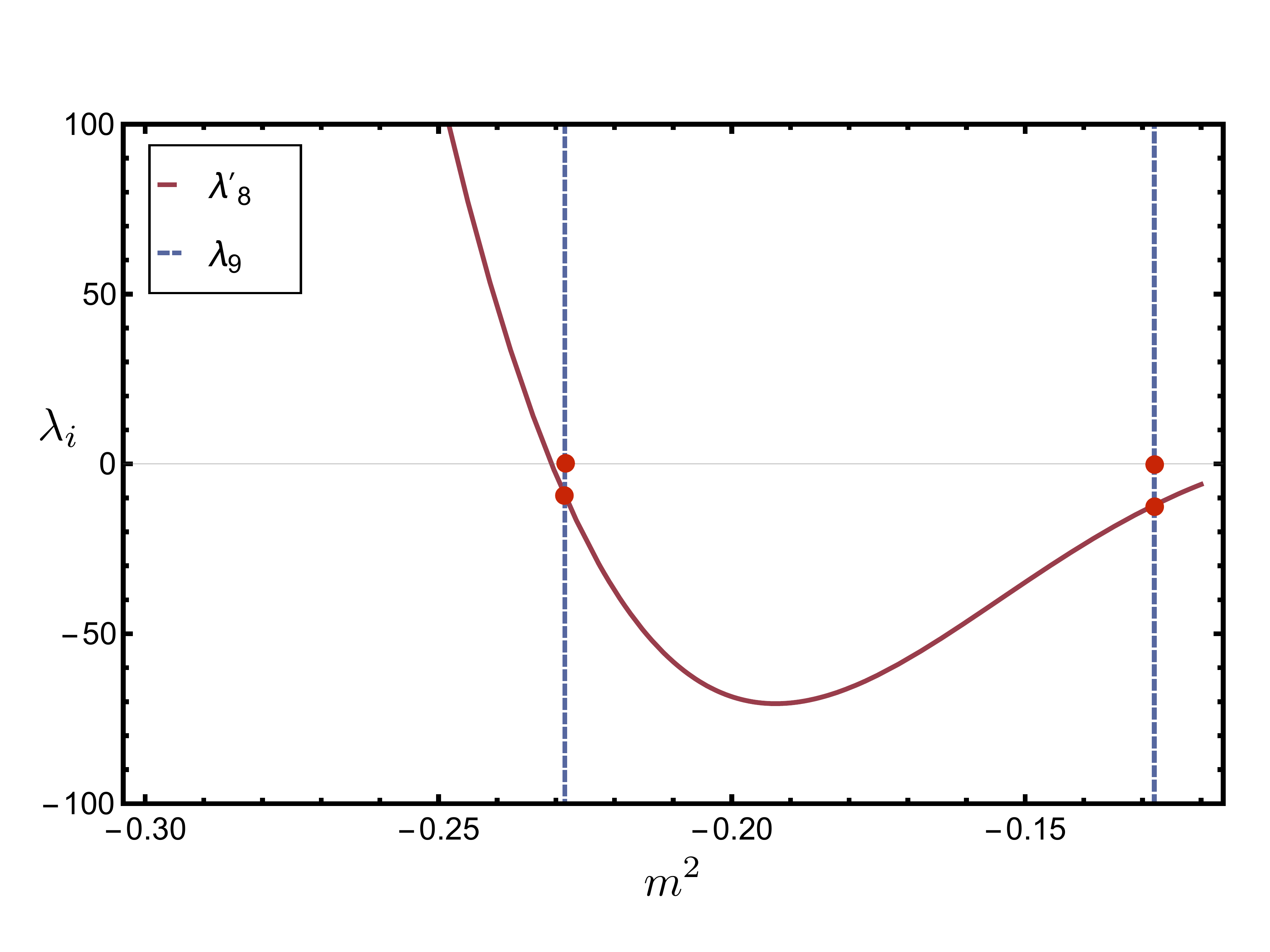}\vspace{-.5cm}
  \caption{The absence of the Wilson-Fisher fixed point at the truncation order $n=8$. We can see that the roots of $\lambda_9$ to the left define a negative
  valued $\lambda_8(m^2_{WF})$, which will provide an unphysical potential. Moreover, the second root to the right does not
  give a sensible potential either. This phenomenon occurs because of the oscillatory nature of $m^2_1$, but at $n\to\infty$ we expect to see a stable
  fixed point potential at the WF fixed point. $\lambda_8$ has been rescaled with a factor of $2\times10^{-7}$.}\label{abs_WF}
  \end{figure}
%%%%%%%%%%%%%%%%%%%%%%%%%%%%%%%%%

%%%%%%%%%%%%%%%%%%%%%%%%%%%%%%%%%
%\vspace*{-.03cm}
%%%%%%%%%%%%%%%%%%%%%%%%%%%%%%%%%%

On the other hand, the roots $m^2_2$, $m^2_3$ also converge to this value, and it is possible to fit a curve on them, but this time the fit is an
exponentially decaying one, as it can be seen from Fig.~\ref{D3N2WF}. The convergence of these roots confirms what we expect from the $M^*$ pattern:
the two roots that will remain in this interval for $n\to\infty$ are $m^2=0$ and $m^2=-0.23$, which correspond to the Gaussian and the Wilson-Fisher fixed point,
respectively. 
One needs to check whether the WF fixed point defines a bounded potential.\\ 
Substituting the value $m^2=-0.23$ into the VBF polynomials 
$\lambda_n$ will give the following result: there are VBF curves for which $\lambda_n(m^2=-0.23)<0$ and for which $\lambda_n(m^2=-0.23)>0$. This is due to the
oscillatory behaviour of the root $m_1^2$, but we know that this will converge to a finite value, and finally the WF fixed point can be defined as the limit of
$m_1^2$. But until this happens, we will find situations which would give the weird result that the Wilson-Fisher fixed point defines an unbounded potential from below; hence,  
strictly speaking, we should not consider it as a physical fixed point. This is the case for example with $\lambda_8$ and $\lambda_9$ (see Fig.~\ref{abs_WF});
hence this tells us that there is no Wilson-Fisher fixed point at the truncation level $n=8$. Even though this is true, we must not forget that the Taylor
expansion is an approximation, so the real fixed point structure of the theory will be found when $n\to\infty$; hence the "absence" of the WF fixed point
can be considered as an artifact of the expansion.
Let us consider now the region $m^2<-1$. A magnified picture of it can be seen in Fig.~\ref{D3N2CO}. Here, we can observe roots which are running into the 
convexity fixed point as $n$ grows. The position of the roots would show a similar pattern to the $M^*$ pattern; however, here we have difficulties with the unbounded 
interval $(-\infty,-1]$. We will show how it is possible to overcome such a situation in the next section; however, there will be some requirements, which are not fulfilled in the 
present case.
Here, we can only assume that all such roots will converge to $m^2=-1$ providing us the IR (or convexity) fixed point in the $n\to\infty$ limit.
For two such roots we can show the convergence by doing a fit. Interestingly, in this case the roots are following a different trend to approach $-1$: the first one is an
exponential function $f(x)=a + c e^{b (x - 12)}$ with $a=-1.014\pm0.005$, $b=-0.39\pm0.01$ and $c=-1.43\pm0.05$; the second is power-law $g(x)=a+c (x -
20)^b$ with $a=-0.885\pm0.003$, $b=1.35\pm0.08$ and $c=-7.2\pm0.06$. Neither of them goes to $-1$ precisely; however, these are the best fits that can be
found. For large $n$ values, when these roots are very close to $-1$, they develop an imaginary part; hence they cannot be considered as true fixed points, strictly
speaking. One has to go so close to $-1$ that this effect can be considered as negligible, taking into account that the VBF curves around $-1$ are extremely flat
thanks to the $(1+m^2)^n$ factor in Eq.~(\ref{genvbf}). Thus, finding roots around $-1$ is not always a reliable thing; it might depend on the precision of the root
finding algorithm, too.\\
The remaining region that we need to consider is the half interval $m^2>0$. The position of the roots for this region is shown in Fig.~\ref{D3POS}. One can find
again some pattern which could remind us of the $M^*$ pattern; however, in this case besides the fact that the interval is unbounded from above, the roots do not heave a clear bound
even from below. The general behaviour of the root positions is that they have the last real value at about $m^2=0.01-0.02$ (indicated by the red lane in Fig.~\ref{D3POS}), 
and below that they will have complex values. Even considering the generated complex roots does not help us to understand this part of the root structure, however, in the case of the four-dimensional O($N$) models we can use these complex roots to capture the real physics, as we will see in Sec. \ref{d4sec}.\\ 
The most important result of this section is the appearance of the Wilson-Fisher fixed point. It may well be possible to find more physical fixed points in other 
fractal dimensions between two and four dimensions for various $N$ values, but it is beyond the scope of the present study.

%%%%%%%%%%%%%%%%%NEW SECTION%%%%%%%%%%

\section{VBF curves for $D\geq 4$}\label{trivi}
So far we considered $D=1,2$ and $3$. In this section we are going to investigate the theories in $D\geq4$. 
Let us look at the first two VBF curves ($\lambda_2$ and $\lambda_3$) in arbitrary dimensions and field components. We already derived their
formulas in Sec. \ref{vbfc}: Eq.~(\ref{lamgen}) and Eq.~(\ref{taugen}). Now, we are interested in the roots of the general expressions:
\begin{eqnarray}\label{triv1}
0&=&-\frac{2 m^2 \left(1+m^2\right)^2}{(2+n) A_D},\nn
0&=&-\frac{2 m^2 \left(1+m^2\right)^3 \left(\text{D} \left(1+m^2\right) (2+N)-4 \left(2+N+2 m^2 (5+N)\right)\right)}{(2+N)^2 (4+N) A_D^2}.
\end{eqnarray} 
for $\lambda_2$ and $\lambda_3$, respectively. The first equation gives zero at $m^2=-1$ and $m^2=0$ independently from the dimension.
Among the roots of the second equation in Eq.~(\ref{triv1}) of course we discover again $m^2=-1$ and $m^2=0$, but factorising with respect to these roots, one will
be left with the equation for the third root, and it will depend on $D$. Let us solve it for $m^2$. The result is the following:
\begin{eqnarray}
m^2=\frac{-DN-2 D+4 N+8}{D N+2 D-8 N-40}.
\end{eqnarray}

\begin{figure}[h!]
  \centering
      \includegraphics[width=0.6\textwidth]{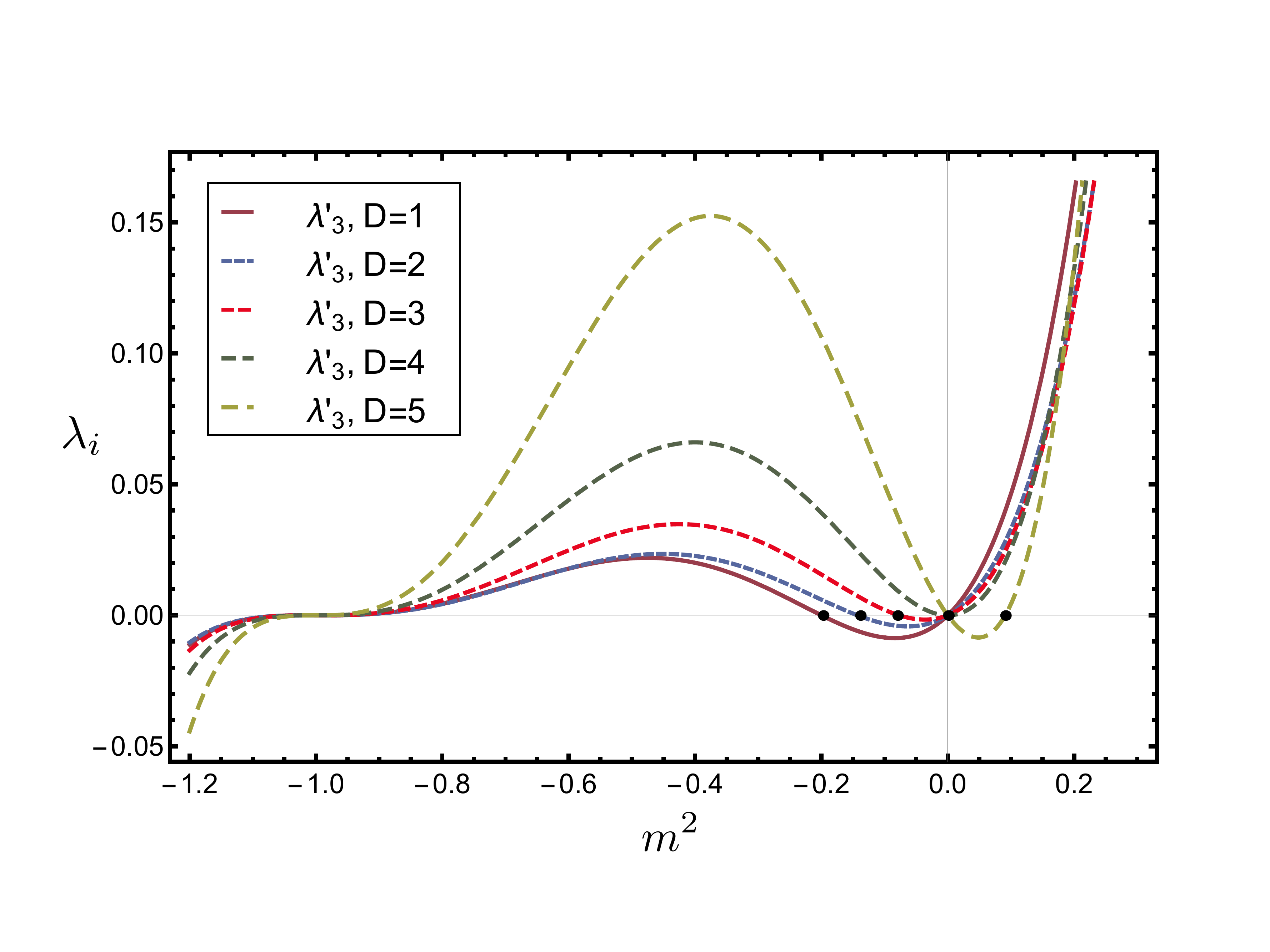}\vspace*{-1cm}
  \caption{In this figure one can see how the $\lambda_3$ VBF curves change with the dimension. Observe that the only root from the $(-1,0)$
  interval runs into zero as $D \to 4$. For $D\geq 4$ the root obtains a value $m_0^2>0$; hence there is no fixed point in the SSB phase but there is
  one in the symmetric phase, which turns out to be unstable because $\lambda_2(m_0^2)<0$. Note that we indicated only integer values for $D$
  but the transformation is continuous in the dimension; hence all the values between the integers would define similar curves. For this illustration
  $N$ was set to 3 and the VBF curves are rescaled as $\lambda'_3\propto\lambda_3 10^{-\alpha}$, where
  $\alpha=0,4,4,5$ and $6$ for $D=1,2,3,4$ and $5$, respectively.  
}\label{triv}
\end{figure}

\begin{figure}[h!]
  \centering
      \includegraphics[width=0.6\textwidth]{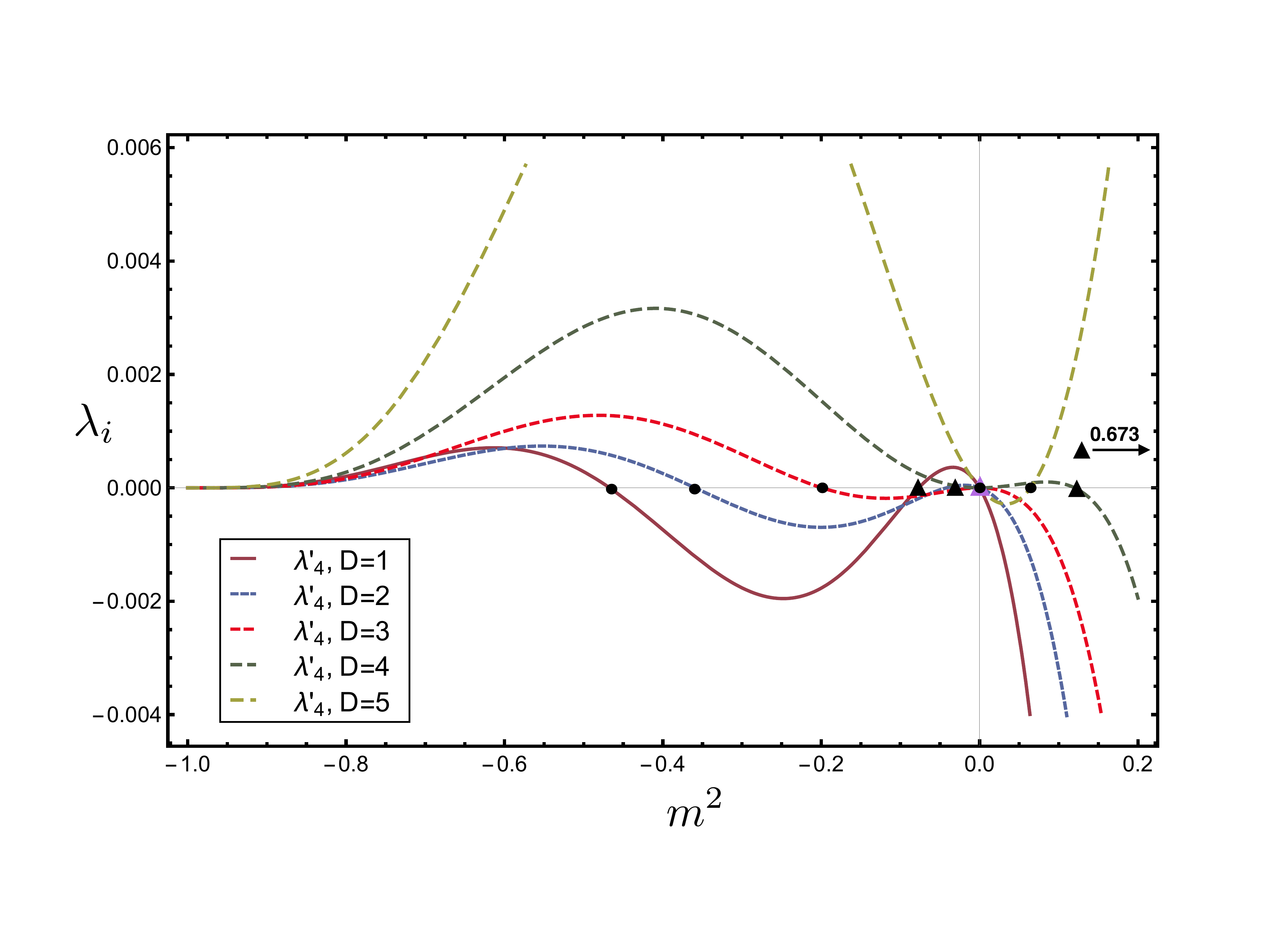}\vspace*{-1cm}
  \caption{In this figure one can see the different $\lambda_4$ VBF curves for different dimensions. Observe that the two roots from the interval 
  $(-1,0)$ runs to the right as $D$ grows. For $D= 4$ the root indicated with a triangle has got already a positive value and the one which was indicated
  by a dot has just melted into zero. For $D=5$ both of the roots have positive values. The triangle is not present in the plot, since it took the value
  $m^2=0.673$. For this illustration $N$ was set to 3 and the VBF curves are rescaled as $\lambda'_4\propto\lambda_4 10^{-\alpha}$, where
  $\alpha=5,6,7,8$ and $9$ for $D=1,2,3,4$ and $5$, respectively.  
 }\label{triv2}
\end{figure}

This expression shows the "running" of the root (the Wilson-Fisher fixed point) with the dimension. For substituting $D=1,2$ and $3$ (and $N=3$),
one would get the results that we obtained in the previous sections. However, evaluating it at $D=4$, we will lose the $N$ dependence
completely. In other words: no matter what O($N$) model we consider, the WF fixed point coincides with the Gaussian in the
$D=4$ case. In Fig.~\ref{triv} one can see the $\lambda_3$ VBF for various dimensions. The observation is the following: the only root in the
interval $(-1,0)$ drifts towards the positive real values. At $D=4$ it merges into zero; hence, besides the convexity and the Gaussian fixed point,
there are no further fixed points present. Above four dimensions the root gets positive values. We can address the question if this is a true fixed
point or not. If one substitutes the value of these positive roots into the VBF $\lambda_2(m^2)$, one will immediately see that 
$\lambda_2(m_0^2)<0$, thus defining an unstable potential at this truncation level. As a consequence, there are no other true fixed points besides the Gaussian and
the IR, hence we can call the system trivial. It is possible to derive the dimension dependence for other $\lambda_i$'s, too, with similar qualitative
behaviour: the roots start from inside the interval $(-1,0)$ with $D=1$ and as the dimension grows, they are tending out from the interval. As $D$
takes the value of 4, the last (i.e. the closest to $m^2=-1$) root of the VBF under consideration merges into $m^2=0$, and all the others have
already obtained positive values; see Fig.~\ref{triv2}, where we illustrate this situation with $\lambda_4$.\\
In the following, we are going to discuss the VBF curves at a higher truncation level for $D=5$ in detail, as a representative of theories for
$D>4$. However, we need to distinguish the case when $D=4$, since showing the triviality in this case is
not that straightforward as it is for $D>4$; hence, the four-dimensional case is going to be presented in a separate section. In the last two sections various examples for theories in dimensions $D\geq4$ are shown.\\
 
%%%%%%%%%%%%%%%%%%%%%%% 
%%%%%%%%%%%%%%%%%%%%%%%

%%%%%%%%%%%%%%%%%%%%%%%%
%%%%%%%%%%%%%%%%%%%%%%%%

%%%%%%%%%NEW SUBSECTION%%%%%%

\subsection{Triviality of the O($N$) model in $D>4$}
For any $D>4$ theories the finding is that the VBF curves have the following structure:
\begin{eqnarray}
\lambda_n \propto(-1)^{n+1} (m^2)(1+m^2)^n\sum\limits_{i=0}^{n-2}g_{i}(N,D)(m^2)^{i}.
\end{eqnarray}
Notice that it has essentially the same structure that we had in the cases $D\leq 2$, but there is one crucial difference. For theories $D\leq 2$ we
found the coefficient functions $g_{i}$'s to be always positive, hence defining only negative (and complex) roots for the polynomial; moreover they were
inside the interval $(-1,0)$. In this case likewise to $2<D<4$ we can obtain negative values for $g_i$'s, too, but contrary to that case, here we do not obtain roots inside
the interval $(-1,0)$: all the real roots in this case are positioned in the disjoint union of the complement set of $(-1,0)$, i.e. in $(-\infty,-1]\cup[0,\infty)$. The first few VBFs are the following:
\begin{eqnarray}
\lambda_{2}&=&-24 \pi ^3 m^2 (m^2+1)^2,\nn
\lambda_{3}&=&\frac{288\pi ^6}{7} m^2 (m^2+1)^3 (39 m^2-5),\nn
\lambda_{4}&=&-\frac{1536 \pi ^9}{7}  m^2 (m^2+1)^4 \left(576 (m^2)^2-425 m^2+25\right),\nn
\lambda_{5}&=&\frac{92160 \pi ^{12} }{539} m^2 (m^2+1)^5 \left(10233 (m^2)^3-176625 (m^2)^2+36125 m^2-1225\right),\nn
...
\end{eqnarray}

%%%%%%%%%%%%%%%%%%%%%%%%%%%%%
\begin{figure}[h!]
  \centering
      \includegraphics[width=0.6\textwidth]{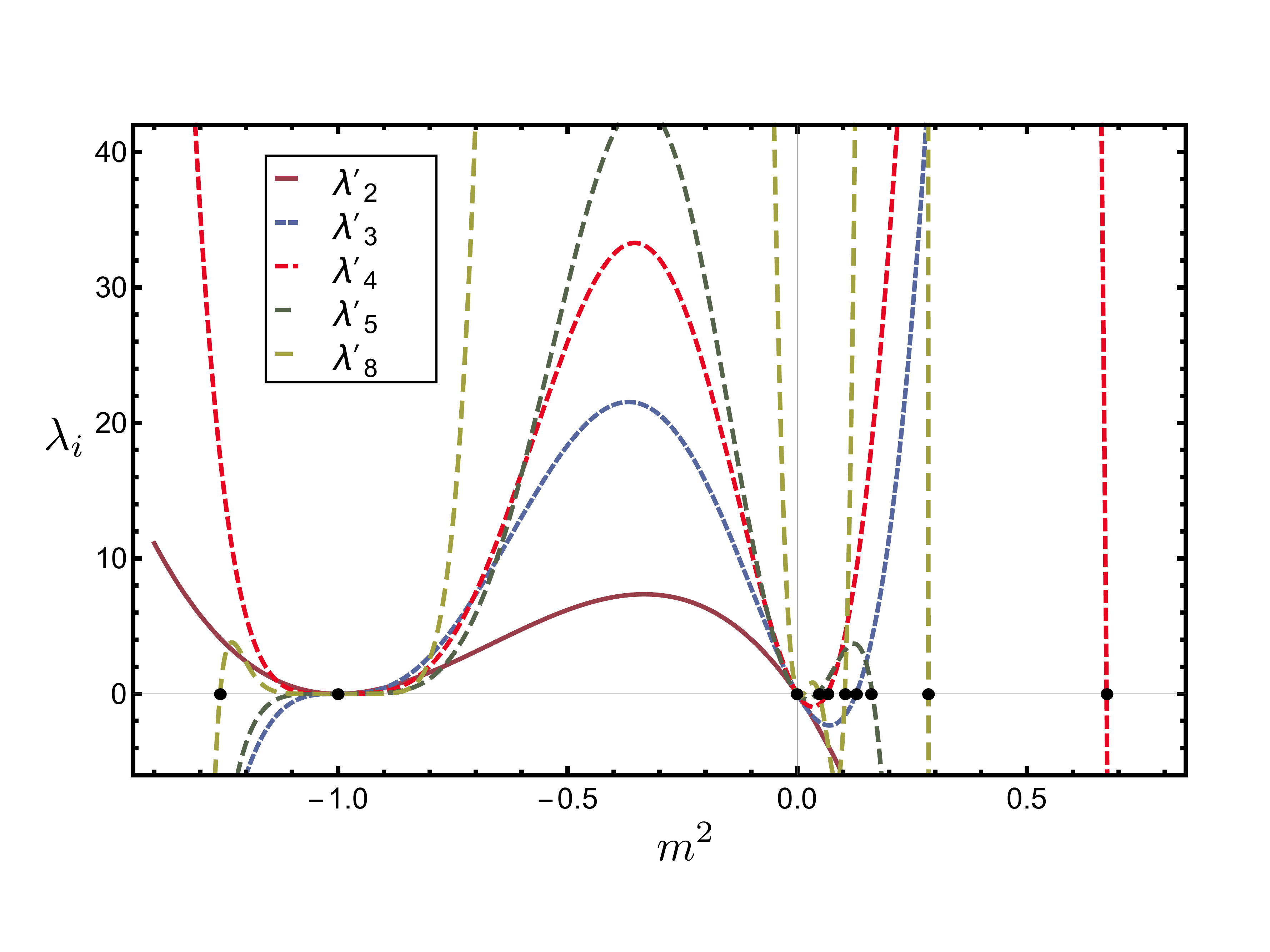}\vspace*{-.5cm}
  \caption{The VBF curves $\lambda'_i$ (i=2,...5,8). Observe that the roots of each curve are always outside the interval $(-1,0)$. For $\lambda_8$ we have negative roots, 
  too. For display purposes the VBF curves are rescaled as $\lambda'_i\propto\lambda_i 10^{-\alpha}$, where
  $\alpha=1,3,6,9$ and $19$ for $i=2,3,4,5$ and $8$, respectively. }\label{trivD5}
\end{figure}
%%%%%%%%%%%%%%%%%%%%%%%%%

\begin{figure}[h!]
\centering
\vspace*{-.7cm}  
\hspace*{-5cm}
\begin{subfigure}{.25\textwidth}
 \centering
 \vspace*{-.15cm} 
  \includegraphics[scale=.24]{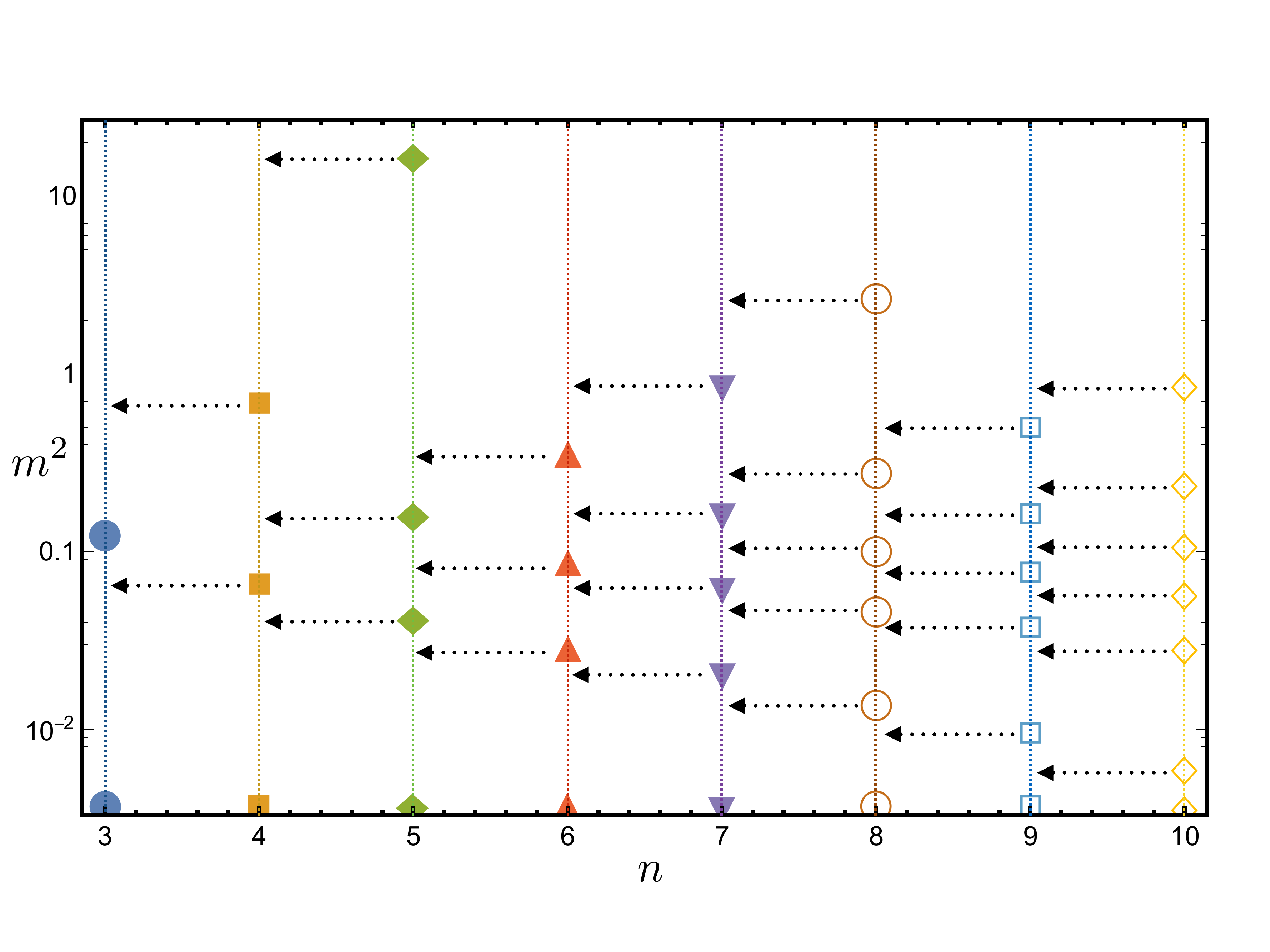}
   \end{subfigure}
\hspace*{4cm}
%\vspace*{.7cm}   
\begin{subfigure}{.25\textwidth}
  \centering
  \includegraphics[scale=.24]{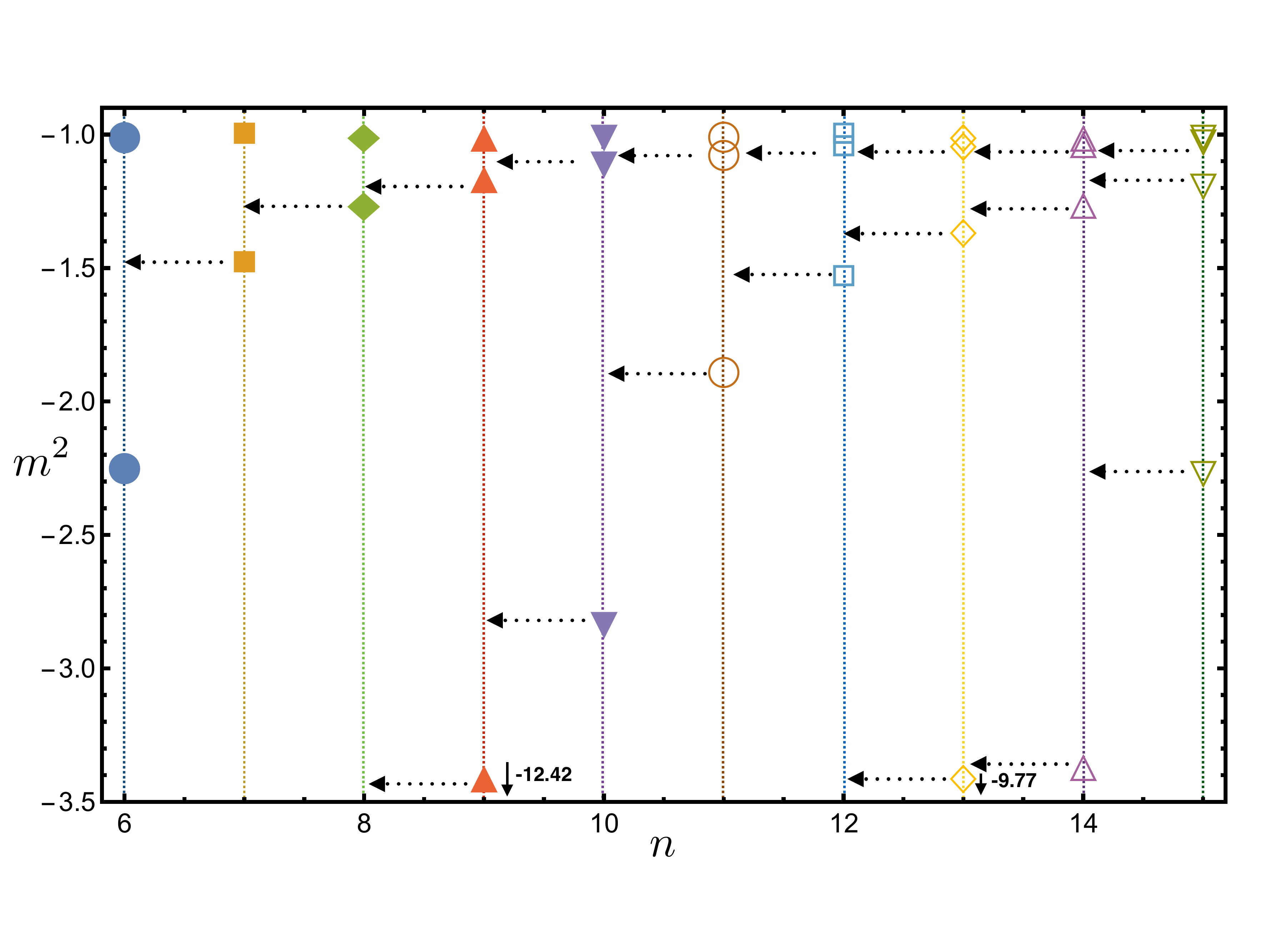}%\vspace*{-.15cm}
  \end{subfigure}%\vspace*{-.5cm}
\caption{In the upper (lower) panel the roots of the VBF curves are shown for $\lambda_3,...,\lambda_{10}$ ($\lambda_6,...,\lambda_{15}$), for
$m^2\geq0$ ($m^2\leq -1$). One can observe a pattern of the roots similar to the $D\leq2$ case: each root of VBF $\lambda_{n+1}$ is surrounded
by the roots of $\lambda_{n}$. The only problem in this case is, since the interval is unbounded, this pattern is not entirely true: indeed, the highest
(lowest) roots of $\lambda_{n+1}$ now do not have an upper (lower) neighbour from the roots of $\lambda_{n}$. We can
overcome this difficulty by one-point compactifying the real line $\mathbb{R}$. For details see the text. }\label{d5roots} 
\end{figure}

These curves are shown in Fig.~\ref{trivD5}. Now, we are going to analyse their root structure. 
In the case for theories in $D>4$ we can clearly identify a pattern of the roots again, just like we did it for $D\leq2$, i.e. the $M^*$ pattern. For the particular
case $D=5, N=3$ one can see the position of the VBF roots in Fig.~\ref{d5roots}. What we can observe is just the reverse of what happened in the O($N$) models for $D\leq2$. The roots are situated only outside the interval $(-1,0)$, and they have a similar pattern to the one we had for the $D\leq 2$ cases; i.e. each root of $\lambda_{n+1}$ is surrounded by the
roots of $\lambda_{n}$.

%%%%%%%%%%%%%%%%%%%%%%%%%%%%%
%%%%%%%%%%%%%%%%%%%%%%%%%%%%%%  

\begin{figure}
\begin{subfigure}{.51\textwidth}
  \centering
  \includegraphics[width=.61\linewidth]{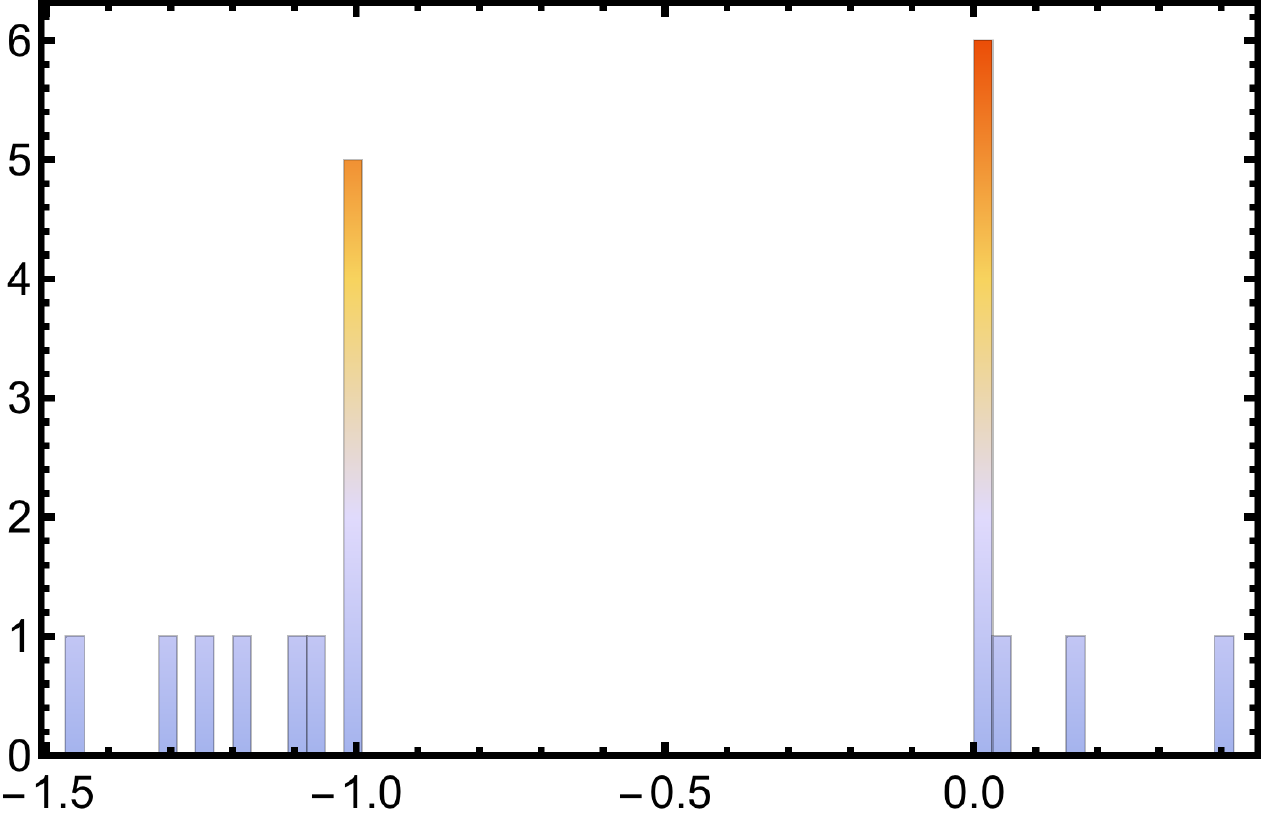}
  \caption{$n=20$}
  \label{fig:sfig1}
\end{subfigure}%
\begin{subfigure}{.51\textwidth}
  \centering
  \includegraphics[width=.61\linewidth]{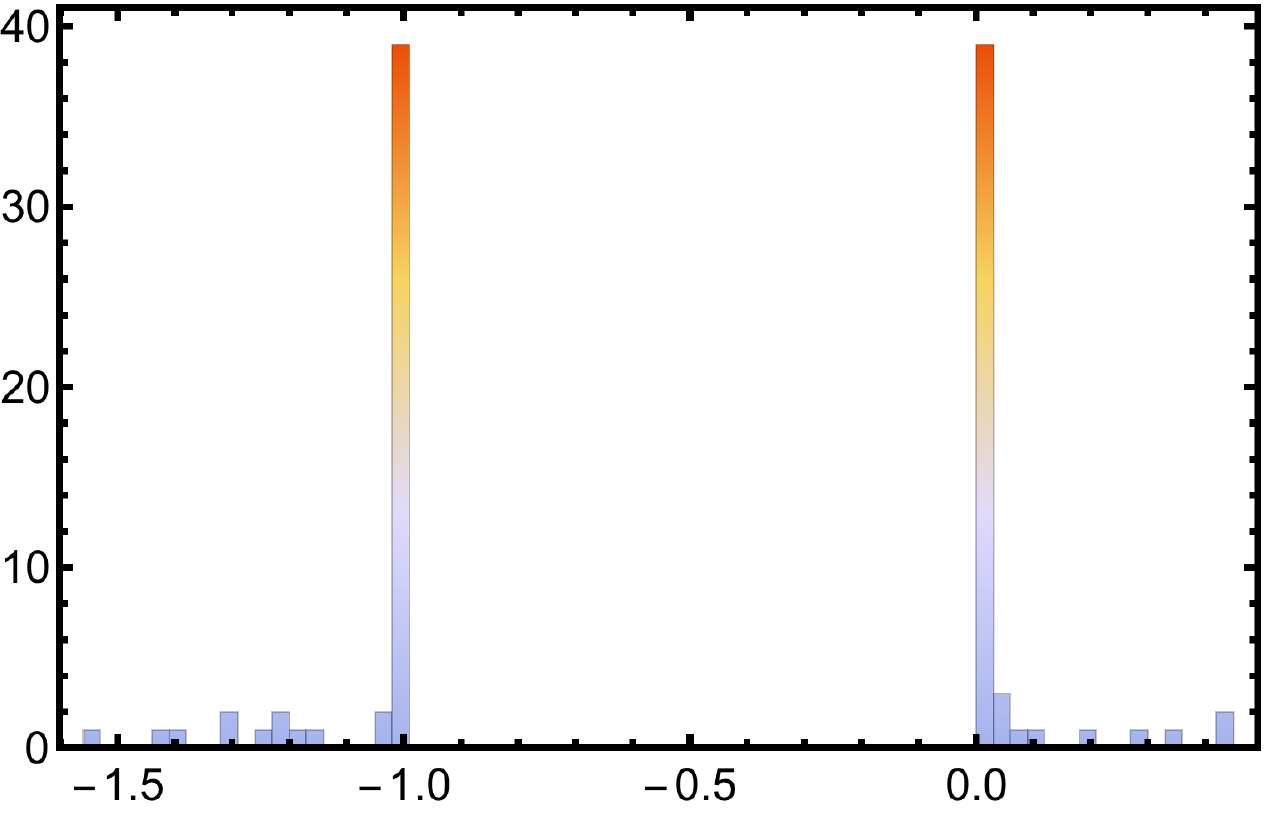}
  \caption{$n=100$}
  \label{fig:sfig2}
\end{subfigure}\vspace{.2cm}
  \centering
\begin{subfigure}{.51\textwidth}
  \centering
  \includegraphics[width=.61\linewidth]{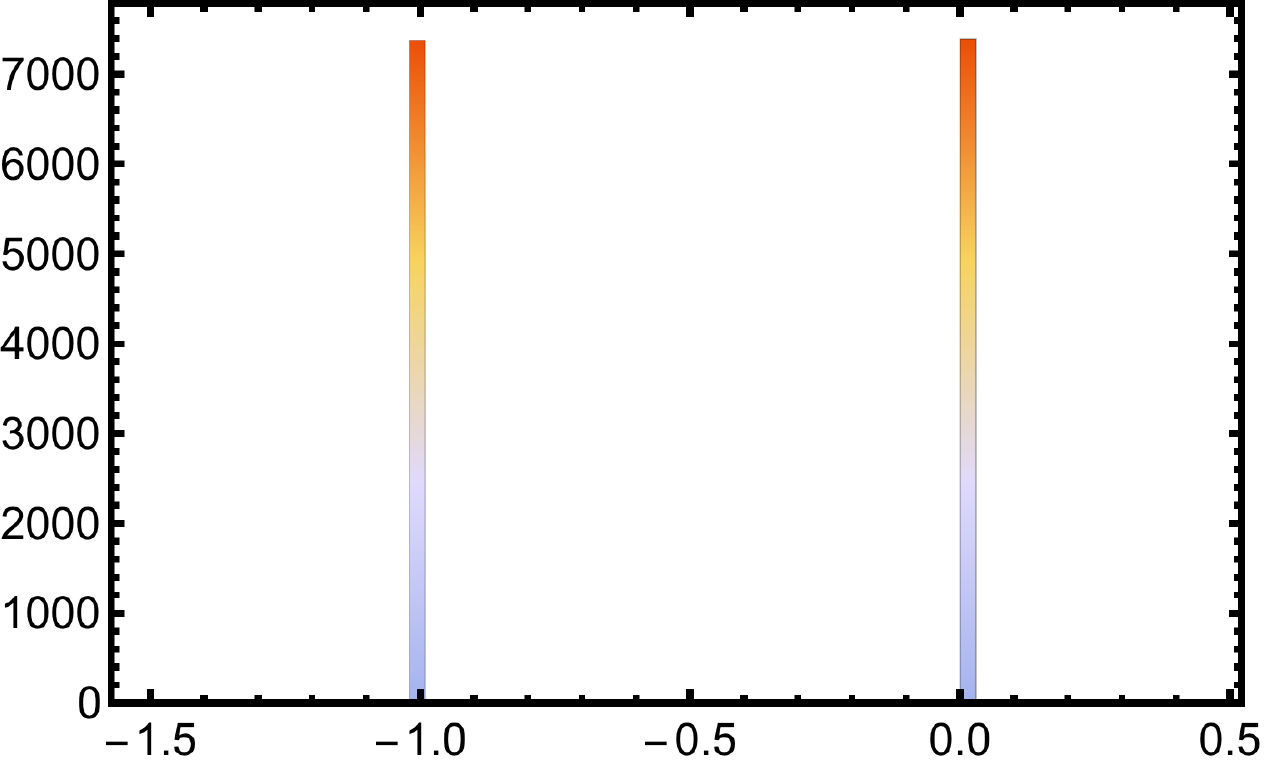}
  \caption{$n=15000$}
  \label{fig:sfig3}
\end{subfigure}
\caption{Histograms of the distribution of the points generated in $X_n$: (a) the distribution of $X_{20}$, (b) the
distribution of $X_{100}$, and (c) the distribution of $X_{15000}$. Observe that the points are accumulating at
$-1$ and $0$ but this time from the outside region.}\label{disd5}
\end{figure}

We can call it only similar, since we clearly have a problem in the present case: we are not able to use our random number
generating model that we did in Section \ref{MW} for simulating the positions of the roots, because the set $(-\infty,-1]\cup[0,\infty)$ is unbounded.
For that reason, the highest (lowest) roots of the VBF $\lambda_{n+1}$ do not have an upper (lower) neighbour from the set of the roots of the
VBF $\lambda_{n}$. However, we are able to do the following trick: let us one-point compactify the real line $\mathbb{R}$, meaning that we "glue" together the 
points $m^2\to\pm\infty$, thus creating a closed set for $[0,-1]$. This may look like an {\textit{ad hoc}} idea but we will see that it works. Let us consider for instance
the orders $n=6$ and $n=7$ in Fig.~\ref{d5roots}. The highest valued root in the region $m^2>0$ from $n=7$ would need a root from $n=6$ to restore the right pattern, but
apparently there is no such root.
However, if we extend our picture with the compactified real line, we will see that in the region $m^2<-1$ the right
root appears just where we would need it at order $n=6$, and thus the pattern goes on. This is true for all the other roots (see Fig.~\ref{d5roots}), thus
the compactification of the real line is well justified in this way, hence providing us with an $M^*$ pattern on the compactified real line.\\ In the interval [0,-1] one can
now use the technique that we introduced in Sec. \ref{MW}. The result of such a random number procedure is shown in Fig.~\ref{disd5}.\\
We can see that the positions of the roots are again accumulating at $-1$ and $0$, but now they are approaching from the complement interval of
$(-1,0)$. In this way we can see how the triviality is emerging in the limit $n\to\infty$.%\vspace*{-.5 cm}

%%%%%%%%%%%%%%%%%%%%%%%%%%%%%%

%%%%%%%%%%%NEW SUBSECTION%%%%%%%%%%%%%%%%%%%%%%%%%%%%%

\subsection{Triviality of the O($N$) model in $D=4$}\label{d4sec}
In this section we are going to discuss the $D=4$ O($N$) models, and special attention will be given to the case $N=1$, which has been in the centre
of interest since triviality is predicted to occur in the $\phi^4$ theory. To have a clue on triviality beyond PT we still need to rely on
lattice simulations, which actually support the trivial behaviour of such theories \cite{triv}. Here, we are going to show a result which suggests that indeed the
trivial scenario holds for such models; i.e. no UV fixed point different from the Gaussian is present in the O($N$) models taking into
account all the symmetry respecting terms in four dimensions (the $\phi^4$ case is shown in Fig.~\ref{triv}). We are going to present the result for $N=1$ but this
holds for general $N$.  
The VBF polynomials in $D=4$ have the form
\begin{equation}
\lambda_n \propto(-1)^{n+1} (m^2)^{1+\Theta(n-3)}(1+m^2)^n\sum\limits_{i=0}^{n-2}g_{i}(N,D=4)(m^2)^{i-\Theta(n-3)}.
\end{equation}
In particular the first few $\lambda_i$ for $N=1$ are
\begin{eqnarray}
\lambda_{2}&=&-\frac{64\pi ^2}{3}  m^2 (m^2+1)^2,\nn
\lambda_{3}&=&\frac{8192 \pi ^4}{5}  (m^2)^2 (m^2+1)^3,\nn
\lambda_{4}&=&-\frac{524288  \pi ^6}{35} (m^2)^2 (m^2+1)^4 (14 m^2-1),\nn
\lambda_{5}&=&\frac{67108864\pi ^8}{315}  (m^2)^2 (m^2+1)^5 \left(168 (m^2)^2-41 m^2+1\right),\nn
...
\end{eqnarray}
By analysing the root structure of these equations one will find that all the roots are situated outside the interval of $(-1,0)$ likewise in the
$D>4$ cases; the only difference is that at the order $n=7$ the VBF curve $\lambda_7$ develops roots on the complex plane, too, with $\text{Re } 
m^2>0$. We have already met such a situation in theories defined in $2<D<4$ dimensions. In that case, we neglected these roots since we 
considered them to be unphysical. In the present case we still stick to our convention; that is, we neglect them as possible fixed point candidates. However, 
we can identify an interesting behaviour for such nonreal roots. Let us look at the root structure of the theory in Fig.~\ref{d4r}, where we present the
position of the real roots for all relevant intervals of $m^2$.

%%%%%%%%%%%%%%%%%%%%%%%
\begin{figure}[h!]
  \centering
      \includegraphics[width=0.5\textwidth]{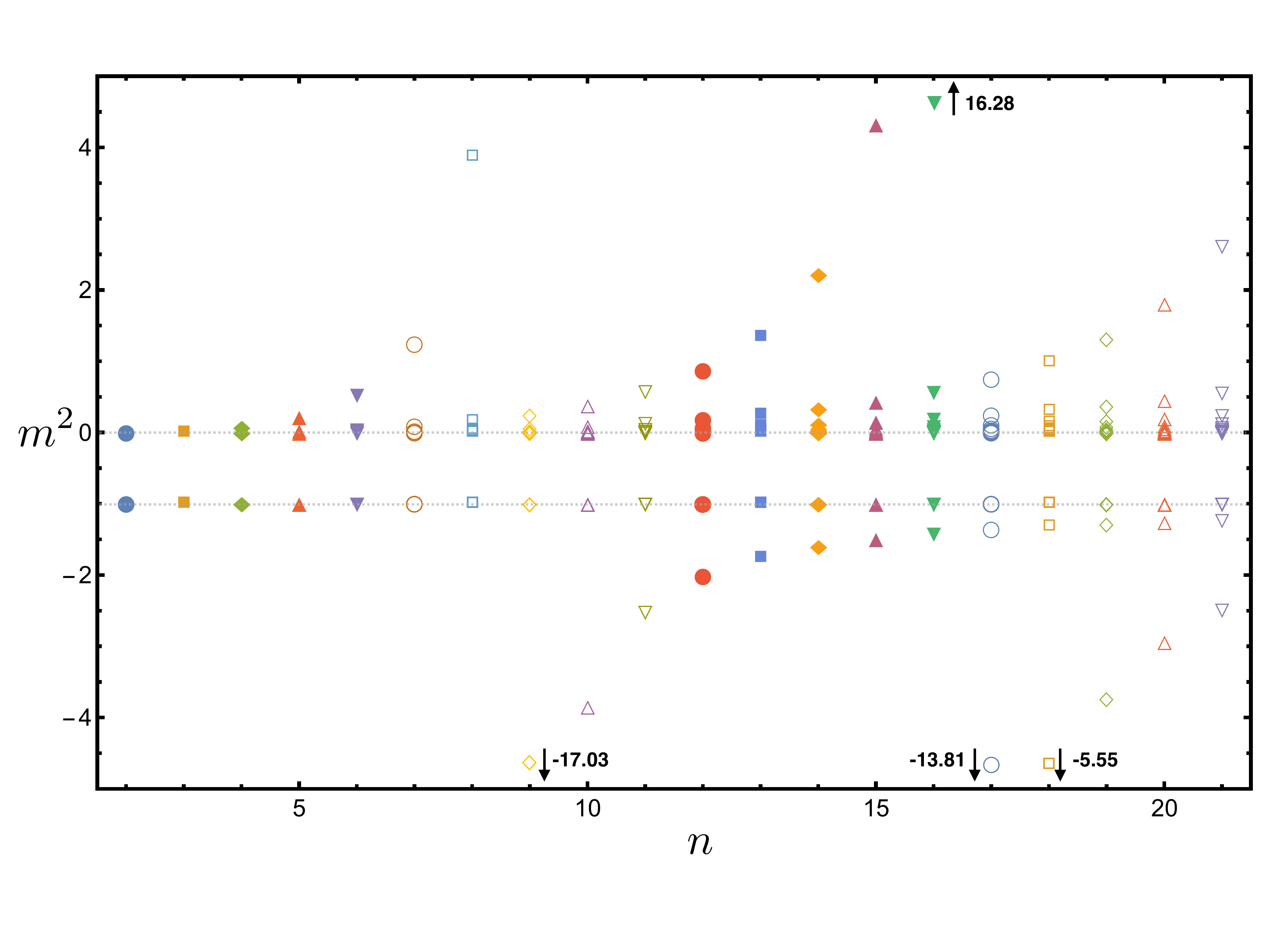}\vspace*{-.8cm}
  \caption{The real roots of the four-dimensional O($N$) model are shown. No root in the interval $(-1,0)$ is found; on the contrary, all roots are outside this
  interval. One can see that, just like in the case of $D>4$, the roots are approaching $-1$ and $0$ from the outside region. However, more careful analysis
  is needed in this case. For details see the text.}\label{d4r}
\end{figure}
%%%%%%%%%%%%%%%%%%%%%%%

What we can observe is that the roots are approaching the points $-1$ and $0$, just like in the case for $D>4$. However, as we indicated
above, complex roots are emerging: the first one and its conjugate from $n=7$, two and their conjugates from $n=11$, three and the conjugates
from $n=15$ and from $n=18$ four complex roots plus conjugates. Among these roots, being complex, we cannot make an ordering; however we
are able to do that for the real part of them. The real parts of the roots are indicated in Fig.~\ref{d4r}. Complex roots with negative real parts do not occur; hence it is enough to consider the ${\text{Re }} m^2>0$.  
%%%%%%%%%%%%%%%%%%%%%%%
\begin{figure}[h!]
  \centering
      \includegraphics[width=0.55\textwidth]{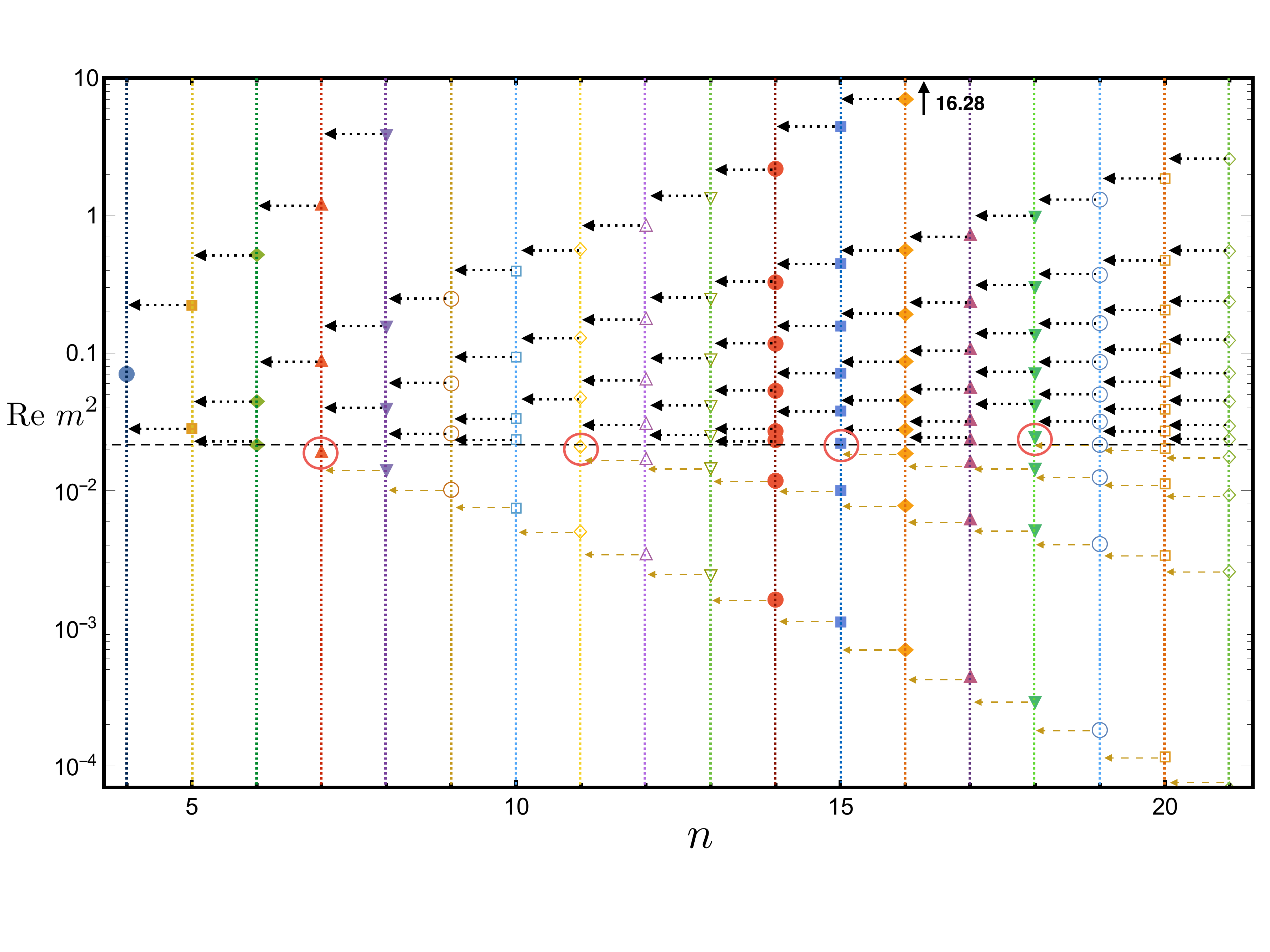}\vspace*{-.8cm}
  \caption{The real parts of the roots. The black dashed line indicates the average threshold between the purely real and complex root values. It is
  at about $m^2=0.022$. For every value of ${\text{Re }} m^2$ under this line we have an imaginary part for the root. New complex roots are
  emerging at orders $n=7$, $n=11$, $n=15$ and $n=18$. At these orders we can see a "violation of the pattern" which is being indicated by the
  arrows: pointed black arrows for the purely real and dashed gold for the complex roots. Apart from the newly emerging complex roots the pattern
  holds. For details see the text.} \label{d4r}
\end{figure}
%%%%%%%%%%%%%%%%%%%%%%%

\begin{figure}[h!]
\centering
\begin{subfigure}{.5\textwidth}
 \centering
% \hspace*{-1cm}
%\vspace*{-1.05cm}
  \includegraphics[scale=.23]{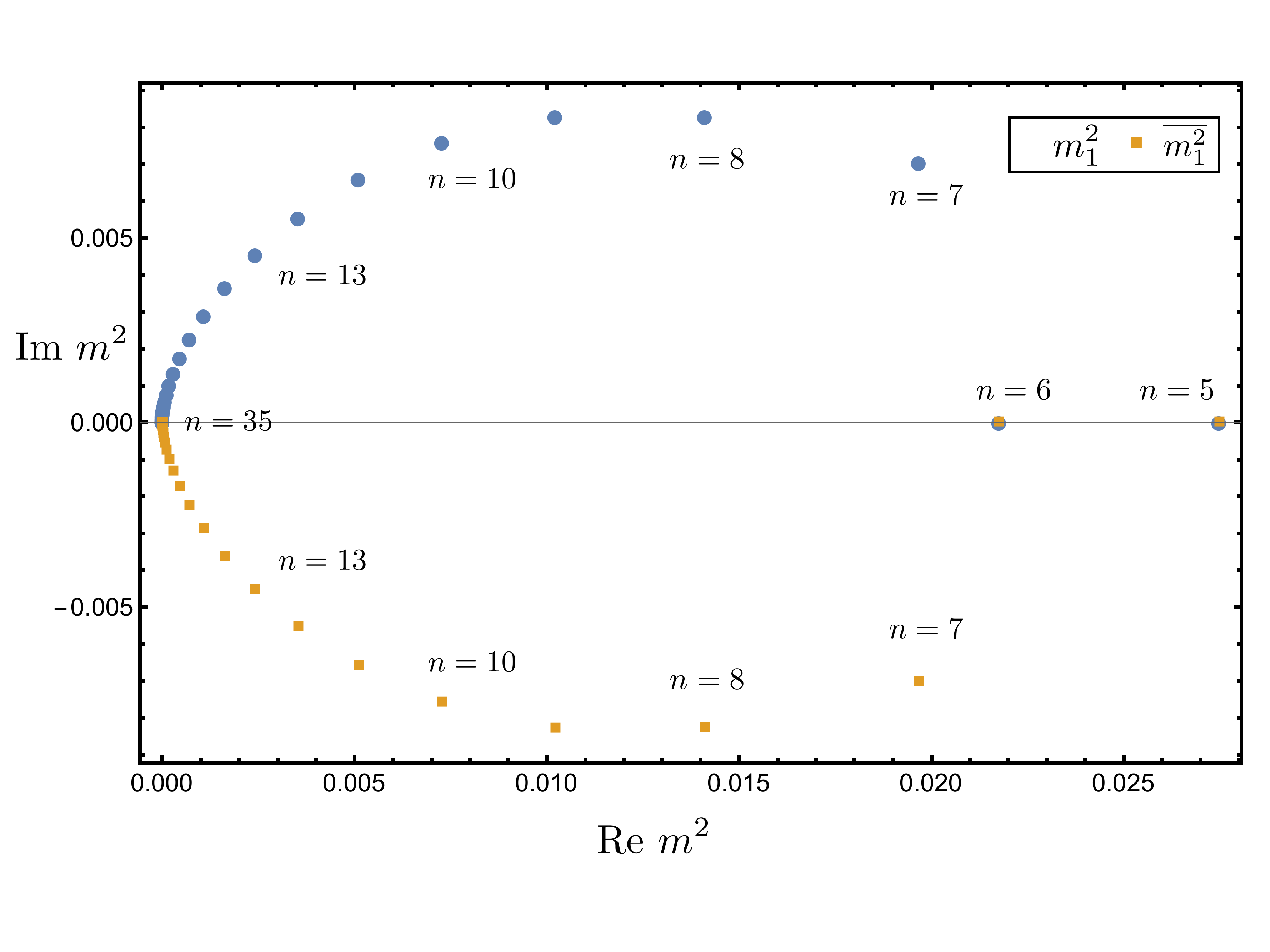}
   \vspace*{.05cm}
   \end{subfigure}%
\begin{subfigure}{.5\textwidth}
  \centering
 % \hspace*{-1cm}
  \includegraphics[scale=.23]{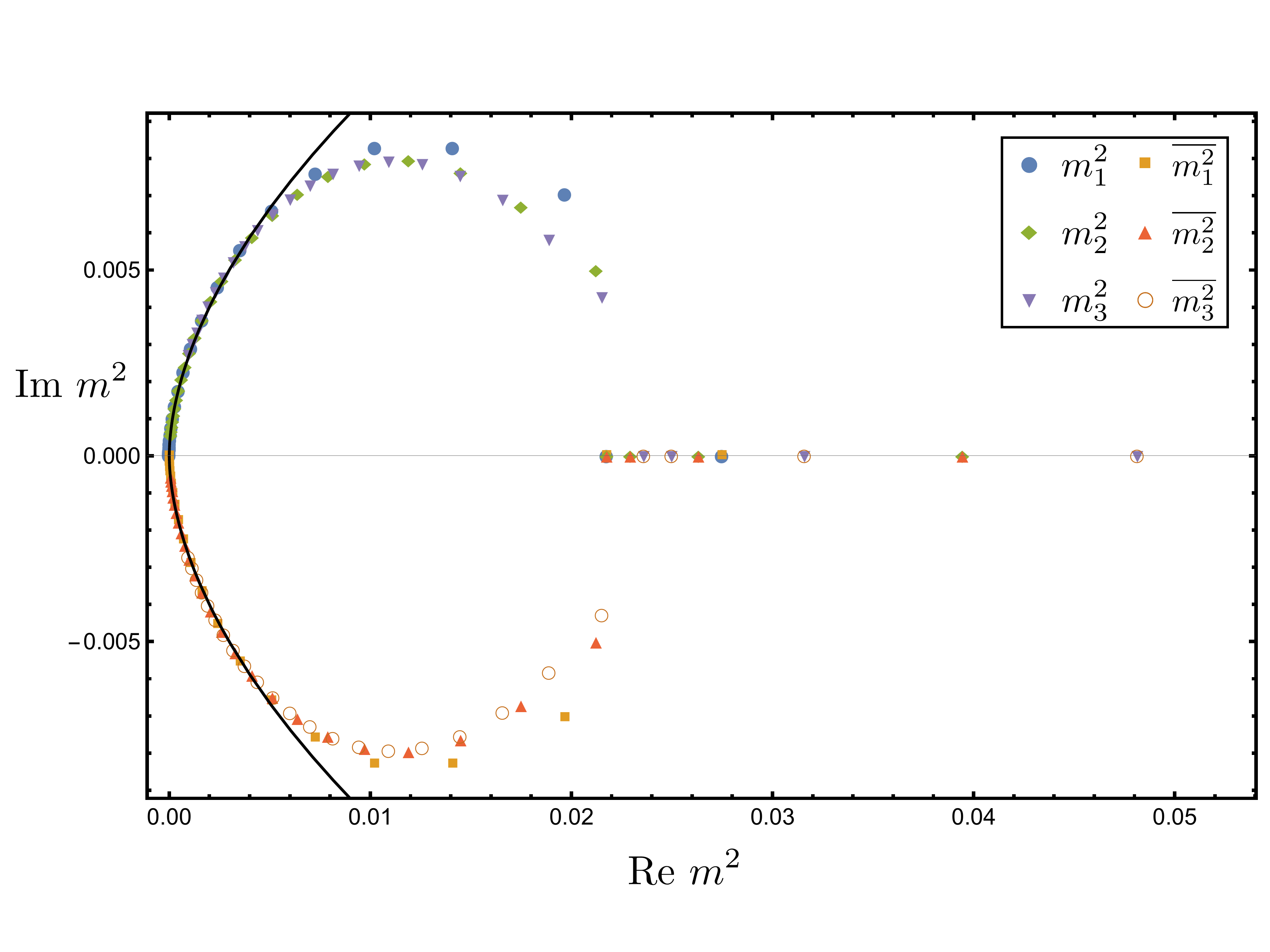}\vspace*{.5cm}
  \end{subfigure}\vspace*{-.7cm}
\caption{The roots on the complex plane. In the upper panel one can see the emergence and disappearance of the first complex root ($m_1^2$). It appears
at $n=7$ and tends to zero as we go to higher orders in $n$. In the lower panel the second and third complex roots are indicated ($m^2_2$ and $m^2_3$).
They appear at orders $n=11$ and $n=15$ respectively, and coincide with the curve defined by the complex root $m_1^2$. The fitted power-law curves are defined in the text, and the fitting procedure was performed on 24 data points.}\label{complex} 
\end{figure}

The figure shows that for the real part of the roots, there is indeed a pattern, namely almost the same that we observed for both the cases $D\leq2$ and $D>4$; the only differences are in the orders where the complex roots appear. Here, we observe that above and below the real part of the newly appearing 
complex roots the pattern continues; only at those particular points do we find the breaking of the pattern (these are indicated by red circles in Fig.~\ref{d4r}): no
root from the real part of the $\lambda_{n+1}$ goes between the corresponding roots of $\lambda_{n}$. Now, if we consider the limiting distribution
of such a pattern of the points, we will find that this anomaly will not have an effect on it: after compactification of the real line, as we did it for the $D>4$,
the points are just accumulating at $-1$ and $0$; hence we will find exactly the same result as in Eq.~(\ref{dist}), but now only for the real part of $m^2$. We are not done yet, 
since we have only considered the real part so far. This result on the limiting position of the roots does not make any sense if we find a finite imaginary part of $m^2$ in this
limit. Let us consider therefore the roots on the complex plane. This is shown in Fig.~\ref{complex}. We can see the developing imaginary part at the
order $n=7$ for the first time.\\ 

%%%%%%%%%%%%%%%%%%%%%%%%%%
%%%%%%%%%%%%%%%%%%%%%%%%%%

Interestingly, the absolute value of the imaginary part will have a maximum at some point and it is tending to zero just like the real part. We can also
express the imaginary part as a power-law of the real part close to the origin: ${\text{Im }}m^2=\pm a ({\text{Re } }m^2)^b$, where the parameters are
$a=0.162 \pm 0.001$ and $b=0.589\pm0.001$. Now, considering the "running" of the second and the third complex roots ($m^2_2,m^2_3$), which are
appearing first at the orders $n=11$ and $n=18$, respectively, we will find that they behave in a very similar way to the first complex root ($m^2_1$);
moreover, they collapse onto each other close to the origin (see Fig.~\ref{complex}), thus approaching $0$ with the same exponent. We can see that
both the real and the imaginary parts of the roots are approaching zero as the order $n$ grows, hence we can say that the roots, although they are not
defined as physical ones when they have complex values, accumulate at $-1$ and $0$ in the $n\to\infty$ limit. It is interesting that only in the limit is the
imaginary part absent entirely, and until we take this limit we will have a gap between $m^2=0$ and the last noncomplex valued root in the particular
order $n$. With this procedure we were able to show that only two fixed points are present (with the Gaussian as the only stable one) in the unexpanded
O($N$) model just like in $D>4$. Altogether this signals the triviality for the O($N$) theories when the dimension is $D\geq 4$. Hence, we were able to show a non-perturbative
evidence of the triviality for theories defined in $D\geq4$.\\
Regarding the $D=3$ case, in Sec. \ref{wf}, the roots on the complex plane do not behave in the same way that they 
do in the four-dimensional case (Fig.~\ref{complex}); hence we cannot obtain the distribution of the roots by using the random
number generating algorithm for the compactified $M^*$ pattern.

%%%%%%%%%%%%%%%%NEW SECTION%%%%%%%%%%%%%%%%%%%%%%%%%%%%%%%%%%%%

\section{The $N$ dependence and the large\,-$N$ limit}\label{Ndepp} 

In this section we are going show what is the effect of changing the number of fields, i.e. $N$, for particular dimensions.\\
We can even compute the limit when $N\to \infty$ (cf.  \cite{tetra1, tetra2}) , which is equivalent to the spherical model \cite{Stanley68}.
To obtain the RG equation for large\,-$N$, we are going to rescale Eq.~\eq{FeE} by $N$, and considering the new variables $\rho\rightarrow \rho/N$ 
and $u \rightarrow u/N$. The derivative of the potential remains invariant under this rescaling $u'\rightarrow \frac{\partial u/N}{\partial \rho/N}=u'$. As a first step, we divide the RG equation \eq{FeE} by $N$:
\begin{align}
\partial_t \frac{u}{ N}  = (D-2)\frac{\rho}{ N} u' - D \frac{u}{ N} + \frac{A_D}{1+u'}-\frac{A_D}{N}\frac{1}{1+u'} + \frac{A_D}{N}\frac{1}{1+u'+2\rho u ''}.
\end{align}
Next we perform the rescaling
\begin{equation}
\partial_t u = (D-2)\rho u' - D u + \frac{A_D}{1+u'} -\frac{A_D}{N}\frac{1}{1+u'} + \frac{A_D}{N}\frac{1}{1+u'+2\rho u ''}.
\end{equation}
By taking the limit $N\rightarrow\infty$ the following terms remain:
\begin{equation}
\label{LN}
\partial_t u = (D-2)\rho u' - D u + \frac{A_D}{1+u'}.
\end{equation}
From this equation one can derive the VBF curves with the usual steps; just the $N$ dependence is absent from the formula Eq.~(\ref{genvbf}).\\
In this section we cannot show of course the full $N$ dependence; these are only a few checks for particular values of $N$. One should derive the root structure for each $N$, independently. However, these few examples could give us some idea what is going on when $N$ is being changed.

%%%%%%%%%Subsection%%%%%%%%%%%%%%%%
\subsection{$N$ dependence for O($N$) theories in $D\leq2$}

\begin{figure}[h!]
\begin{subfigure}{.55\textwidth}
  \centering
   \hspace*{-1cm}
  \includegraphics[width=.8\linewidth]{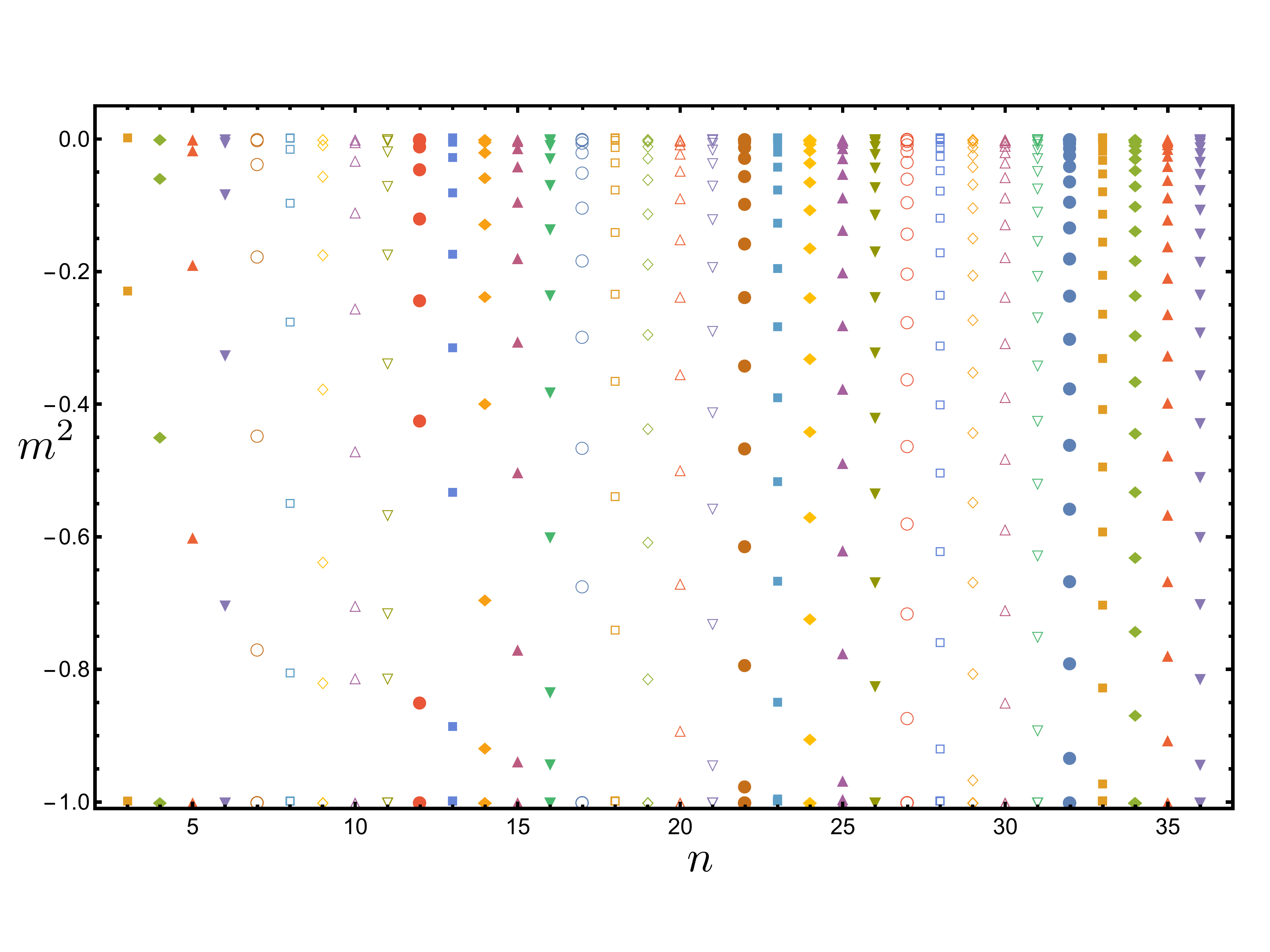}\vspace*{-.5cm}
  \caption{$N=7$}
 \end{subfigure}%
  \hspace*{-1cm}
\begin{subfigure}{.55\textwidth}
  \centering
  \includegraphics[width=.8\linewidth]{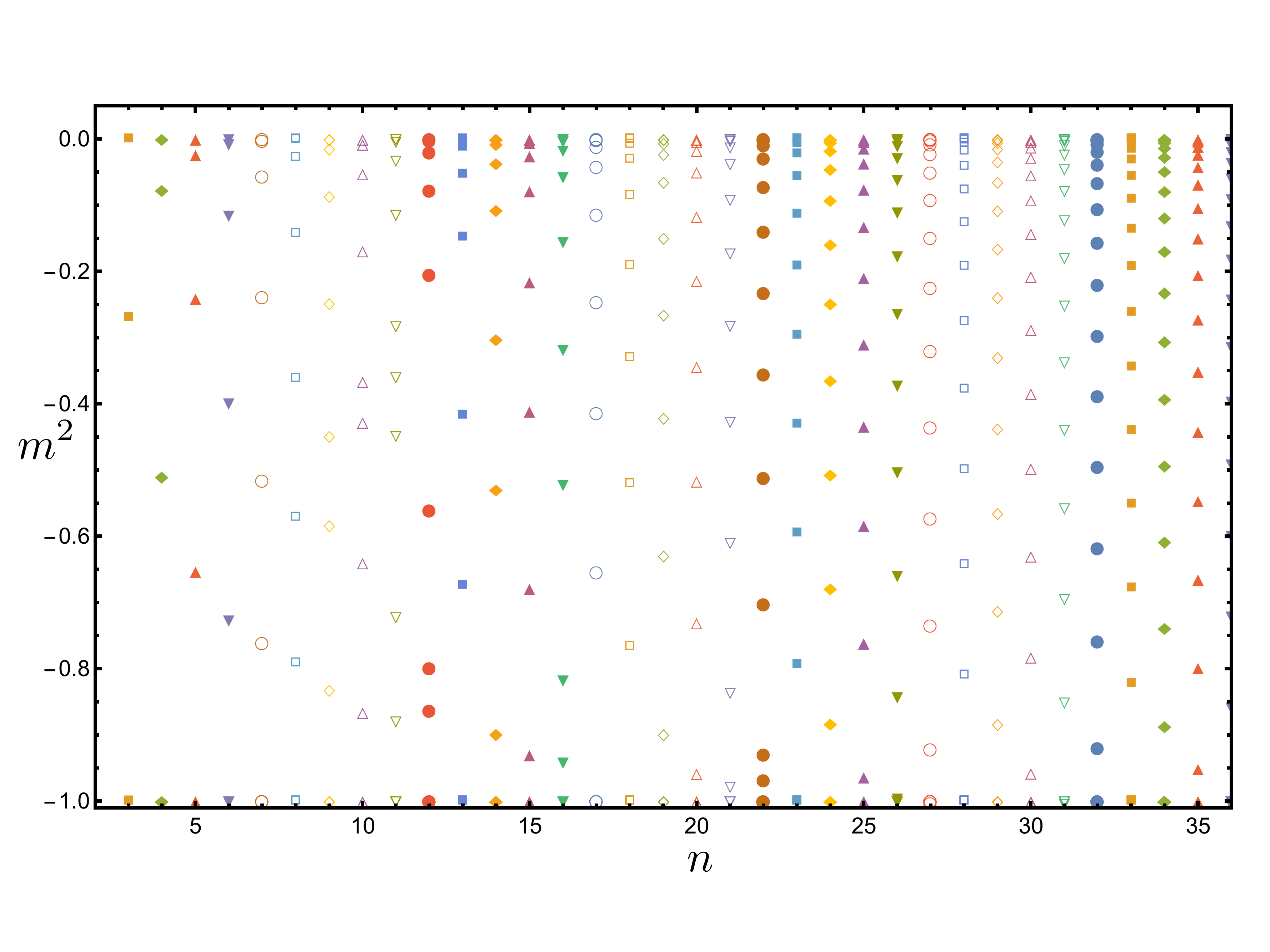}\vspace*{-.5cm}
  \caption{$N=15$}
  \end{subfigure}
  \centering
\begin{subfigure}{.55\textwidth}
  \centering
  % \hspace*{-.5cm}
  \includegraphics[width=.8\linewidth]{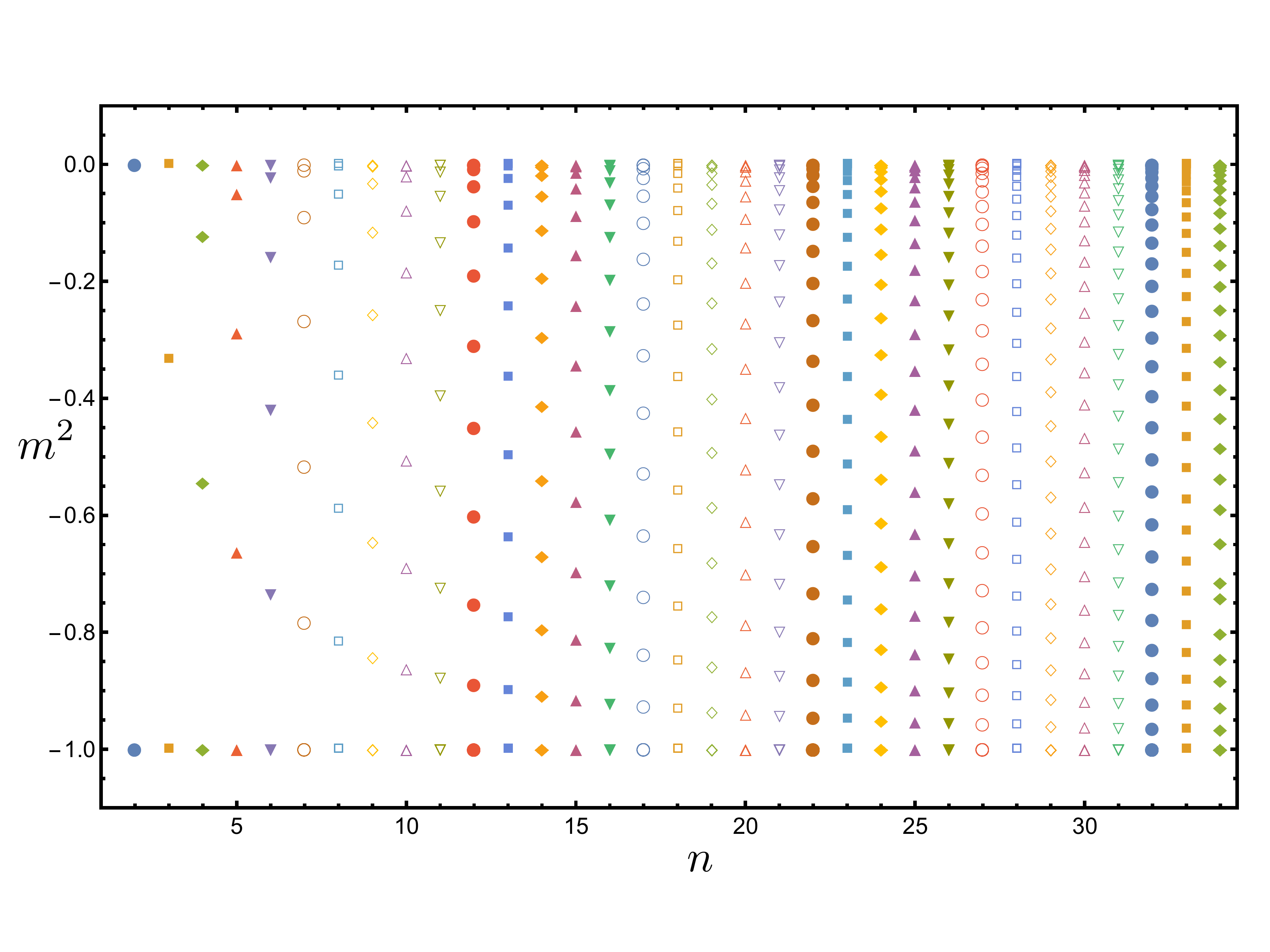}\vspace*{-.5cm}
  \caption{$N\to\infty$}
\end{subfigure}\vspace*{-.2cm}
\caption{The $N$ dependence of the root positions in $D=2$. We can see that the $M^{*}$ pattern restores at larger order of the expansion for various $N$. Sometimes one 
will find complex roots close to $-1$ which can be considered unphysical. In this way it does not actually matter if we consider the limit of  the distribution of
the roots or it gets complex providing an unphysical situation. When the $n\to\infty$ limit is taken the only two roots that can be found are $-1$ and $0$. In
the large\,-$N$ limit no complex valued root has been found till the order $n=34$.}\label{Ndep}
\end{figure}
%%%%%%%%%%%%%%%%%%%%%%%%%%%%%%%%%%%

Let us consider three different two-dimensional theories with field components $N=7$ and $N=15$ and $N\to\infty$. 
The positions of their roots are shown in Fig.~\ref{Ndep}. We can see in the two finite $N$ cases that there seems to be a
problem in the low orders of the expansion: the position of the roots will not satisfy the requirement of the $M^*$ pattern; however, as the order grows the pattern is restored.
We need to mention that when a particular root gets very close to $-1$ in the next order sometimes it becomes complex, but its real part still stays around $-1$;
hence one cannot see this accumulation exactly, but rather we can say this root "melted" into $-1$. This might be related to the fact that we already
explained for the $D=3$ case: the VBF polynomials get extremely flat close to $-1$ because of the $(1+m^2)^n$ factor in Eq.~(\ref{genvbf}); hence it could mean some problem for the root finding algorithm to provide the right value.
One way or another, a complex root does not define a physically sensible theory; thus we can still look at this pattern which will provide Eq.~(\ref{dist}) as the 
probability density of the root positions when $n\to\infty$.\\
However, no complex roots occur for the large\,-$N$ case where we know that only the 
symmetric phase must be present, as it was proven analytically in \cite{ssb} in the framework of FRG. The fact that a root gets complex or accumulates at one of the stable fixed 
points $-1,0$ is irrelevant from the point of view of the physics: in both of the cases only $-1$ and $0$ survive when $n\to\infty$. For $D=1$ we can find similar
results. These were, of course, only a few checks on the $N$ dependence and it should be considered case by case for every $N$; however, from our
experience we could assume that in $n\to\infty$ roots inside the interval $(-1,0)$ either get complex or they are arbitrarily close to $-1$ or $0$ for any field component $N$.\\     
%%%%%%%%%%%%%%%New Subsection%%%%%%%%%%%%%%%%%%%%%%%%%%%%%%%
\subsection{$N$ dependence in $2<D<4$}
\begin{figure}[h!]
  \centering
      \includegraphics[width=0.55\textwidth]{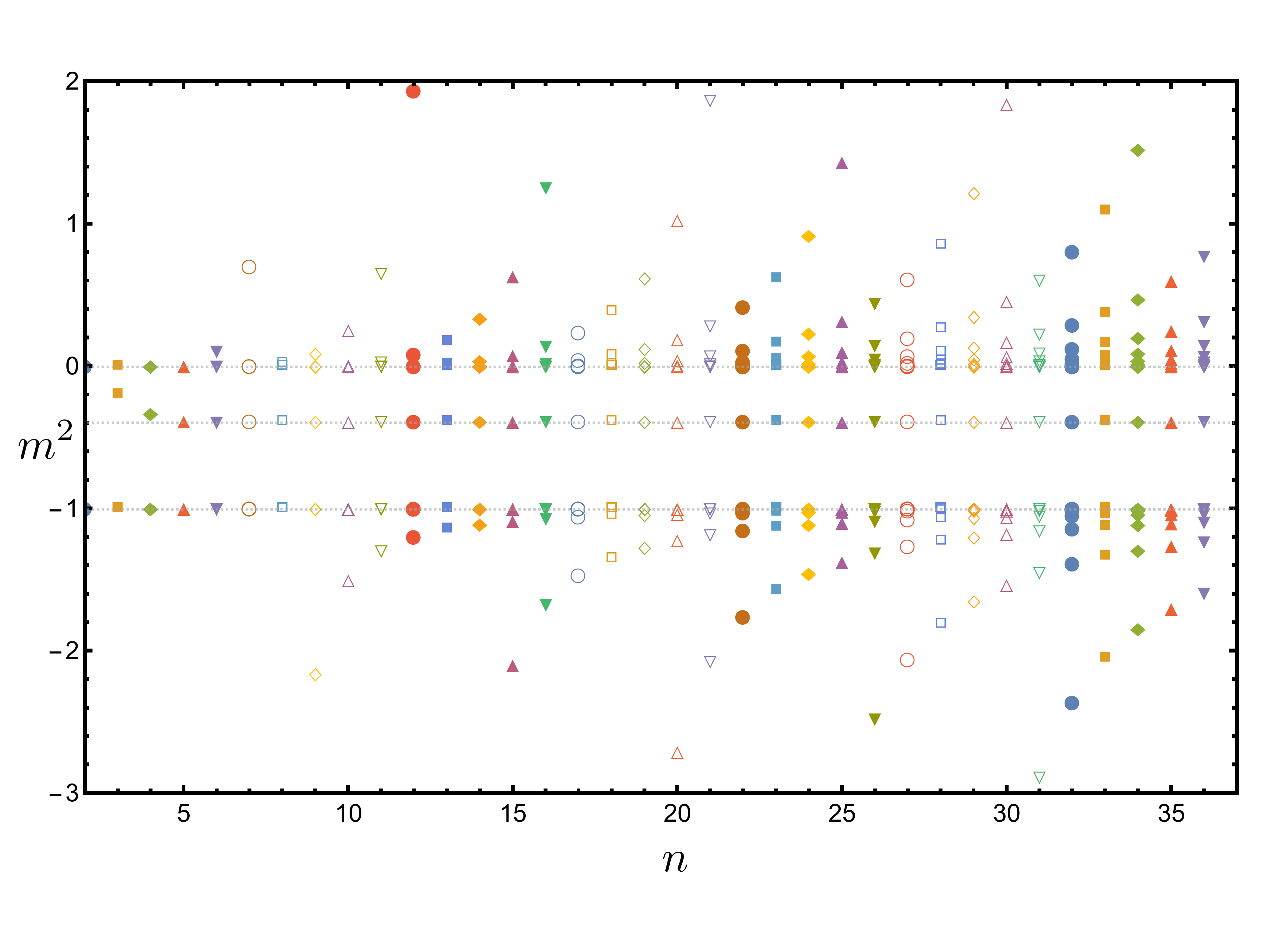}\vspace*{-.4cm}
  \caption{The roots of the VBF in $D=3$, $N\to\infty$ are shown. Comparing it to Fig.~\ref{rootsd3} the most striking difference is that there are no 
  "fake" roots between the root which is indicating the Wilson-Fisher ($m^2=-0.388$) and the one which is for the Gaussian, i.e. $m^2=0$. It is equally 
  interesting that  in this case we do not find any threshold beyond which the roots become complex in the $m^2>0$ region.} \label{3dLN}
\end{figure}

In dimensions $2<D<4$ for finite $N$ we will get a very similar result to the one which we obtained in Sec. \ref{wf}. For these values of the dimension,
one has the richest fixed point structure of all; hence it should be checked case by case. However, here we will only focus on the integer valued
dimension $D=3$. For the large\,-$N$ in $D=3$ the positions of the VBF roots show a significant difference comparing them to the finite $N$ (compare Fig.
\ref{3dLN} and Fig.~\ref{rootsd3}). We can identify the Wilson-Fisher fixed point just as we did it in Sec. \ref{wf}, but in this case we cannot see the additional
"fake" fixed points at any truncation level. Also for the $m^2>0$ roots we can see differences: they start to be organised in the pattern which leads us to
the distribution found in Sec. \ref{MW}, Eq.~(\ref{dist}). In this case it is also remarkable that no complex roots are found unlike for finite $N$.\\
Comparing our results to \cite{tetra1,tetra2}, where in the large\,-$N$ the fully analytic solution was derived without relying on the polynomial 
expansion, we can see that in the case of the analytic solution a line of fixed points starting from the Gaussian is found,  whereas in our case, using the VBF, seemingly, we do not have that. However, we can observe a unique behaviour of the VBF curves, namely, that the multiplicity of the Gaussian fixed point grows with the truncation order. More precisely, we have
\begin{equation}%\label{12dvbf}
\lambda_n \propto(-1)^{n+1} (m^2)^{\gamma(n)}(1+m^2)^n\sum\limits_{i=0}^{n-2}g_{i}(m^2)^i,
\end{equation} 
where $\gamma(n)$ is an integer function of the truncation order $n$: 
\begin{eqnarray}
\gamma(n)=\left\{ 
\begin{array}{lll}
		1  & \mbox{if } n=2,\,3 \\
		2 & \mbox{if } n=4 \\
		\frac{[n+1]}{2} & \mbox{if } n\geq5.
\end{array}
\right.
\end{eqnarray}
This is the only case (that is found), which has the property of growing multiplicity of the root $m^2=0$. That might be a fingerprint of the line of fixed points which can be found in the full analytical solution presented in the aforementioned papers.

%%%%%%%%%%%%%%%%%%%%%%%%%%%%%%%%%%%%%%%%%
%%%%%%%%%%%%%%%%%%%%%%%%%%%%%%%%%%%%%%%%%

%%%%%%%%%%%%%%%New Subsection%%%%%%%%%%%%%%%%%%%%%%%%%%%%%%%
\subsection{$N$ dependence in $D\geq 4$}
\begin{figure}[h!]
\begin{subfigure}{.55\textwidth}
  \centering
  \hspace*{-2cm}
  \includegraphics[width=.8\linewidth]{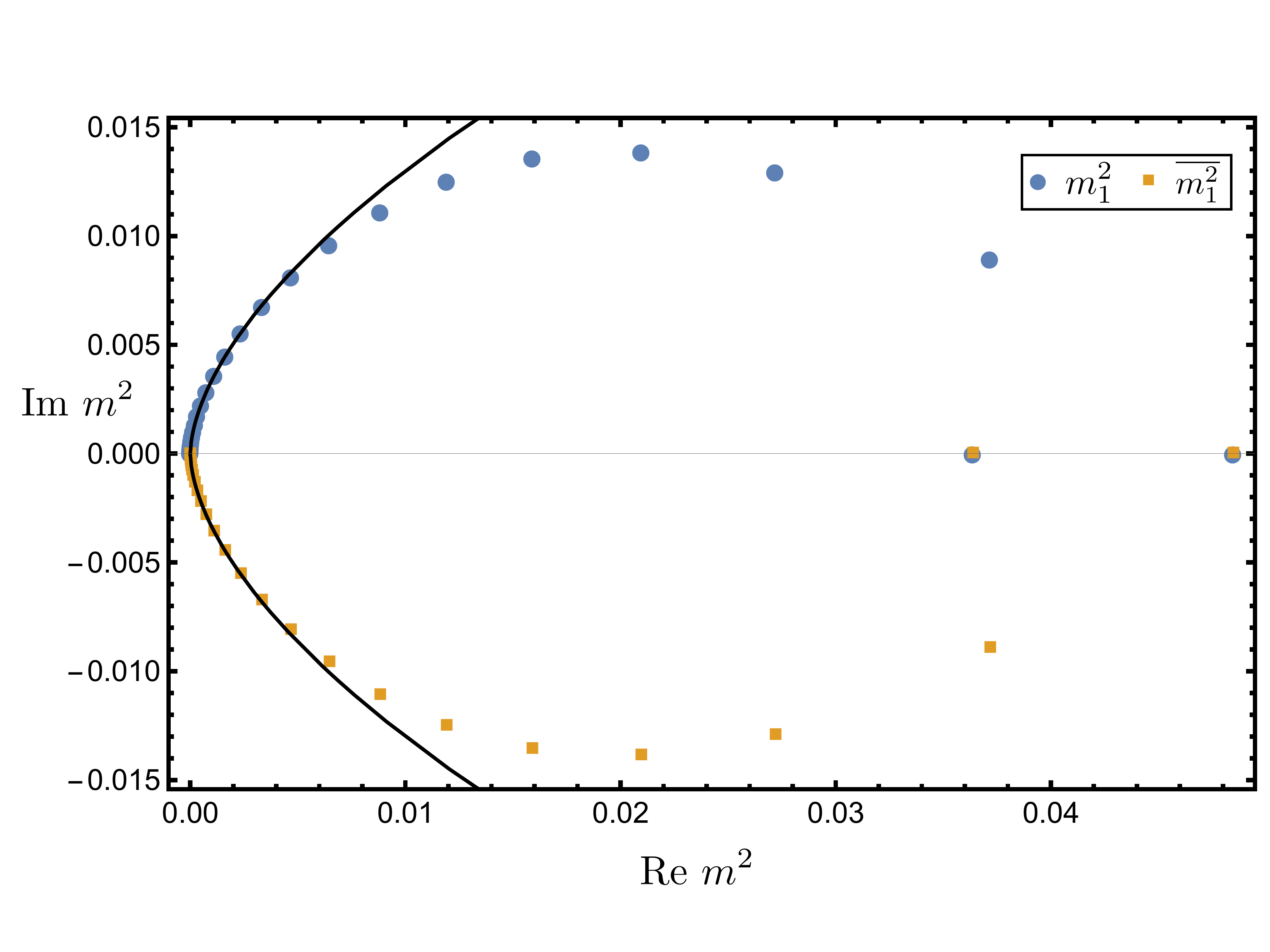}
  \caption{$N=4$}
 \end{subfigure}%
  \hspace*{-2cm}
\begin{subfigure}{.55\textwidth}
  \centering
  \includegraphics[width=.8\linewidth]{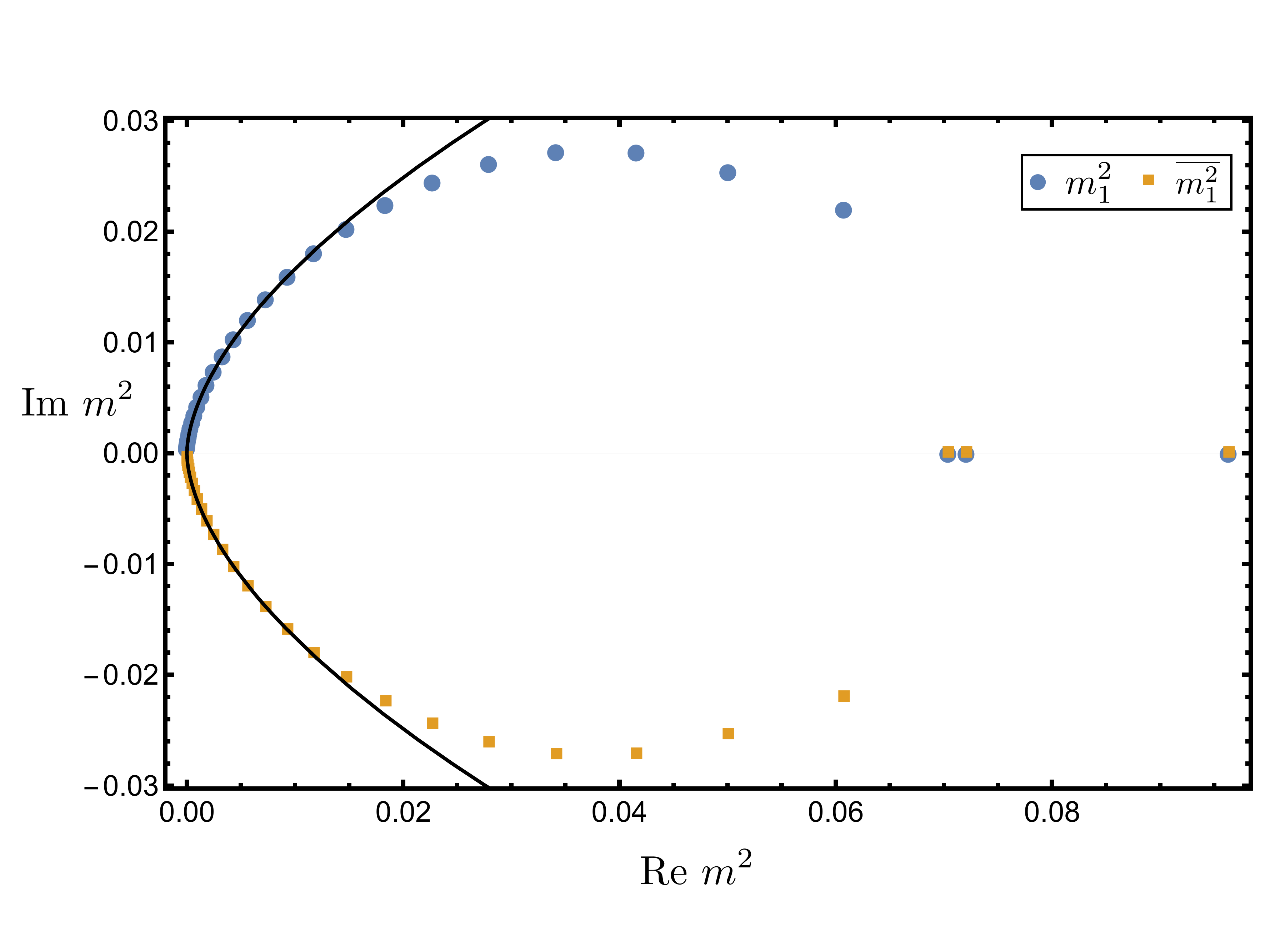}
  \caption{$N=15$}
  \end{subfigure}
\begin{subfigure}{.55\textwidth}
  \centering
  \hspace*{-2cm}
  \includegraphics[width=.8\linewidth]{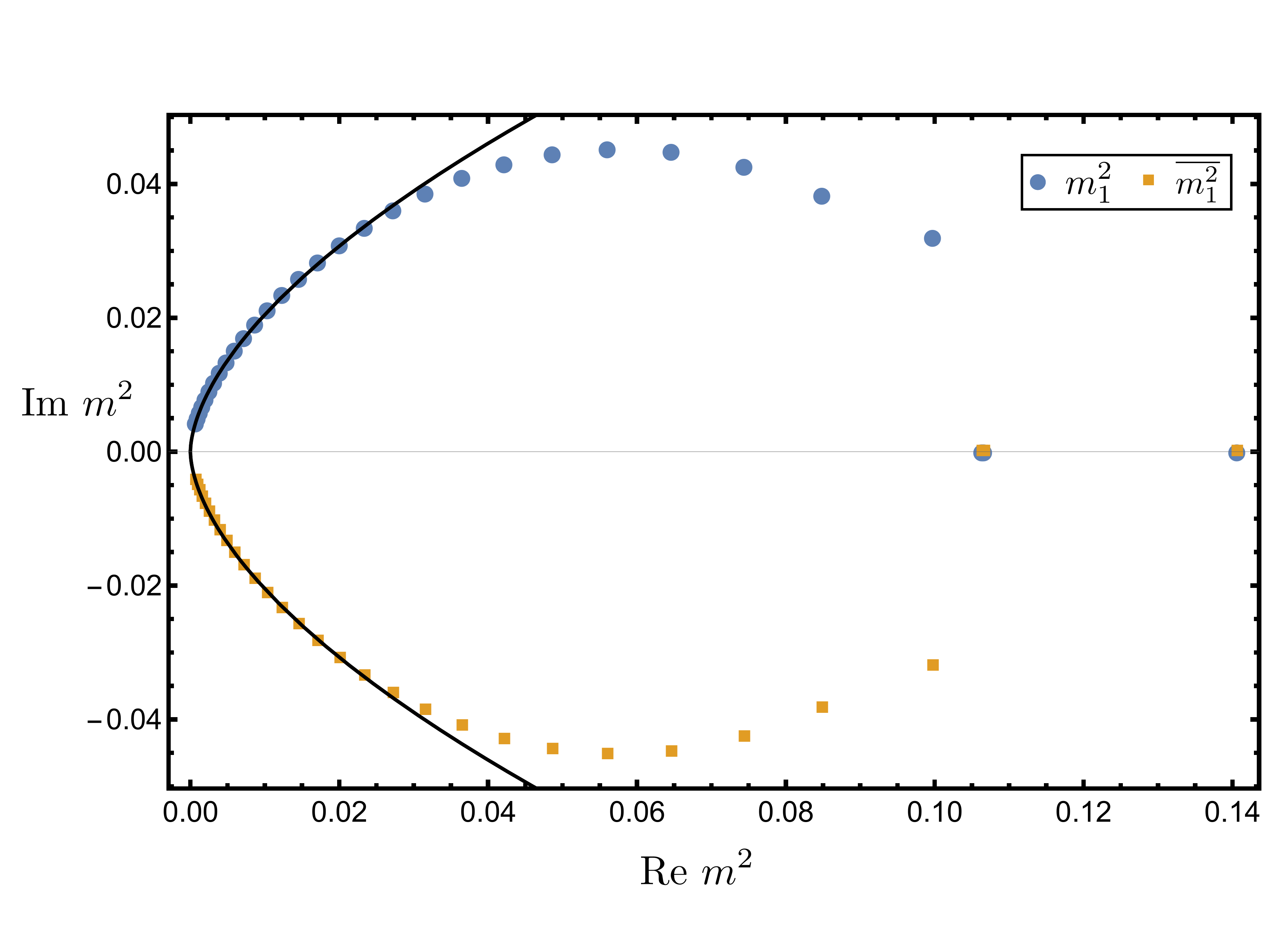}
  \caption{$N=40$}
\end{subfigure}%
\hspace*{-2cm}
\begin{subfigure}{.55\textwidth}
  \centering
  \includegraphics[width=.8\linewidth]{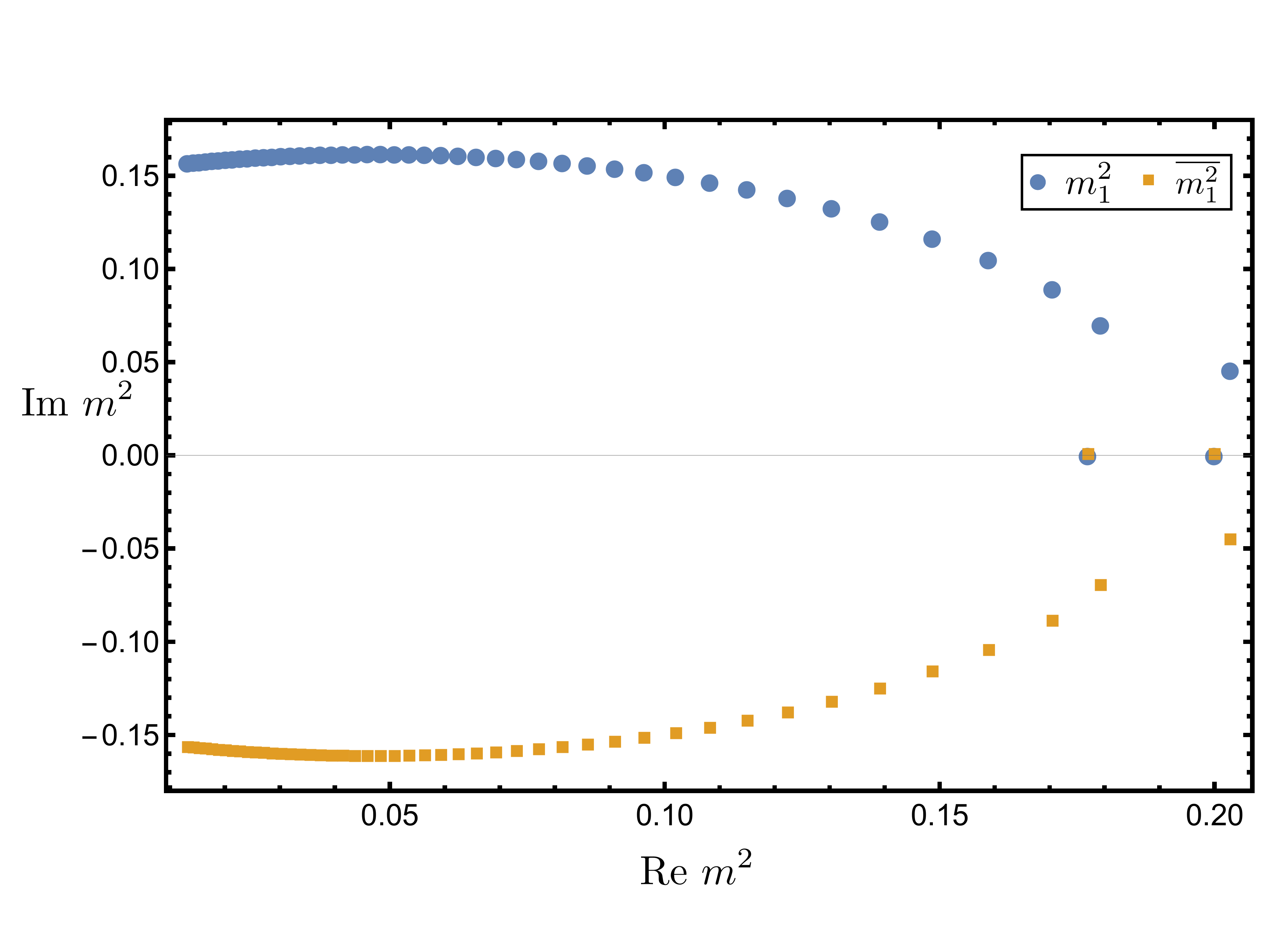}
  \caption{$N\to\infty$}
\end{subfigure}
\caption{The convergence to the origin on the complex plane is shown for (a) $N=4$,  (b) $N=15$ and (c) $N=40$. On (d) the $N\to\infty$ is
presented; in this case the number of the data points is not enough to fit a curve on the points near the origin (the expansion order in this case is $n=53$, and for
the finite $N$ cases $n=35$). However, the same behaviour is expected: the convergence to the origin slows down as $N$ grows. For the finite $N$ 
cases a power-law fit can be found around the origin (the fit was performed on 24 data points), for which the exponent seems to be universal. For details see the text.}\label{NdepCom}
\end{figure}

One of the most important results for theories $D\geq4$ was the triviality (Sec. \ref{trivi}). Our finding is that the modification of the number of the field
components $N$ does not give any qualitatively different result for any such theory, when $N$ kept finite. However, we can discover differences
between the results when $N<\infty$ and $N\to\infty$ for theories in $D>4$.\\
In four dimensions we will see the "compactified" $M^*$ pattern of the real part of the roots as in Fig.~\ref{d4r}. The only question is whether the convergence on the
complex plane towards the origin still holds. We present the results in Fig.~\ref{NdepCom}, for $N=4$, $N=15$, $N=40$ and $N\to\infty$. We can observe that the convergence to the origin slows down as $N$ grows, however, for the $N<\infty$ cases, it is possible to fit a curve on
them near the origin. The same form of the power-law is found for each $N$, that is, $f(x)=a x^b$, just like in the case of $N=1$ in Sec. \ref{trivi}. The
parameters are the following: $a=0.193\pm0.001$, $b=0.587\pm 0.002$ for $N=4$; $a=0.249\pm0.002$, $b=0.586\pm 0.002$ for $N=15$; and
$a=0.302\pm0.006$, $b=0.584\pm 0.004$ for $N=40$. It is interesting that close to the origin the exponent of the power-law seems to take the value $0.586\pm0.002$, and thus one can speculate whether it is universal.
%%%%%%%%%%%%%%%%%%%%%%%%%%%%%%%%%%%%%%%%%

\begin{figure}[t!]
\begin{subfigure}{.47\textwidth}
  \centering
  \includegraphics[width=.9\linewidth]{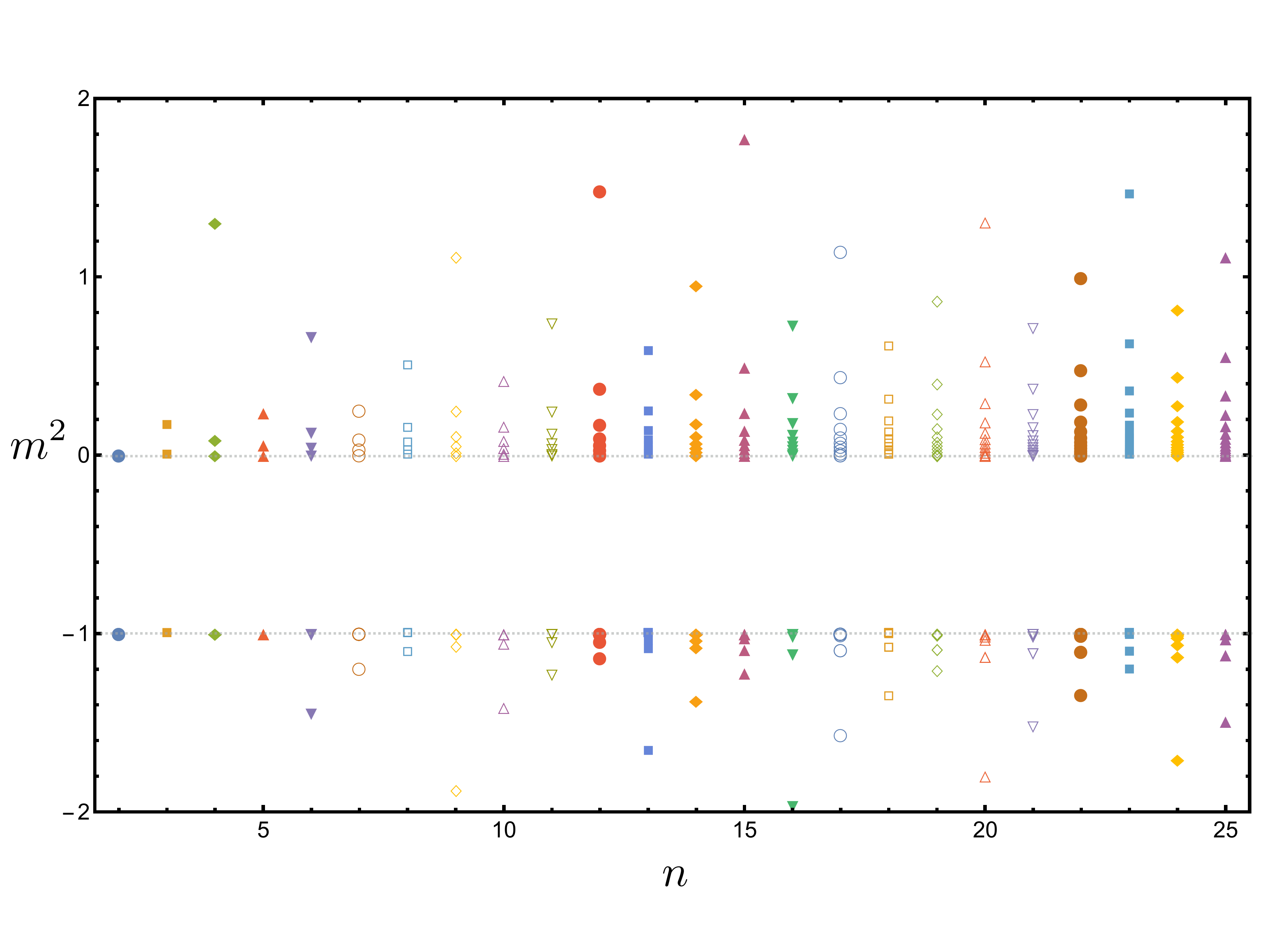}
  \caption{$D=5$, $N=6$}
 \end{subfigure}%
\begin{subfigure}{.47\textwidth}
  \centering
  \includegraphics[width=.9\linewidth]{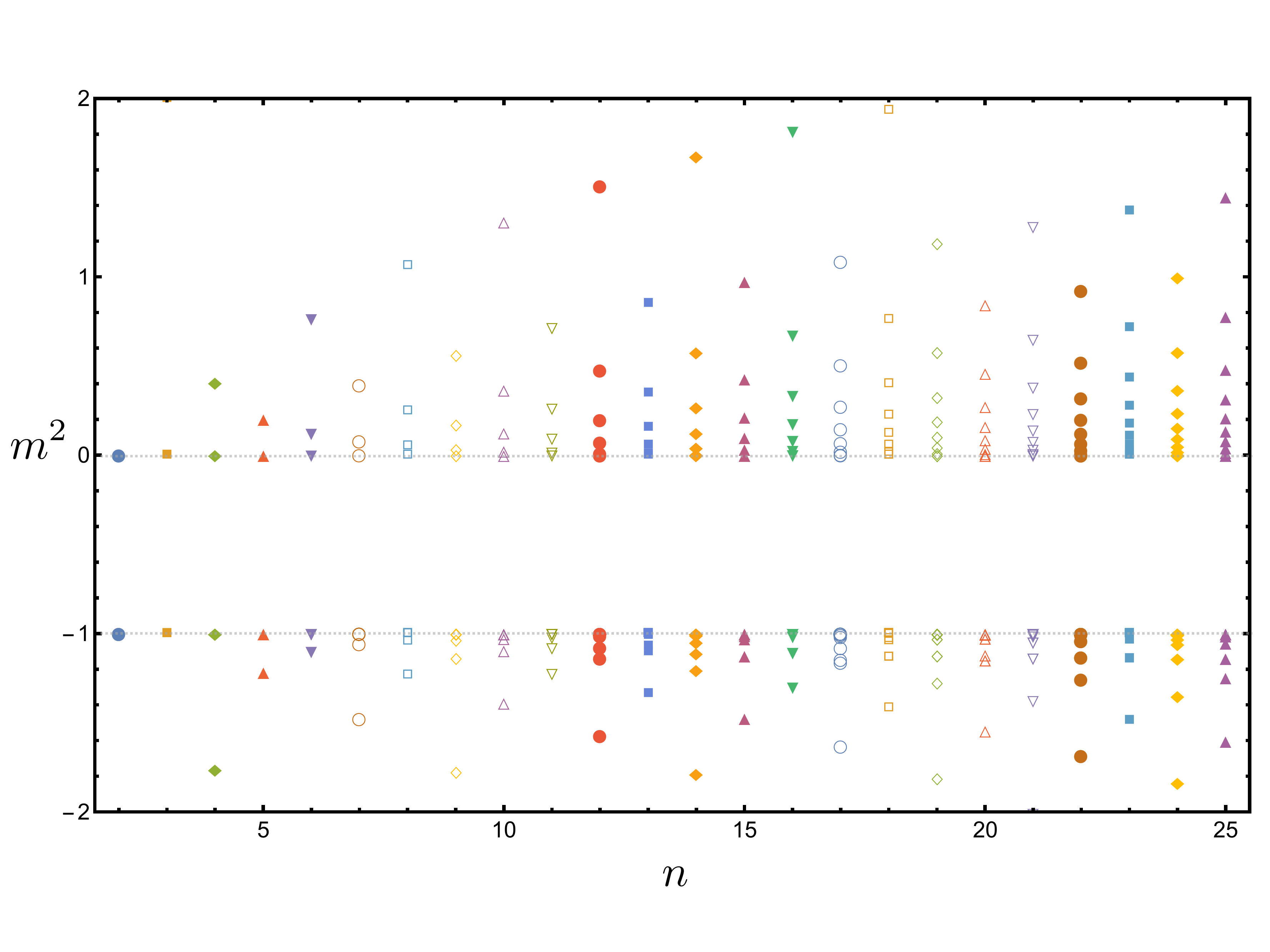}
  \caption{$D=8$, $N=10$}
  \end{subfigure}
 % \hspace*{3.5cm}
\begin{subfigure}{.47\textwidth}
   %\hspace*{4cm}
  \centering
  %\hspace*{4cm}
  \includegraphics[width=.9\linewidth]{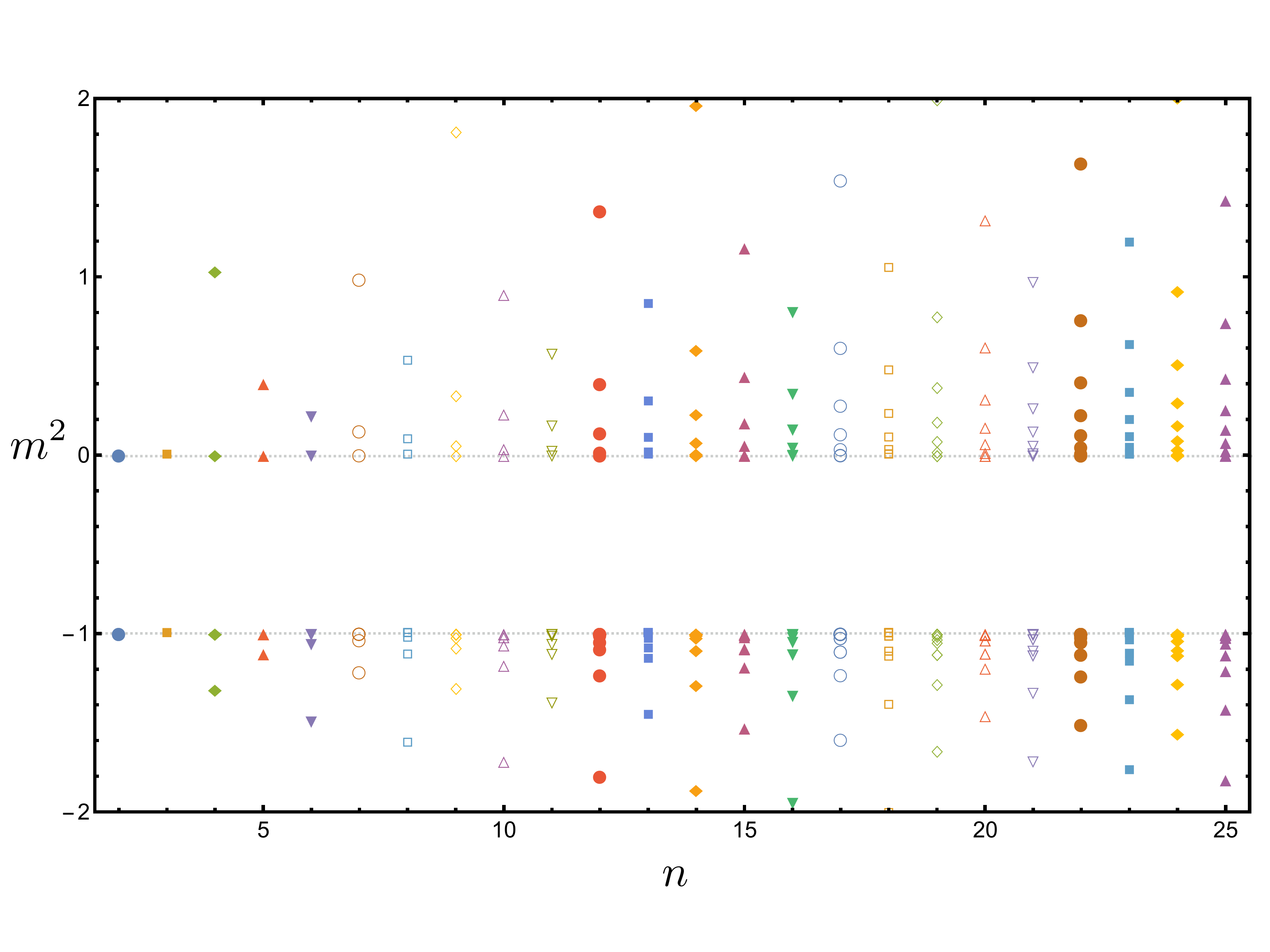}
   %{\hspace*{4cm}
   %{\hspace*{-.5cm}
  \caption{$D=11$, $N=12$}%}
\end{subfigure}
\caption{The VBF roots of the O($N$) model in (a) $D=5$ and $N=6$, (b) $D=8$ and $N=10$, and (c) $D=11$ and $N=12$. On each plot the accumulation
of the roots at $m^2=-1$ and $m^2=0$ can be observed. For the roots $m^2<-1$ there can be small deviation from the $M^*$ pattern (most likely due to the extreme
flatness of the VBFs around $-1$), but at some point the pattern restores completely.}\label{NdepG4}
\end{figure}
Regarding the $N\to\infty$ case, we did not reach the region where we could do the fit; however, the absolute value of the imaginary part reached its maximum and started to decrease. We expect the same behaviour (maybe with a different exponent) as for the finite $N$ cases; however, to see that we would need to go to a higher expansion order than the one used in this case ($n=53$).\\
For theories in greater than four dimensions the fixed point structure does not change much compared to the results of Sec. \ref{trivi}; only quantitative
differences can be observed. We will present a few plots on the root positions in Fig.~\ref{NdepG4}. Arbitrarily we have chosen the cases
$\{D=5,N=6\}$, $\{D=8,N=10\}$ and $\{D=11,N=12\}$. The roots are positioned in such a way that they accumulate more and more at the two stable roots $-1$ and $0$,
which is essential for triviality. There can be minor deviation from the pattern $M^*$; however, at some point the pattern is restored completely, and one can obtain the triviality in the same way that we did it for the case $D=5$, $N=1$.\\

\begin{figure}[htb]
\begin{subfigure}{.4\textwidth}
 \centering
\hspace*{-.5cm}
% \vspace*{-.6cm}
  \includegraphics[scale=.2]{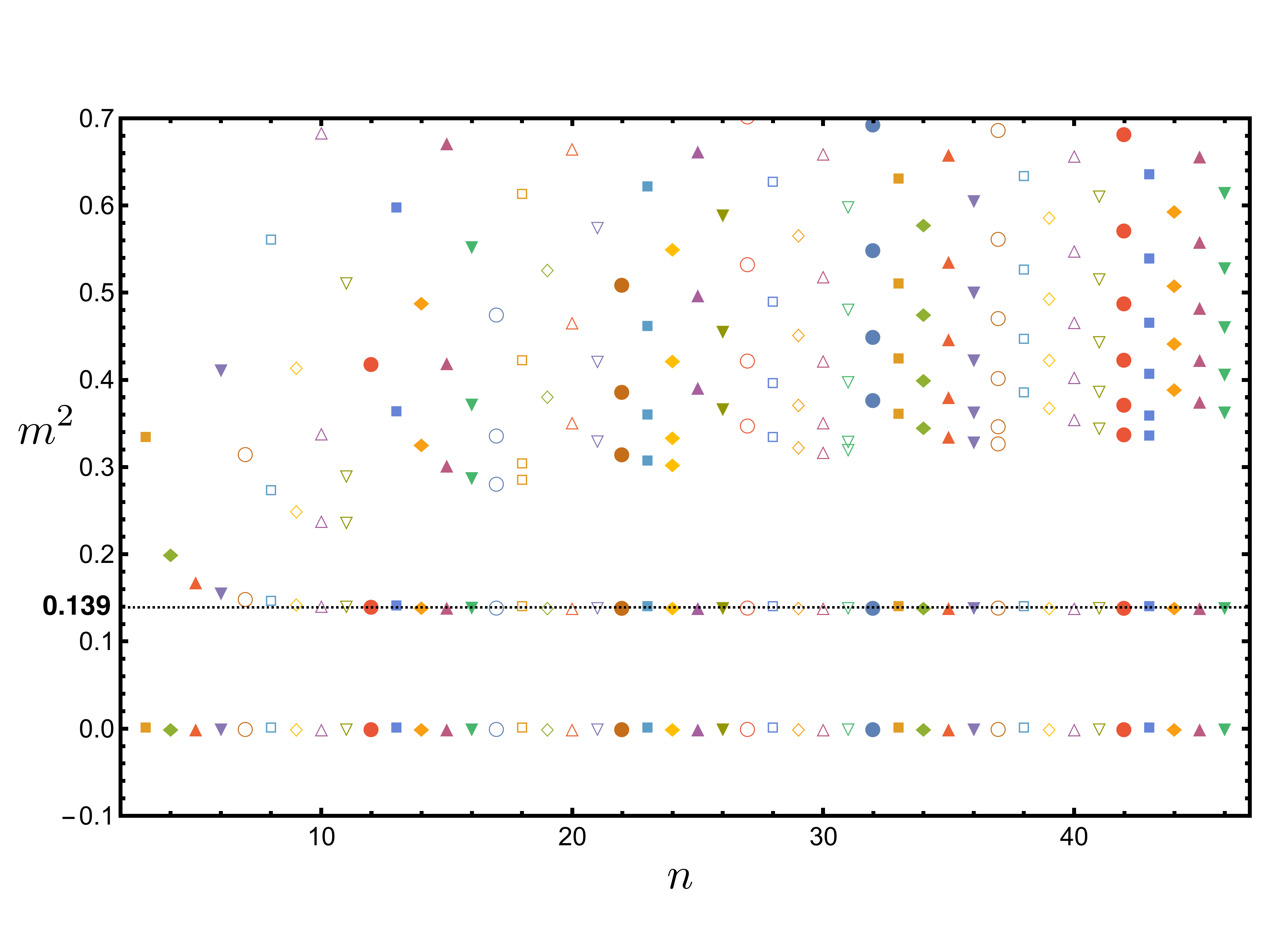}
   \end{subfigure}
   %\vspace*{-.6cm}
\begin{subfigure}{.4\textwidth}
  \centering
 % \hspace*{1.3cm}
 %\vspace*{.6cm}
  \includegraphics[scale=.2]{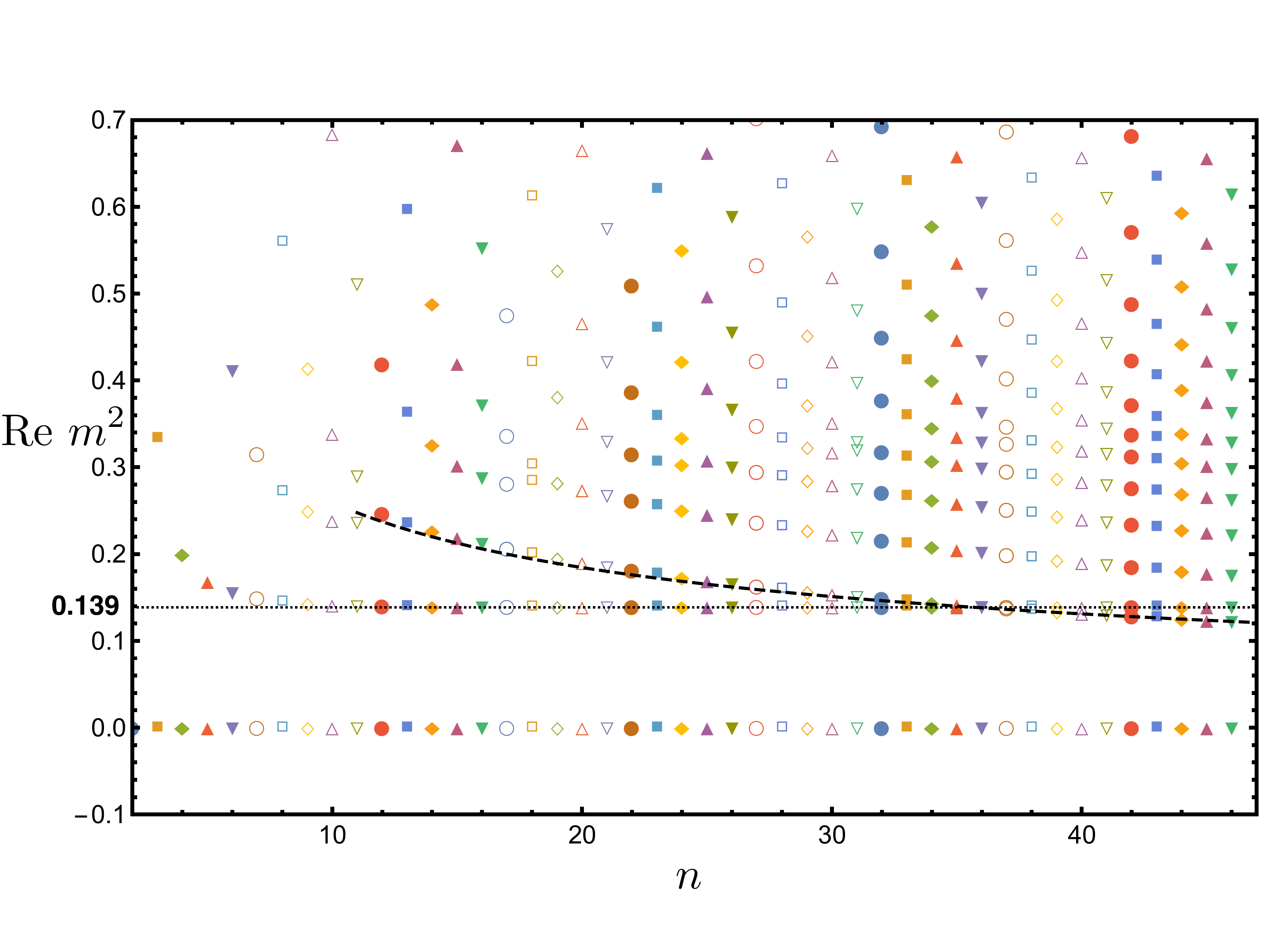}
  \end{subfigure}
\caption{In the left panel the root structure provided by the real roots is shown. A stable line of the roots can be observed at the value $m^2=0.139$. In the right panel the same root structure is shown, but now the real parts of the complex roots have been taken into account, too. The real parts of the complex roots show
the $M^*$ pattern, and also a power-law curve can be fitted on them: $f(x)=a x^b$, with $a=0.807\pm0.005$ and
$b=-0.492\pm0.002$. In the $n\to\infty$ we expect to find triviality again. For details see the text.}\label{D5LNtriv} 
\end{figure}

A more interesting case that can be found is in the large\,-$N$ limit. From Eq.~(\ref{LN}) one can derive again the VBF curves in $D>4$. In the present
section we are going to study only the $D=5$ theory for the large\,-$N$ limit. In Fig.~\ref{D5LNtriv} one can see the root positions for this model. The first thing
that we can notice is that there is a stable line at $m^2=0.139$ which would signal a new fixed point solution. There has been a recent work 
\cite{Percacci} on the topic of whether such a fixed point exists in models $4<D<6$ and the answer was that, if there is such a fixed point it will provide an unphysical 
fixed point potential. In an earlier work \cite{Xav} related to holography a similar conclusion was drawn.\\
In the papers \cite{Klebanov} an IR fixed point was found
for an O($N$) symmetric theory with $N+1$ scalars in $D=6-\epsilon$ for sufficiently large $N$, including a cubic interaction, too. It was argued that this IR fixed point
of the cubic O($N$) theory with $N+1$ scalars is equivalent to a UV fixed point of the O($N$) model in the large\,-$N$, which would imply an unexpected
asymptotically safe behaviour of such theories. Now, since we have found such a fixed point, we need to
check whether it gives a stable fixed point potential.   
Substituting the value $m^2=0.139$ into the VBF polynomials will provide the exact value for the corresponding coupling at that fixed point, hence
defining the fixed point potential. The finding is the following: the highest order in the expansion is $n=46$, and what we get is $
\lambda_n(m^2=0.139)\leq0$ for $2\leq n\leq 46$. This signals that the fixed point potential defined by this root is unbounded from below in agreement with 
the findings in \cite{Percacci}. Since the RG equation of the effective potential can be considered to be exact in the large\,-$N,$ this unstable fixed point should be found also beyond LPA. In Fig.~\ref{D5LNtriv} one can also observe that there is a gap between the roots corresponding to this newly found unstable 
fixed point, which can be filled in by considering the real part of the roots. In the same figure, considering the lower panel, we can see how the real parts of such 
roots behave. We can fit a power-law decaying curve again on the real part of the first root. This curve goes through the line $m^2=0.139$ and in the 
asymptotic limit goes to zero. The imaginary part behaves pretty much in the same way as in the $D=4$ case (see Fig.~\ref{NdepCom}). Here, we 
can use the same argument as for $D=4$, $N=1$: one needs to consider the positions of the roots on the complex plane in order to catch the underlying physics. According to our findings, the real parts of these roots exhibit the same $M^*$ pattern; hence for $n\to\infty$ we can expect them to accumulate 
at $m^2=0$, leaving us with three fixed points, which are defined by the roots: $m^2=-1$ (since for $m^2<-1$ the root structure is the same as it was for the 
finite  $N$ case), $m^2=0$ and $m^2=0.139$. Out of these three roots only the Gaussian ($m^2=0$) seems to provide a stable fixed point potential; thus triviality 
was found again.
%%%%%%%%%%%%%%%%%%%%%%%%%%%%%%%%%%

%%%%%%%%%%%%%%%%%%%%%%%%%%%%%%%%%%%
%%%%%%%%%%%%%%%%%NEW SECTION%%%%%%%%%%
\section{The fractal dimensions}\label{fractal} 
\begin{figure}[h!]
\begin{subfigure}{.4\textwidth}
  \centering
  \includegraphics[width=1\linewidth]{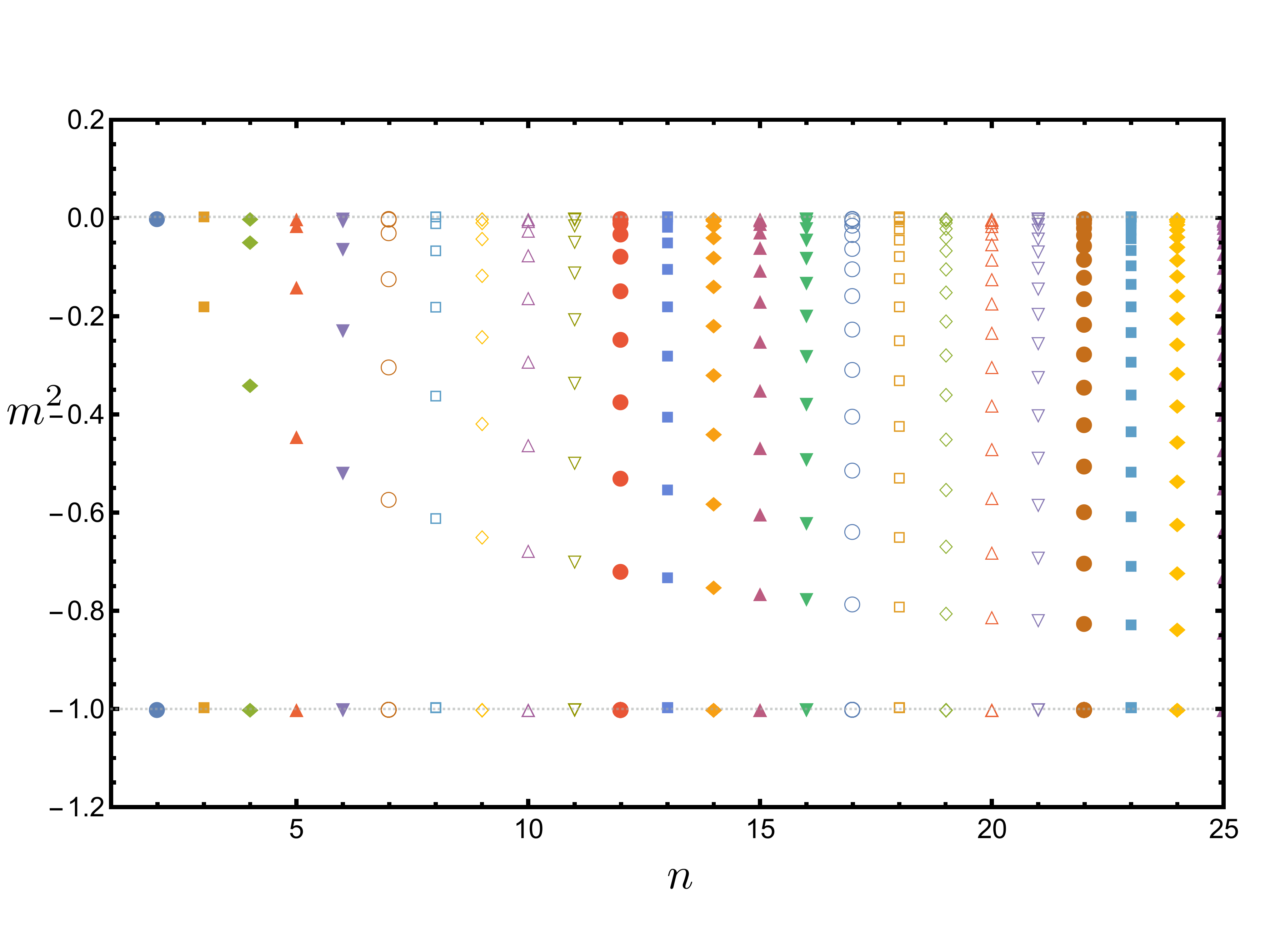}
  \caption{$D=1.3,N=1$}
 \end{subfigure}%
\begin{subfigure}{.4\textwidth}
  \centering
  \includegraphics[width=1\linewidth]{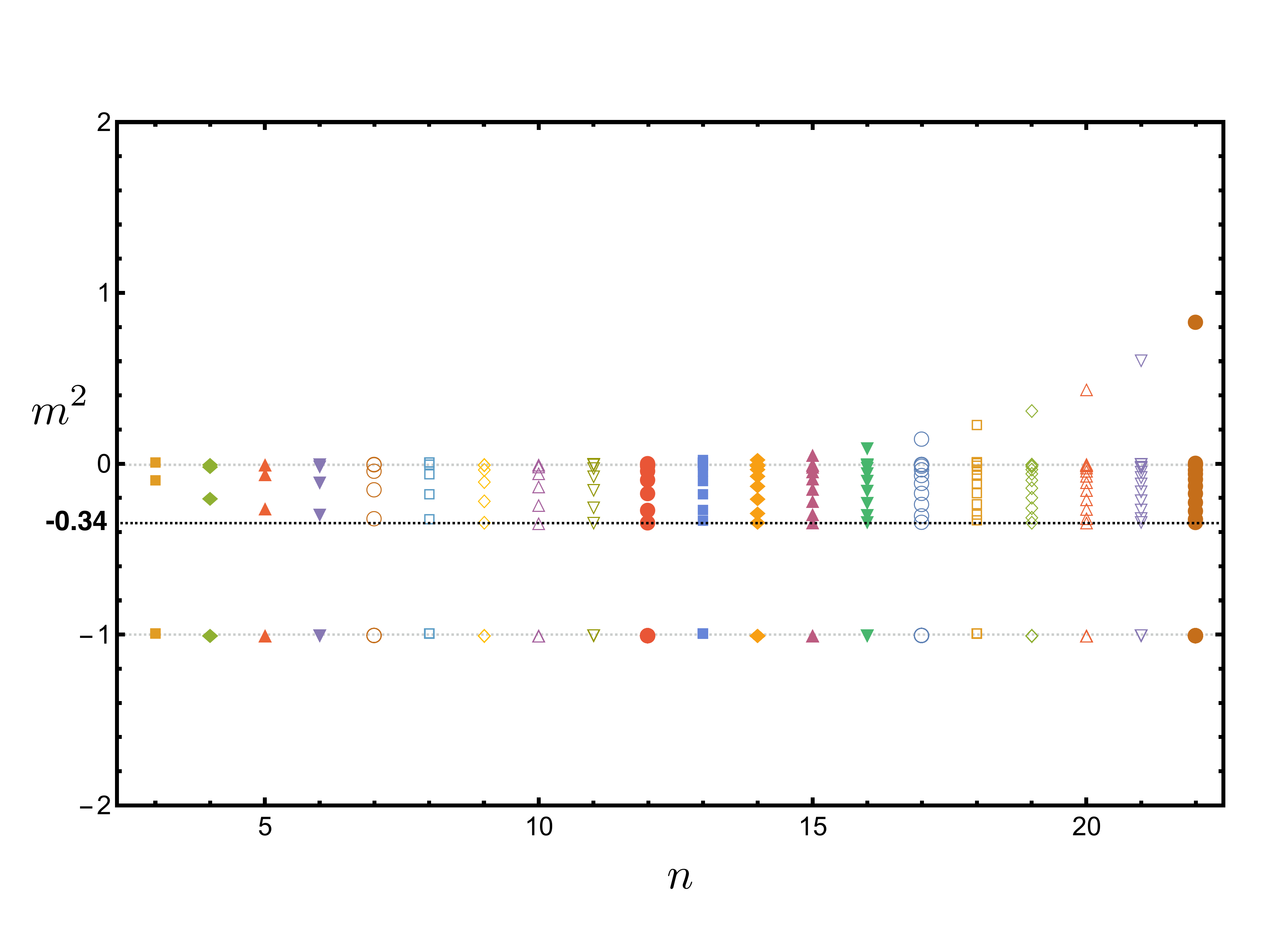}
  \caption{$D=2.6,N=1$}
  \end{subfigure}
\begin{subfigure}{.4\textwidth}
  \centering
  \includegraphics[width=1\linewidth]{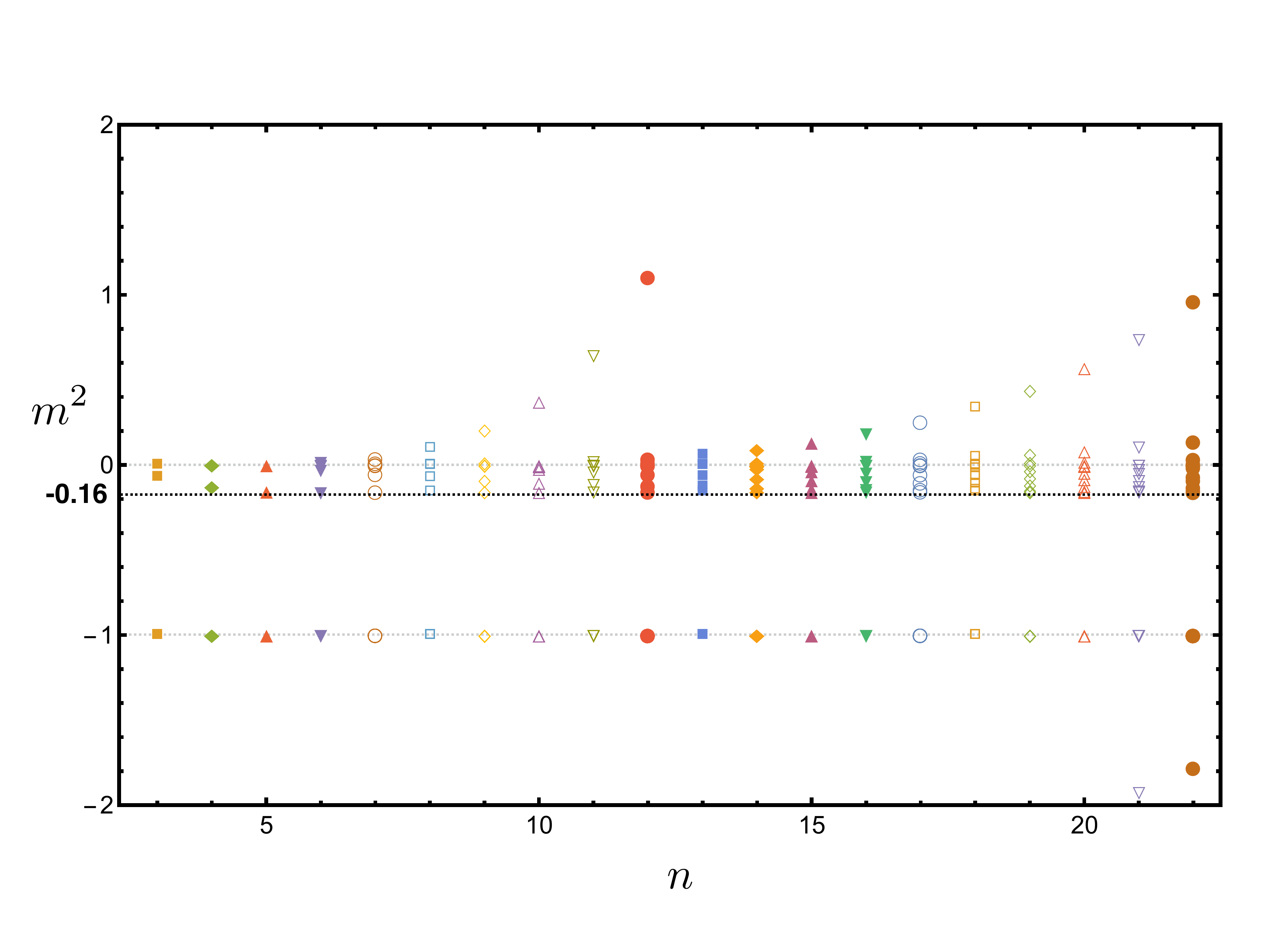}
  \caption{$D=3.1,N=1$}
\end{subfigure}%
\begin{subfigure}{.4\textwidth}
  \centering
  \includegraphics[width=1\linewidth]{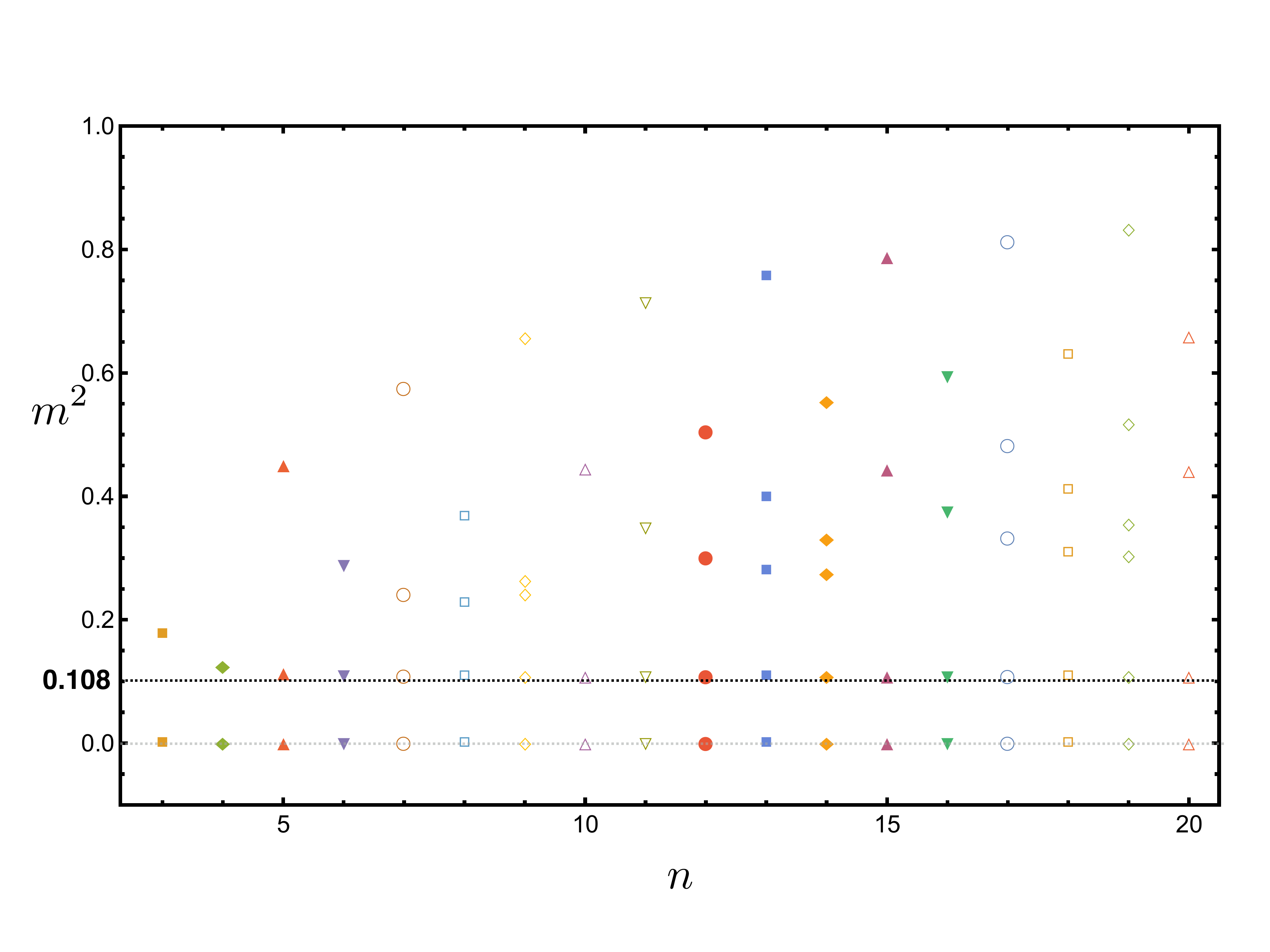}
  \caption{$D=4.6,N\to\infty$}
\end{subfigure}
\caption{The root structure is shown for some fractal-dimensional cases. The usual fixed points are found for $4<D$; thus the most remarkable is (d), where we
obtain a fixed point candidate for $D=4.6$ in the large\,-$N$. However, it turned out to define an unstable fixed point potential just like it is shown in \cite{Percacci}.}
\label{frac}
\end{figure}
We present a few examples of the fractal-dimensional cases in the intervals of the dimension that we have investigated. In Fig.~\ref{frac} the results are
shown for \{$D=1.3, N=1$\}, \{$D=2.6, N=1$\}, \{$D=3.1, N=1$\} and \{$D=4.6, N\to\infty$\}. In the first three cases the usual fixed points are found; however it could be possible to find additional fixed points for $2<D<4$, but that would require a more detailed study of this interval of the dimension. For the two $D$ values,
which are in this interval, we found the WF fixed point, too, and it can be observed that they are situated at different values of $m^2$: for $D=2.6$ $m^2_{WF}
=-0.34$ and for $D=3.1$ it oscillates around $m^2_{WF}=-0.16$. The most remarkable case is when we set $N\to\infty$ and choose $D=4.6\in(4,6)$. Just
as we saw it for the large\,-$N$ in $D=5$, here, we find again a fixed point candidate at the value $m^2=0.108$. However, again in agreement with
\cite{Percacci}, this fixed point is found to be an unphysical one, i.e. the root $m^2=0.108$ defines an unstable fixed point potential. (More precisely, it is a metastable potential.)

%\vspace*{1.5cm}
\section{Conclusion} 
We investigated the fixed point structure of the O($N$) model for various dimensions and filed components. In this paper we considered only the 
integer-dimensional cases in detail, but results for fractional dimensions are presented in the last section, too. For our analysis we used the LPA with a Taylor-expanded potential around zero VEV of the field. In this case from the fixed point equation one is able to express all the fixed 
point couplings through a polynomial in the mass of the theory; see Eq.~(\ref{genvbf}). Each $n$th polynomial is obtained from the fixed point equation of the $
(n+1)$th coupling; that is, the equation for $\partial_t \lambda_{n+1}(\lambda_n,m^2)=0$ can be solved for $\lambda_n=\lambda_{n}(m^2)$, hence the name 
VBF (vanishing beta function) curve. A general property of these polynomials is that they always have a root at $m^2=-1$ and $m^2=0$ corresponding to 
the convexity (or IR) and the Gaussian fixed point, respectively.
These polynomials can be obtained for arbitrary components of the field $N$ and dimension $D$. To find a true fixed point at a 
given level of truncation, we established the following rule: let $m_0^2$ be a root of the polynomial $\lambda_n(m^2)$; then a physically well-defined (i.e. 
bounded from below) fixed point potential is given by the set of couplings $\{\lambda_{n-1}(m_0^2),\lambda_{n-2}(m_0^2),...,\lambda_1(m_0^2)=m_0^2\}$, 
provided $\lambda_{n-1}(m_0^2)\geq0$ [if it happens to be a root for the $(n-1)$th polynomial, too, then the same rule holds for the $(n-2)$th polynomial, and 
so on]. Using this rule to find fixed points at a truncation level $n$ gave us a nice opportunity to make considerations about theories in different dimensions.\\
First, we considered theories for $D\leq 2$, and in particular, we analysed the case of $D=2$ in detail. The finding is that for all the theories we considered
up to and including dimension 2 they behave qualitatively in the same way for $N\geq2$. In such theories for continuous symmetries the Mermin-Wagner theorem must 
hold as it was shown in \cite{ssb}. We provided a statistical approach to show the validity of the theorem by considering the statistics of the root patterns. It was found that for such systems ($D\leq 2$) all the roots are in the closed interval $[-1,0]$. Since at every truncation level a new root appears, it seems to be paradoxical to prove the 
Mermin-Wagner theorem: we would not expect any root inside the interval but practically there appears (countably) infinitely many. A true fixed point in this 
interval signals a SSB phase; hence the MW theorem seemed to be violated. However, it turned out that this is only the case for finite $n$ truncation. As $n\to \infty$ 
we can recover the MW theorem by simulating the position of the roots and derive their distribution in the infinite limit; see Sec. \ref{MW}. For discrete 
symmetry $N=1$ the O($N$) model is equivalent to the Ising model which has a SSB phase in $D=2$. This result is also shown as a part of this section.\\
In Sec. \ref{wf} we investigated the theory for $D=3$ with $N=2$, and in $2<D<4$ the finding is that the theories that have been considered have similar 
properties. We found an additional fixed point beside the Gaussian and the IR fixed point, namely, the Wilson-Fisher fixed point. Although these fixed points 
have been found, we could not clearly analyse the root structure for the $m^2>0$ region. It is also possible that one could find additional fixed points in $2<D<4$ for fractional dimensions, but it would require more detailed study. The main result of this section was that we were able to show the appearance of the Wilson-Fisher fixed point.\\
For $D\geq4$ we found two qualitatively different results (see Sec. \ref{trivi}). However, essentially they lead us to the same physics, namely, to the
triviality of the model; that is, only the Gaussian fixed point exists for such theories. For theories $D>4$ we analysed the $D=5$ case, where the fixed point structure shows similarity to the $D=2$ case, but now the roots only appear outside of the interval $(-1,0)$; however, for the root pattern the statistical simulation can be used again if we compactify the real line of $m^2$. In this way triviality can be shown. The situation for $D=4$ is different: here we need to consider the complex roots that
appear during the calculations, but it can be shown that both the real and the imaginary part converges to the origin on the complex plane; hence leading to the triviality of the model.\\ 
In Sec. \ref{Ndepp} we showed some results considering different values of $N$. The most interesting outcome is connected to the recent results in 
\cite{Percacci,Klebanov}, where it was argued that in the O($N$) model for $N\to\infty$ if a non-trivial fixed point exists in dimensions $4<D<6$, then it defines 
an unstable fixed point potential. From our analysis we found such a fixed point, and indeed, it can be shown that it defines an unbounded effective potential.
In the last section we gave some results for fractional dimensions, obtaining similar behaviour to the integer-dimensional cases.\\ 
According to our findings, we can draw the conclusion that although we used the local potential approximation during our computations, that seems to be enough to
obtain the right qualitative physical results (Mermin-Wagner theorem, the presence of the Wilson-Fisher fixed point and triviality), when one extrapolates to infinite order the results obtained in a given order of the Taylor expansion. On the other hand, one has to be careful with such expansions, since as we saw, at finite level of the truncation it can generate fake fixed points from the physical point of view.  

%-------------------------------------------
\section*{Acknowledgements}
%-------------------------------------------
I would like to thank Istv\'an N\'andori and Antal Jakov\'ac for fruitful discussions. I am also grateful to Zsolt Pajor-Gyulai, Matteo Giordano, Zolt\'an Sz\H or 
and \'Akos Nagy for very useful comments.  This research was supported by the European Union and the State of Hungary, co-financed by the European 
Social Fund in the framework of TAMOP-4.2.4.A/ 2-11/1-2012-0001 'National Excellence Program' and by the Hungarian National Fund OTKA-K104292.

%---------------------------------------------
\appendix 
%-------------------------------------------
\section{The proof of the nested formula}\label{nested app}

We would like to prove the nested formula for the coupling constants in  Eq.~\eq{nestedf}.
The easiest to consider first is the case for the large\,-$N$, $D=2$, because the RG equation has the simplest form for 
these parameters [see Eq.~(\ref{LN})], however, it can be generalised to arbitrary $N$ and $D$, which we will present after this simpler example.
Without the loss generality, we set the constant $A_D=1$.
Our starting point is the RG equation, and we start to differentiate it with respect to $\rho$:
\bea
\partial_t u&=&-2 u(\rho )+\frac{1}{u'(\rho )+1},\nn
\partial_t u'&=&-2 u'(\rho )-\frac{u''(\rho )}{\left(u'(\rho )+1\right)^2},\nn
\partial_t u''&=&-2 u''(\rho )+\frac{2 u''(\rho )^2}{\left(u'(\rho )+1\right)^3}-\frac{u^{(3)}(\rho )}
{\left(u'(\rho )+1\right)^2},\nn
\partial_t u'''&=&-2 u^{(3)}(\rho )-\frac{6 u''(\rho )^3}{\left(u'(\rho )+1\right)^4}+\frac{6 u^{(3)}(\rho ) 
u''(\rho )}{\left(u'(\rho )+1\right)^3}-\frac{u^{(4)}(\rho )}{\left(u'(\rho )+1\right)^2},\nn
...
\eea
Let us rewrite the last three equation in the following form:
\bea
\partial_t u'&=&F_1(u')+g(u')u'',\nn
\partial_t u''&=&F_2(u',u'')+g(u')u^{(3)},\nn
\partial_t u'''&=&F_3(u',u'',u''')+g(u')u^{(4)},
\eea
where we defined
\bea
F_1(u') && \equiv -2 u'(\rho ),\nn
F_2(u',u'')&& \equiv -2 u''(\rho )+\frac{2 u''(\rho )^2}{\left(u'(\rho )+1\right)^3}-\frac{u^{(3)}(\rho )}
{\left(u'(\rho )+1\right)^2},\nn
F_3(u',u'',u''')&&\equiv-2 u^{(3)}(\rho )-\frac{6 u''(\rho )^3}{\left(u'(\rho )+1\right)^4}+\frac{6 u^{(3)}(\rho ) 
u''(\rho )}{\left(u'(\rho )+1\right)^3},\nn
g(u')&& \equiv- \frac{1}{\left(u'(\rho )+1\right)^2}.
\eea
We can find the following relations between $F_i$'s:
\bea\label{relat}
F_2(u',u'')&&=\frac{\d F_1}{\d \rho}+\frac{\d F_1}{\d u'}u'' + \frac{\d g(u')}{\d u'} u''^2 ,\nn
F_3(u',u'',u''')&&=\frac{\d F_2}{\d \rho}+ \frac{\d F_2}{\d u'}u''+\frac{\d F_2}{\d u''}u'''+ \frac{\d g(u')}{\d u'} u'' u'''.
\eea
Of course, $\d F_i/\d\rho=0$ since in this case there is no explicit $\rho$ dependence, but for the sake of consistency, we will indicate this term, as well.
We can make the following statement for $n\geq 1$:
\bea\label{ind}
\partial_t u^{(n)}&=&F_n(u',u'',u''',...,u^{(n)})+g(u')u^{(n+1)},\nn
F_{n}(u',u'',u''',...,u^{(n)})&=&\frac{\d F_{n-1}}{\d u'}u''+\frac{\d F_{n-1}}{\d u''}u'''+...+\left(\frac{\d F_{n-1}}{\d u^{(n-1)}}+ \frac{\d g(u')}{\d u'} u''\right)u^{(n)}.\nn
\eea
We are going to show this by induction. Let us suppose Eq.~(\ref{ind}) is true. We are going to show that it holds for $n+1$, too. Let us differentiate Eq.~(\ref{ind}) once with respect to $\rho$. It yields
\bea
\partial_t u^{(n+1)}&=&\frac{\d F_n}{\d \rho} + \frac{\d F_n}{\d u'}u''+\frac{\d F_n}{\d u''}u'''+...+\frac{\d F_n}{\d u^{(n)}}u^{(n+1)}  + \frac{\d g(u')}{\d u'} u'' u^{(n+1)} + g(u')u^{(n+2)},\nn
\partial_t u^{(n+1)}&=&\frac{\d F_n}{\d \rho} + \frac{\d F_n}{\d u'}u''+\frac{\d F_n}{\d u''}u'''+...+\left(\frac{\d F_n}{\d u^{(n)}}+ \frac{\d g(u')}{\d u'} u''\right)u^{(n+1)} + g(u')u^{(n+2)},\nn
\partial_t u^{(n+1)}&=& F_{n+1}(u',u'',u''',...,u^{(n+1)}) + g(u')u^{(n+2)},\nn
\eea
where
\bea
F_{n+1}(u',u'',u''',...,u^{(n+1)})=\frac{\d F_n}{\d u'}u''+\frac{\d F_n}{\d u''}u'''+...+\left(\frac{\d F_n}{\d u^{(n)}}+ \frac{\d g(u')}{\d u'} u''\right)u^{(n+1)}.\nn
\eea
In this way it was shown that the rhs always depends on the highest derivative of $u(\rho)$ linearly, e.g. $u^{(n+1)}$ in Eq.~(\ref{ind}). Now, if we look for the scaling solution then the lhs vanishes; hence from Eq.~(\ref{ind}),
\bea
u^{(n+1)}=-\frac{F_n(u',u'',u''',...,u^{(n)})}{g(u')}.
\eea
Evaluating this expression at $\rho=0$ gives
\bea
\lambda_{n+1}=-\frac{F_n(\lambda_1,\lambda_2,\lambda_3,...,\lambda_n)}{g(\lambda_1)}.
\eea
From this expression the nesting is straightforward:
\bea\label{nest}
\lambda_{n+1}=-\frac{F_n(\lambda_1,\lambda_2,\lambda_3,...,\lambda_n)}{g(\lambda_1)}=-
\frac{F_n(m^2)}{g(m^2)}.
\eea 
Here, we used the notation $m^2\equiv\lambda_1$. Thus, the formula of the nested couplings in Eq.~\eq{nestedf} is proved for $N\to\infty$ and $D=2$.\\
Note that in the finite $N$ case there are terms in the initial RG flow like $\rho u'$ and $\rho u''$. 
Evaluating the equation at $\rho=0$ is crucial to be able to neglect these terms, which would prevent 
us from performing the nesting. Thus, expanding into Taylor series around zero is the only case when we can define VBFs.\\
In what follows, we will give a proof of the nested formula in Eq.~\eq{nestedf} for arbitrary field components $N$ and dimensions $D$.
Again, we set the constant $A_D=1$.
Let us start again by differentiating the RG equation [in this case Eq.~\eq{FeE}] with respect to $\rho$:
\bea
\partial_t u&=&(D-2) \rho  u'(\rho )-D u(\rho )+\frac{N-1}{u'(\rho )+1}+\frac{1}{2 \rho  u''(\rho )+u'(\rho )+1},\nn
\partial_t u'&=&(D-2) \rho  u''(\rho )+(D-2) u'(\rho )-D u'(\rho )-\frac{(N-1) u''(\rho )}{\left(u'(\rho )+1\right)^2}-\frac{2 \rho  u^{(3)}(\rho )+3 u''(\rho )}{\left(2 \rho  u''(\rho )+u'(\rho )+1\right)^2},\nn
\partial_t u''&=&(D-2) \rho  u^{(3)}(\rho )+2 (D-2) u''(\rho )-D u''(\rho )+(N-1) \left(\frac{2 u''(\rho )^2}{\left(u'(\rho )+1\right)^3}-\frac{u^{(3)}(\rho )}{\left(u'(\rho )+1\right)^2}\right)\nn
&&+\frac{2 \left(2 \rho  u^{(3)}(\rho )+3 u''(\rho )\right)^2}{\left(2 \rho  u''(\rho )+u'(\rho )+1\right)^3}-\frac{2 \rho  u^{(4)}(\rho )+5 u^{(3)}(\rho )}{\left(2 \rho  u''(\rho )+u'(\rho )+1\right)^2}.
\eea
Let us express the equations above in the following way:
\bea
\partial_t u'&=&F_1(u',\rho u'')+g(u')u''+h(u',2\rho u'')(3u''+2\rho u^{(3)}),\nn
\partial_t u''&=&F_2(u',u'',\rho u^{(3)})+g(u')u^{(3)}+h(u',2\rho u'')(5u^{(3)}+2\rho u^{(4)}),\nn
\eea
where 
\bea
F_1(u',\rho u'')&\equiv&(D-2) \rho  u''(\rho )+(D-2) u'(\rho )-D u'(\rho ),\nn
F_2(u',u'',\rho u^{(3)})&\equiv&(D-2) \rho  u^{(3)}(\rho )+2 (D-2) u''(\rho )-D u''(\rho )\nn
&&+(N-1) \frac{2 u''(\rho )^2}{\left(u'(\rho )+1\right)^3}+\frac{2 \left(2 \rho  u^{(3)}(\rho )+3 u''(\rho )\right)^2}{\left(2 \rho  u''(\rho )+u'(\rho )+1\right)^3},\nn
g(u')&\equiv&-\frac{N-1}{\left(u'(\rho )+1\right)^2},\nn
h(u',2\rho u'')&\equiv&-\frac{1}{\left(2 \rho  u''(\rho )+u'(\rho )+1\right)^2}.
\eea
We can establish the relation again just like in Eq.~(\ref{relat}):
\bea
F_2(u',u'',\rho u^{(3)})=\frac{\d F_1}{\d \rho}+\frac{\d F_1}{\d u'}u''+\frac{\d F_1}{\d u''}\rho u^{(3)} + \frac{\d g(u')}{\d u'} u''u''
+\frac{dh(u',2\rho u'')}{d\rho}\left(2 \rho  u^{(3)}(\rho )+3 u''(\rho )\right).\nn
\eea
We can make the following statement for $n\geq1$:
\begin{eqnarray}\label{tru}
\partial_t u^{(n)}&=&F_n(u',u'',...,\rho u^{(n+1)})+g(u')u^{(n+1)}\nn
&&+h(u',2\rho u'')\left((2n+1)u^{(n+1)}+2\rho u^{(n+2)}\right),\nn
F_n(u',u'',...,\rho u^{(n+1)})&=&\frac{\d F_{n-1}}{\d \rho}+\frac{\d F_{n-1}}{\d u'}u''+...+\frac{\d F_{n-1}}{\d u^{n}}\rho u^{(n+1)}\nn
&& + \frac{\d g(u')}{\d u'} u''u^{(n)}+\frac{dh(u',2\rho u'')}{d\rho}\left(2 \rho  u^{(n+1)}(\rho )+(2n-1) u^{(n)}(\rho )\right).\nn
\end{eqnarray}
Let us suppose that Eq.~(\ref{tru}) is true for the $n$th term. Now, we will show that it is true for the $(n+1)$th term:
\begin{eqnarray}
\partial_t u^{(n+1)}&=&\frac{\d F_n}{\d \rho}+\frac{\d F_n}{\d u'}u''+...+\frac{\d F_n}{\d u^{n+1}}\rho u^{(n+2)}+\frac{\partial g(u')}{\partial u'}u''u^{(n+1)}\nn
&&+g(u')u^{(n+2)}+\frac{d h(u',2\rho u'')}{d \rho}\left((2n+1)u^{(n+1)}+2\rho u^{(n+2)}\right) \nn 
&&+ h(u',2\rho u'')\left((2n+3)u^{(n+2)}+2\rho u^{(n+3)}\right),\nn
&=&F_{n+1}(u',u'',...,\rho u^{(n+2)})+g(u')u^{(n+2)}+ h(u',2\rho u'')\left((2n+3)u^{(n+2)}+2\rho u^{(n+3)}\right),\nn
\end{eqnarray}
where
\bea
F_{n+1}(u',u'',...,\rho u^{(n+2)})&=&\frac{\d F_n}{\d \rho}+\frac{\d F_n}{\d u'}u''+...+\frac{\d F_n}{\d u^{n+1}}\rho u^{(n+2)}\nn
&&+\frac{\partial g(u')}{\partial u'}u''u^{(n+1)}+\frac{d h(u',2\rho u'')}{d \rho}\left((2n+1)u^{(n+1)}+2\rho u^{(n+2)}\right).\nn
\eea
In this way, by induction, we could show that the statement of Eq.~(\ref{tru}) is true. Here (in the case of finite $N$), we have to set $\rho=0$ and only then is it 
possible to do the nesting like in Eq.~(\ref{nest}). From this point it is straightforward to show this.\\

\end{document}